\documentclass[review]{elsarticle}
\journal{IJMF}
\usepackage{graphicx}
\usepackage{placeins}
\usepackage{tabularx}
\usepackage{rotating}
\usepackage{xcolor}
\usepackage{textcomp}
\usepackage{amssymb,amsmath,amsbsy}
\usepackage{latexsym}
\usepackage{amsthm}
\usepackage{eucal}
\usepackage{bm}
\usepackage{bbm}
\usepackage{rotating}
\usepackage{array}
\usepackage{subcaption}

\newcolumntype{L}[1]{>{\raggedright\arraybackslash}p{#1}}
\newcolumntype{C}[1]{>{\centering\arraybackslash}m{#1}}
\newcolumntype{R}[1]{>{\raggedleft\arraybackslash}b{#1}}
\usepackage{lineno,hyperref}
\modulolinenumbers[1]

\usepackage{hyperref}
\hypersetup{
	colorlinks=true,
	linkcolor=blue,
	anchorcolor=black,
	citecolor=blue,
	urlcolor=blue,
}
\usepackage[utf8]{inputenc}
\usepackage{epsfig}
\usepackage{amssymb}
\usepackage{amsmath}
\usepackage{empheq}
\usepackage{booktabs}
\usepackage{float}
\usepackage{lineno,hyperref}

\usepackage{numcompress}
\biboptions{authoryear}

\begin{document}
	\begin{frontmatter}
		\title{Intermittency acceleration of water droplet population dynamics inside the interfacial layer between cloudy and clear air environments
		}
		\author[addr1]{Mina Golshan}
		\author[addr1]{Shahbozbek Abdunabiev}
		\author[addr1]{Mattia Tomatis}
		\author[addr1]{Federico Fraternale\texorpdfstring{\fnref{myfootnote}}}
		\author[addr1]{Marco Vanni}
		\author[addr1]{Daniela Tordella\texorpdfstring{\corref{mycorrespondingauthor}}}
		\cortext[mycorrespondingauthor]{Corresponding Author}
		\ead{daniela.tordella@polito.it}
		\address[addr1]{Dipartimento di Scienza Applicata e Tecnologia, Politecnico di Torino, 10129 Torino, Italy}
		\fntext[myfootnote]{Current Address: Center for Space Plasma and Aeronomic Research, University of Alabama in Huntsville, Huntsville 35805 (AL), USA.}
				
		\begin{abstract}
			The formation and life-span of clouds as well as the associated unsteady processes concerning the micro-physics of the water phases they may contain are open questions in atmospheric physics.				
			We here use three-dimensional direct numerical simulation to analyse the temporal evolution of a small portion of the top of a cloud. The Eulerian description of the turbulent velocity, temperature and vapor fields is combined with the Lagrangian description of two different ensembles of cloud droplets, that is, with a monodisperse and a polydisperse size distribution. A shear-free turbulent mixing layer is used to model the background air flow of the cloud top. This flow is considered appropriate because clouds cannot stand the presence of shear, which inevitably destroys them quickly. Luke-warm clouds are generally found at an altitude of 1000-2000 meters, live for a few hours or up to 1-2 days, continuously change shape, and have typical dimensions of some hundreds of meters. The global time-scale of these changes is recognized as being of the order of 100 seconds (Warhaft 2009). From the formation phase to the dying out phase, clouds live under a continuous sequence of transients that are slightly different one from the other.
				
			In this study, we have tried to reduce the simplification level with respect to the real warm cloud situation as much as possible. We have included the same level of supersaturation of warm clouds, the same amount of liquid water content, and thus, the same numerical number of water droplets, and finally, a typical unstable perturbation of the density stratification and a typical kinetic energy cloud / clear air ratio (order of 10). We have considered an observation duration of the order of a few seconds (about 10 initial turnaround times). During this time, the kinetic energy decays throughout the system by 95\%. It should be recalled that the kinetic energy inside the interfacial layer (the shear-free turbulent mixing layer that matches the cloud region to the ambient air region) also decays spatially, by nearly 85\%. We observed, with respect to the cloud region, in the interfacial layer, a five times faster achievement of a common value of standard deviation for the probability density of both the monodisperse and poly-disperse populations. This acceleration of the dynamics is remarkable and is somewhat counterintuitive. It is closely correlated with the intermittency of the small scale of the air flow and of the supersaturation fluctuation. We give information on the size distribution of both the positive and negative droplet growth and on the drop size and the corresponding numerical concentration value of the distribution peak as time passes. Finally, we comment on the extension of the concept of the collision kernel for an unstable and inhomogeneous system in which turbulence decays faster than the time scales of the involved aqueous phases.			
			
		\end{abstract}
	\end{frontmatter}

	\vspace{-2mm}
	\section{\textit{\textbf{Introduction}}}
	Atmosphere clouds are fascinating systems that host a rich and complex physics not yet completely known. They are still one of the major uncertainty affecting  reliable weather and climate forecasts. Many different methods of investigation are used to understand the multiple physical phenomenologies that regulate the life of clouds. The methods are in a continuous phase of development all over the world, which gives the index of the liveliness of research in this area. Whether it is field studies, or laboratory studies, or studies conducted by means of numerical simulations on machines capable of hosting High-Performance Computing, at state of the art, studies can only focus on sections or subsections of the physics involved.
	One aspect not yet understood is the fact that inside clouds, the kinetic energy is larger than in the clear air outside. Clouds behave as energy traps. The energy can be developed by inner physical-chemical processes as latent heat release by water drops nucleation and condensation or by turbulent energy amplification induced by unstable density stratification. The energy captured from acoustic-gravity waves propagating into clouds from below or above cloud layers, or from cosmic rays during their interaction with water drops, or from electromagnetic radiations from the Earth or from outside the atmosphere should be also taken into account. However, the introduction into a numerical simulation of all these facts is yet very difficult. For instance, compressibility must be included to account for internal acoustic,  gravity waves and baroclinicity effects, but efficient techniques to carry out compressible simulation of cloud at the evanescent relevant values of the Mach number have not been developed yet.
	
	Drops nucleate in clouds when gaseous water vapour condenses on a substrate into water. Usually, they have diameters of less than 30 microns and follow air streamlines. In any case, droplets interact with each other with a low probability of collision. 	The range of scales involved in the dynamics of clouds cannot yet be covered by fully resolved numerical simulations \cite{Atkinson1996}. The complexity of the multiscale cloud dynamics becomes fully apparent at the cloud boundary where air, water vapour, and droplets and less humid air, usually named as clear air, interact in a situation where turbulence is highly intermittent and anisotropic. Direct numerical simulations (DNS), which resolve the turbulence down to the finest scales, can help to associate turbulence dynamics to a simplified cloud microphysics model that includes droplet formation, growth, and interaction. In particular, inside 
	an atmospheric cloud, the shear-free mixing layer – one of the simplest set-ups of inhomogeneous turbulence - is considered a good model flow for their edges.  This layer forms when two homogeneous and turbulent regions with different mean kinetic energies are brought together and was studied in laboratory experiments, starting with \cite{Gilbert1980}, \cite{Veeravalli1989}, as well as in direct numerical simulation,  \cite{Knaepen2004}, \cite{Briggs1996} or \cite{Tordella2006}, \cite{Tordella2011}.
	
	In past literature, most simulations of lukewarm clouds, on average, assumed static and homogeneous conditions. We are interested in simulating more realistic regimes of warm clouds that actually are systems that live in perpetual transitional situations.
	
	In our simulation, cloud boundaries (called interfaces in the following) are modeled through the shear-less turbulent mixing matching two interacting flow regions - a small portion of cloud and an adjacent clear air portion of equivalent volume - at different turbulent intensity. An initial condition reproduces local mild unstable stratification in density and temperature. The droplets model includes evaporation, condensation, collision, and coalescence. 	We  investigate the effect of transient anisotropic  turbulence on two different populations of water droplets initially randomly positioned in the cloud region. We implement both a mono-disperse and a poly-disperse population of particles. For the collision model, unlike \cite{franklin2005} (phantom collision model), we use a geometrical collision model combined with condensation- evaporation growth-decay.
	The paper is organized as follows: Section 2 provides a general description of the physical model for cloud droplets and cloud turbulence and the methodology used for this study. Section 3 describes the statistical results concerning the drop size distribution temporal evolution. Section 4 presents a preliminary investigation on the workability of obtaining from the numerical simulation of a fast time decaying turbulent shear-free layer a collision kernel. Conclusions and outlook follow in Section 5.
	
	\vspace{-3mm}
	\section{\textit{\textbf{The physical system}}}
	
	\subsection{\textit{\textbf{Turbulent air flow, temperature and water vapor mixing ratio fields}}}
\begin{figure}[bht!]
		\centering 
		\begin{subfigure}[t]{1.0\textwidth}
			\includegraphics[width=\linewidth]{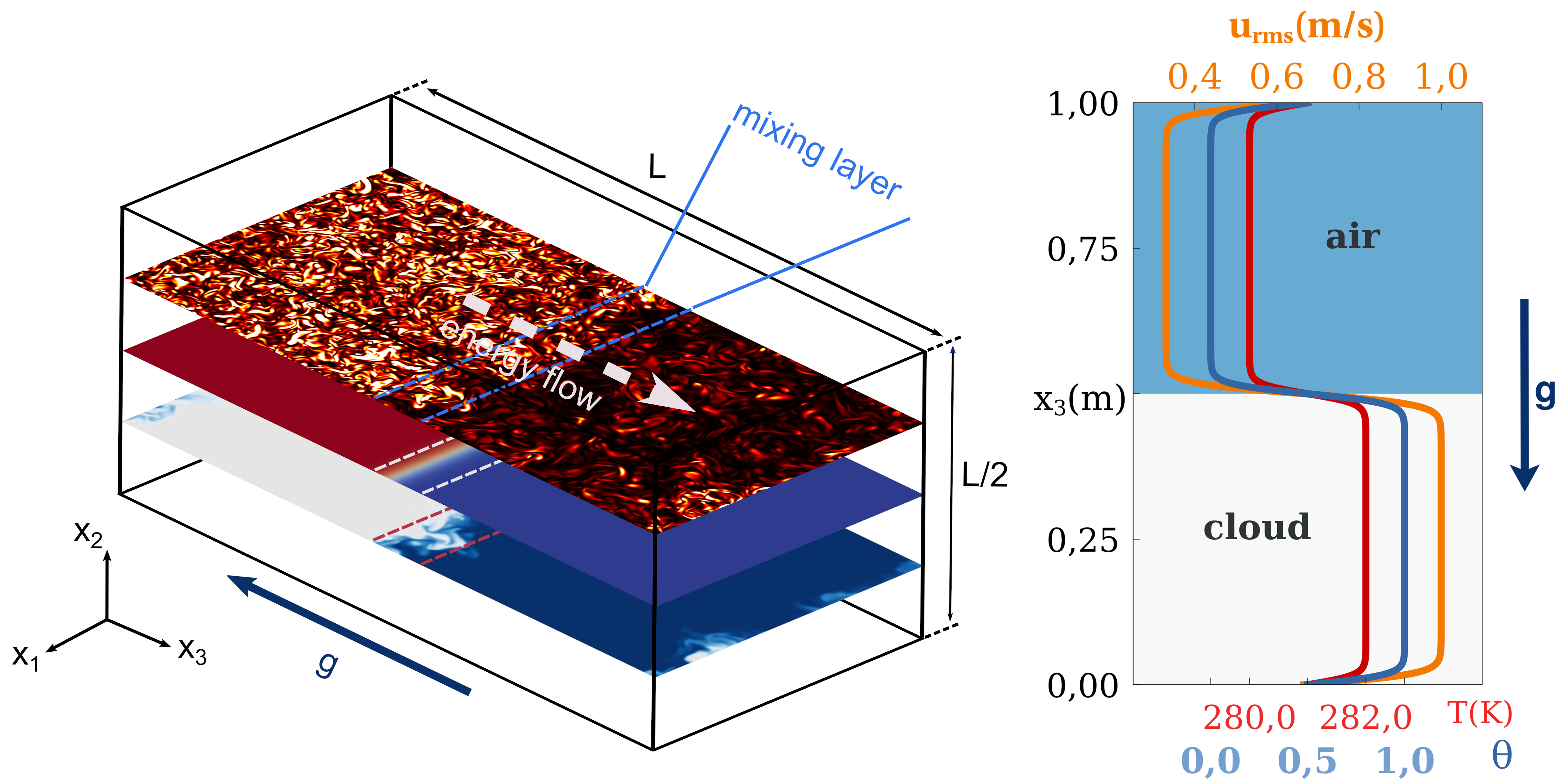}
			\caption{Schematics of the simulation domain (left panel) and of the initial profiles of the {\it rms} velocity (orange), temperature (red) and vapour content (blue) (right panel). The turbulent kinetic energy flow is from botton to top along $x_3$ direction,  $E_1/E_2=10$.} 
		\end{subfigure}
		\vspace{3mm}
		\begin{subfigure}[t]{1.0\textwidth}
			\includegraphics[width=0.8\linewidth]{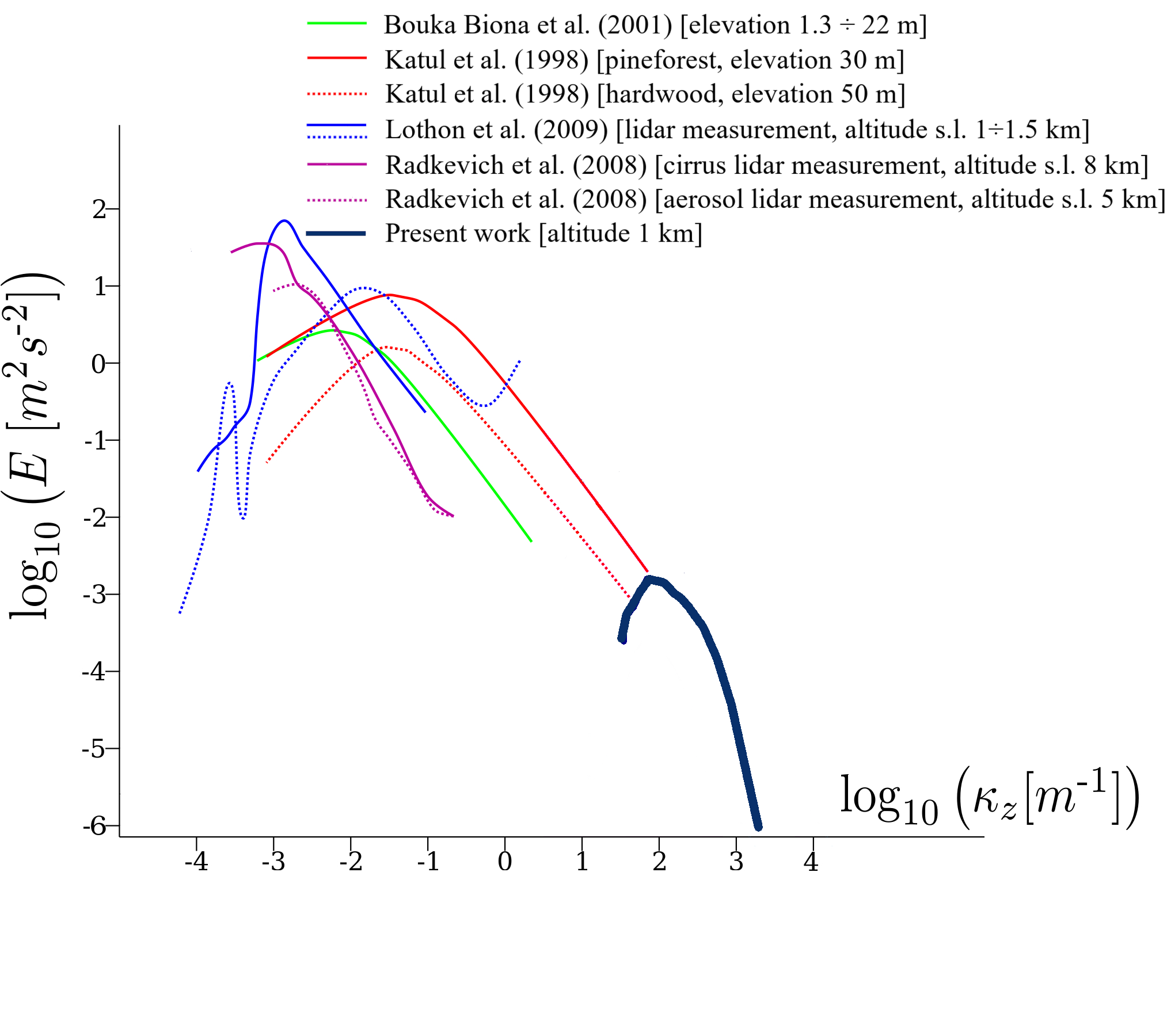}
			\caption{Three dimensional kinetic energy spectrum of the atmosphere turbulence observed in a set of infield measurement campaigns. In blue, to the extreme right, the part of the spectrum relevant to the present simulations.}
		\end{subfigure}
		\caption{\bf Overview of the physical system, cloud -  clear air transient interaction, and of a few relevant averaged and spectral physical properties.}
		\label{figure1}
	\end{figure}

	Our simulations focus on regimes of warm cumulus clouds, which systems that are in constant transition. Cloud boundaries are represented through a shear-less turbulent mixing. This flow is considered a good model for several reasons:  as clouds, it is intrinsically  non-steady, it may accommodate an integral scale gradient parallel to that of kinetic energy and enstrophy, its   intrinsic  anisotropy includes the small scales of the turbulence. In fact, the moment tensors of the velocity fluctuation derivative have main diagonals with different values of their terms (\cite{Tordella2011}). The decaying shearless mixing is fundamentally simple because it is free of the turbulence production due to the presence of a mean shear, which is a typical situation of the life of clouds. The presence of a mean shear in fact 
	causes atmospheric clouds to dissolve. 	For the flow schematic, please, see Figure \ref{figure1}. 
	
	Shearless velocity fluctuation mixings are easily generated in 2D and 3D numerical simulations by exploiting periodical boundary conditions. In practice, they are produced by the interaction of two initially homogeneous isotropic turbulent flows (HIT) with different levels of (i) turbulent kinetic energy (\cite{Knaepen2004}, \cite{Briggs1996}, \cite{Tordella2006}, \cite{Tordella2008}, \cite{Tordella2011}), (ii) temperature \cite{iovieno2014}, \cite{kumar2014}), (iii) intertial particles (\cite{ireland2012}), also in the presence of supersaturation  (\cite{Gotzfried2017}). This configuration has been studied also in laboratory experiments, starting with \cite{Gilbert1980} and \cite{Veeravalli1989}, where only mono-phase fluid turbulence was considered, to configurations where inertial particles were present (\cite{good2012}, \cite{gerashchenko2011entrainment}).
	
	The  simulation parameters  match those of cloud regions close to  borders, see Tables \ref{tab:Table1} and \ref{tab:Table2}. 
	 The governing equations are the incompressible Navier-Stokes ones, used with the Boussinesq approximation for both temperature and vapour density, and active scalar transport equations for the water vapour and the thermal energy. Inertial water drops are represented via a Lagrangian approach, including Stokes drag and gravitational settling. This model is coupled to the vapor and temperature equations through their respective evaporation-condensation source terms. We follow the drop position, velocity, and radius. This is a one-way coupling approach and does not include feedback from droplets to the fluid airflow field.
	
\begin{figure}[bht!]
		\centering
		\begin{subfigure}[t]{0.47\textwidth}
			\includegraphics[width=\linewidth]{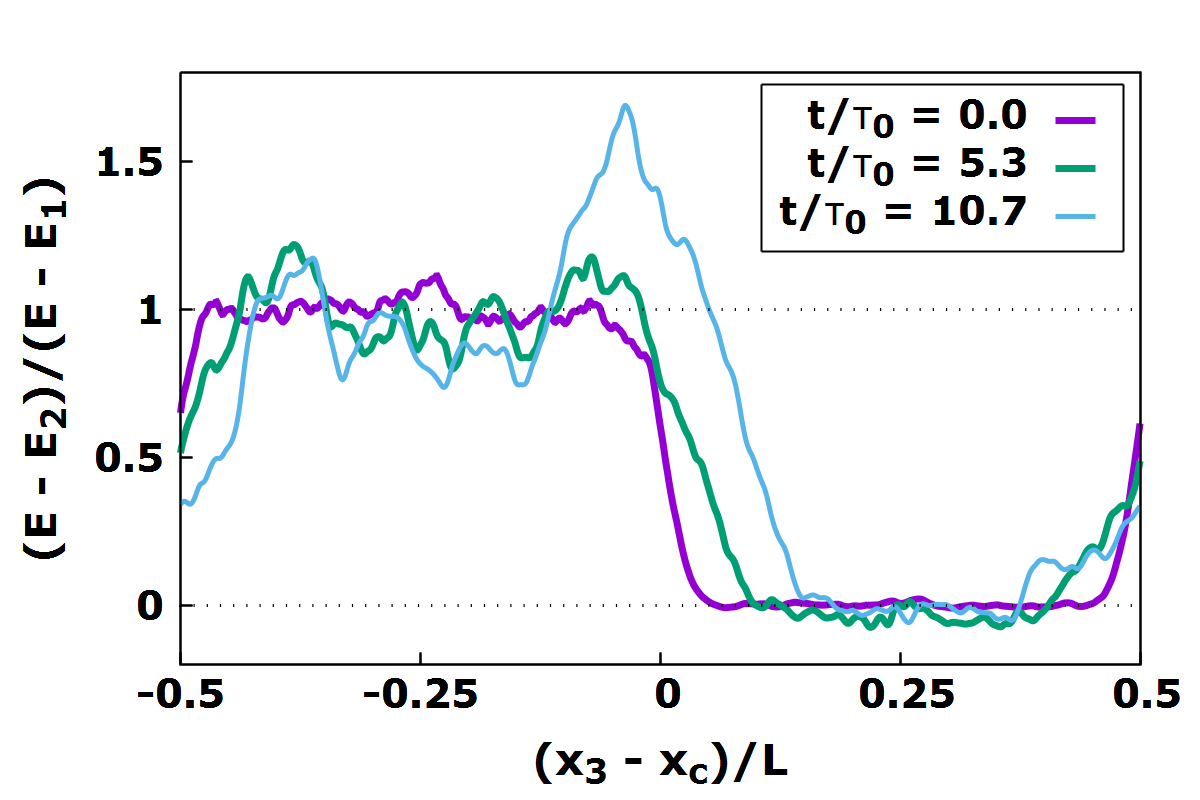}
			\caption{}
		\end{subfigure}
		\hspace{0.03\textwidth}
		\begin{subfigure}[t]{0.47\textwidth}
			\includegraphics[width=\linewidth]{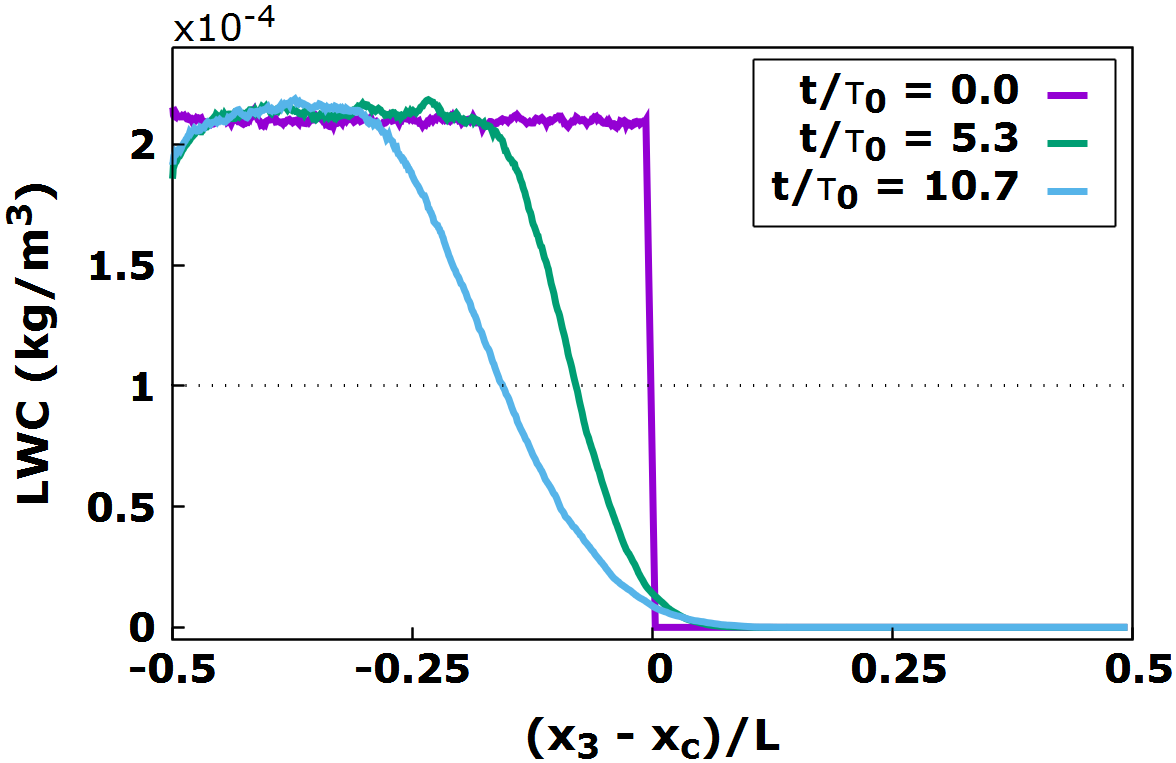}
			\caption{}
		\end{subfigure}
		
		\smallskip
		
		\begin{subfigure}[t]{0.47\textwidth}
			\includegraphics[width=\linewidth]{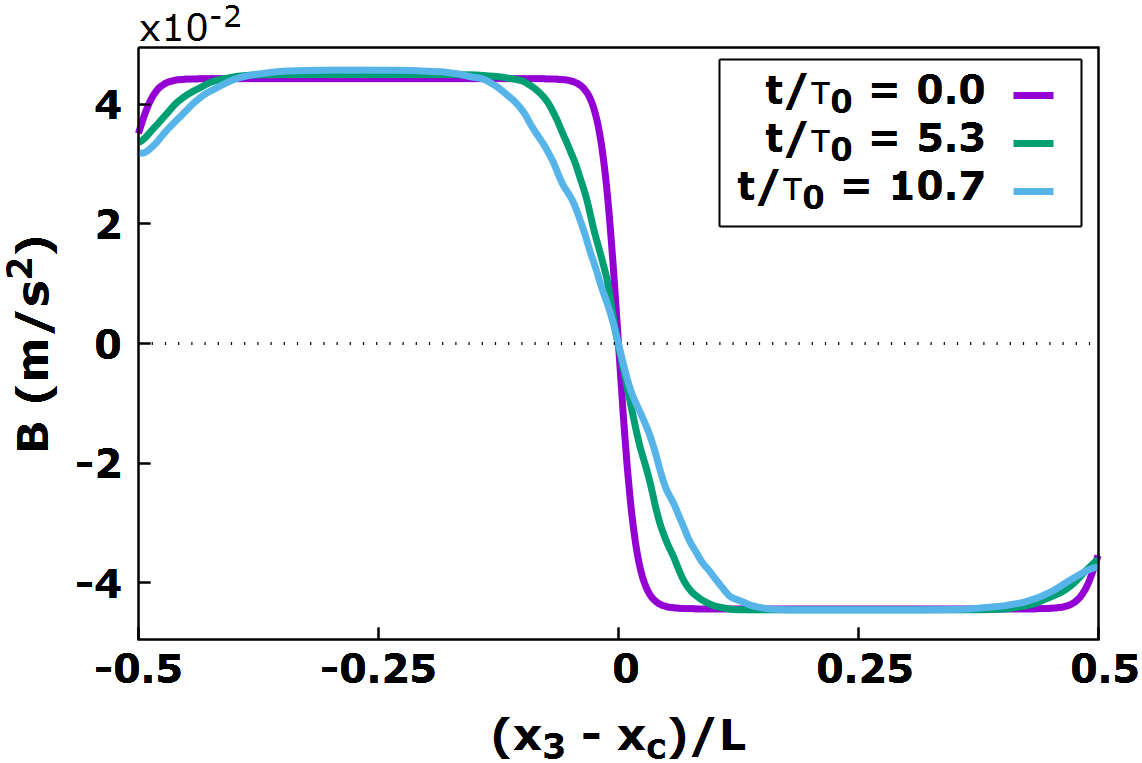}
			\caption{}
		\end{subfigure}
		\hspace{0.03\textwidth}
		\begin{subfigure}[t]{0.47\textwidth}
			\includegraphics[width=\linewidth]{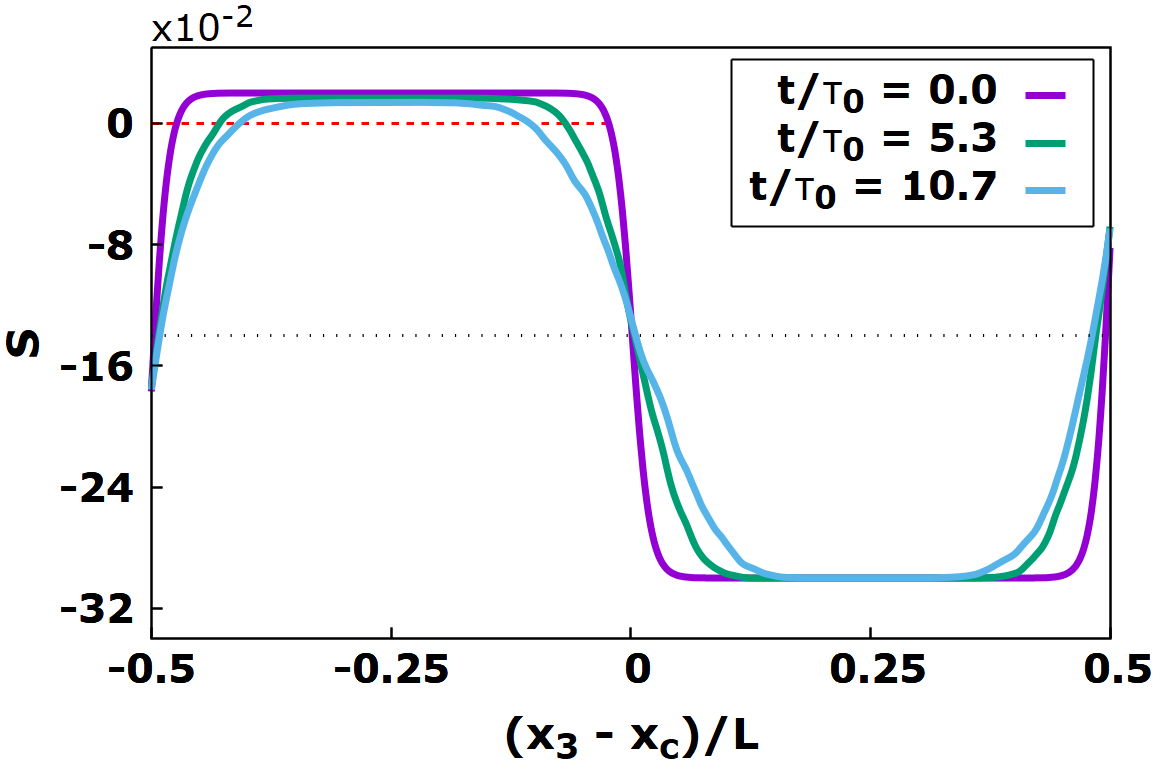}
			\caption{}
		\end{subfigure}
		
		\begin{subfigure}[t]{0.47\textwidth}
			 \includegraphics[width=\linewidth]{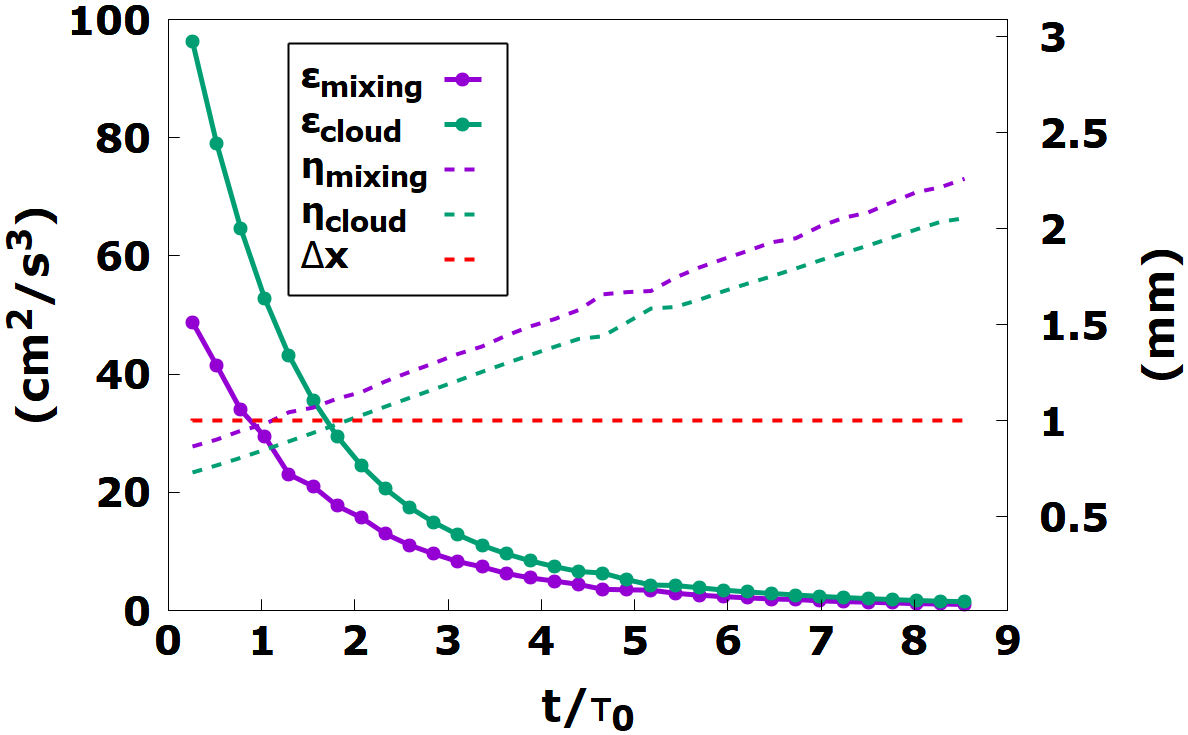}
			\caption{}
		\end{subfigure}
		\hspace{0.03\textwidth}
		\begin{subfigure}[t]{0.47\textwidth}
			\includegraphics[width=\linewidth]{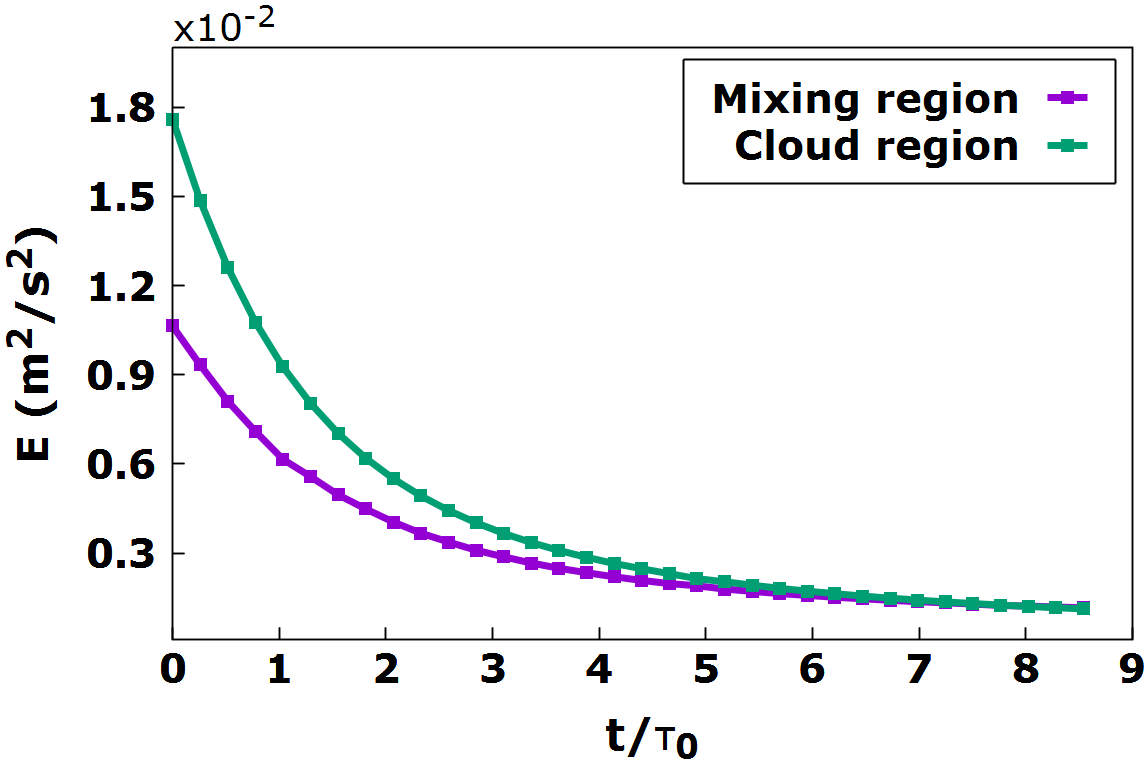}
			\caption{}
		\end{subfigure}
		\caption{\bf Kinetic energy, Liquid Water Content $(LWC)$ for the polydisperse droplet population, buoyancy and supersaturation mean values along the in-homogeneous direction at three stages along the temporal evolution.  In physical non normalized terms, the evolution lasts a few seconds ($\tau_0=0.42$ s, see Table 2). The top left panel shows the turbulent energy excess with respect to the clear-air part, normalized with the difference difference between the two regions ($E_1 = 10 \cdot E_2$) at $t = 0$.}
		\label{fig:initial_conditions}
	\end{figure}

	The size of the computational domain is $0.512 m \times 0.512 m \times 1.024 m$ and is discretized by using $512 \times 512 \times 1024$ grid points.  
	Since the turbulence intensity, and thus the dissipation rate, decay in time, the small scales, in particular the Kolmogorov scale $\eta_k$ grow in time, see Fig. \ref{fig:initial_conditions} . This allows the grid size of 1 mm to be well below  $\eta_k$ during most part of the transient decay, and nearly equal to $\eta_k$ at the  simulation begin, in particular inside the two first eddy turn over times.
	
	A synthetic divergence-free field with a $-1.67$ slope power spectrum in the inertial range and an exponential tail in the dissipation range (random phases) is used to build the initial condition for the velocity field. A preparatory simulation  with an initial dissipation of $\epsilon\approx500$ cm$^2$/s$^3$ for the cloud region was set and the field  was let to evolve for one eddy turn over  time (1100 iterations) until it reached the dissipation of $\epsilon\approx130$ cm$^2$/s$^3$. This field was then used to build the initial condition where an energy ratio of 6.7 was arranged between the cloud and clear air regions, as well as different levels of temperature and supersaturation, see Table \ref{tab:Table1}.
	
	Model equations for the fluid flow are solved using the Fourier-Galerkin (FG) pseudo-spectral method. The temporal integration uses a four-stages  fourth-order explicit Runge-Kutta scheme in the low storage version by Jameson, Schmidt and Turkel (1981) with exponential integration of the diffusive terms, see \cite{iovieno2001}. The numerical code uses a one-dimensional slab parallelization and Message Passing Interface (MPI) libraries.
	
	Similar to previous Direct Navier Stokes numerical simulation models, \cite{kumar2014}, \cite{Gotzfried2017}, our code is neglecting compressible effects and is based on the incompressible Navier-Stokes equations under the Boussinesq approximation, where both the vapor mixing ratio field $q_v(x_j,t)$, which is given by the vapour density $\rho_v$ referred to the dry air density, $q_v(x_j,t) = \rho_v / \rho_0$, and the temperature $T(x_j,t)$ are transported as passive scalars. The governing equations are given for the turbulent velocity field $u_i(x_j,t)$, the temperature field $T(x_j,t)$, the pressure field $p(x_j,t)$ and the vapor mixing ratio $q_v(x_j,t)$. In the following, indices $i,j,k$ are used within the Einstein convention.
	\begin{align}
	&\frac{\partial u_i}{\partial x_i}=0
	\label{eq.cont}\\
	&\frac{\partial u_i}{\partial t}+u_j \frac{\partial u_i}{\partial x_j} = - \frac{1}{\rho_0} \frac{\partial p}{\partial x_i}+ \nu \frac{\partial^2 u_i}{\partial x_j^2} - \mathcal{B} \, \delta_{z i}, 
	\label{eq.qdm}\\
	&\frac{\partial T}{\partial t}+u_j \frac{\partial T}{\partial x_j} = \kappa \frac{\partial^2 T}{\partial x_j^2} + \frac{\mathcal{L}}{c_p} C_d
	\label{eq.T}\\
	&\frac{\partial q_v}{\partial t}+{\bm u_j}\frac{\partial q_v}{\partial x_j} = \kappa_v \frac{\partial^2 q_v}{\partial x_j\partial x_j} - C_d 
	\label{eq.rhov}
	\end{align}
	
	\noindent Here, $\nu$ is the kinematic viscosity of air, $g$ the gravitational acceleration, $\rho_0$ is the reference value for the density of dry air, $c_p$ the specific heat at constant pressure, $\mathcal{L}$ the latent heat ($2.48\cdot10^{6}$  $\rm{J\,kg^{-1}}$), $k$ the temperature diffusivity, $D$ the diffusivity of the vapor mixing ratio. $C_d$ and $\mathcal{B}$  are the condensation rate field and buoyancy field, respectively.
	
	Upon the introduction of the volume average $ < \dot >$, an average computed on the slice of domain normal to the vertical direction, i.e. the $x_3$ direction, and thickness equal to the distance between two consecutive grid steps, the temperature fluctuations $T'$ are given by
	
	\begin{equation}
	T'(x_j,t)=T(x_j,t)-<T(x_3, t)>
	\end{equation}
	
	\noindent where  the volume averaged temperature is actually constant in time and equal to the sum of the temperature $T_0$, the average   over the entire domain, and a linear background negative variation   which sets the unstable stratification, thus $<T(x_3)> = T_0 + G x_3$, see Table \ref{tab:Table1}. 
	The initial temperature field term $T'$ depends only on the $x_3$ coordinate and has an hyperbolic tangent representation. For details, see Section 2.3.
	
	
	The vapor mixing ratio fluctuation is are given by
	\begin{equation}
	q'_v(x_j,t)=q_v(x_j,t) - <q_v(t)>.
	\end{equation}
	\noindent In this case,  the volume average is time dependent.
	
	The buoyancy field $\mathcal{B}$ in the momentum equation (\ref{eq.qdm}) depends on the temperature field $T(x_j,t)$ and the vapor mixing ratio field $q_v(x_j,t)$ and is defined as:
	
	\begin{equation}
	\mathcal{B} = g [T'/<T> + \alpha q'_v]
	\label{eq.buoyancy}
	\end{equation}
	where $\alpha = M_a /M_v - 1 = 0.608$ and $M_a$ and $M_v$ are the dry air and  vapor molar masses, respectively, see  \cite{Saito2018}.
	
	In this model, droplets affect the evolution of the fluid motion through the condensation term $C_d$ in \ref{eq.T} and \ref{eq.rhov}. The condensation rate field $C_d = C_d (x_i, t)$ is defined as time derivative of the mass of liquid water, $m_l$, contained within each $\Delta^3$ volume cell surrounding the grid point $x_i$, 
	referred to the mass of dry air $m_a$, 
	\cite{vaillancourt2001, Vaillancourt2002}. Since cloud droplets are advected by the turbulent flow, $C_d$ must be determined in the Lagrangian frame of reference used for the liquid water mixing ratio, which is described below in sub-section 2.2. 
	
	However, for the use in equations \ref{eq.T} and \ref{eq.rhov}, $C_d$ must be in turn rendered into the Eulerian frame of reference. The condensation rate field is determined as:
	
	\begin{align}
	C_d(x_i,t) = \frac{1}{m_a}\frac{d m_l(x_i,t)}{d t} = \frac{4 \pi \rho_l K_s}{\rho_0 \Delta^3} \sum_{j=1}^{N_\Delta} \mathrm{R}_j(t) S(\bm{X_j}(t),t)
	\label{eq.cond-rate}
	\end{align}
	
	\noindent where $m_a$ and $m_l$ are the air mass and liquid mass per grid cell, $\rho_l$ is density of water, $\rho_0$  is reference density of dry air, $\mathrm{R}_j(t)$ and $\bm X_j(t)$ are the radius and vector space coordinate of the $j-th$ drop contained inside the grid cell, respectively. $N_\Delta$ represents the number of drops inside each grid cell, $S$ is supersaturation described below, see equation \ref{eq:Tetens}, and $K_s$ is a temperature and pressure dependent diffusion coefficient that includes the self-limiting effects of latent heat release. In literature, for typical warm cloud conditions where the characteristic heat flux due to latent heating from a small variation in the droplet temperature is of the same order of the heat flux due to thermal conduction for the same temperature difference, this diffusion coefficient is  considered to be constant because its temperature dependence is weak ($K_s$ value in m$^2$ s$^{-1}$ ranges from $5.07 \cdot 10^{-11}$ at $T=270$ K, to $1.17 \cdot 10^{-10}$ at  $T=293$ K), see for instance \cite{Rogers1989}, \cite{gotoh2016}
	\cite{kumar2014}). In agreement to our volume averaged initial temperature of $281$ K, we used the value $8.6 \cdot 10^{-11}$ m$^2$ s$^{-1}$. The interpolation of Eulerian field values at grid points to the position occupied by the water droplets inside the cell is done via second order Lagrange polynomials. An inverse procedure is then used for the calculation of the condensation rate, which is determined at a first step at each droplet position and then relocated to the closest among the eight grid vertices. A collision is supposed to occur when the distance betweeoplet centers is equal or below to the sum of their radii. Collisions are assumed to be completely inelastic.
	
	\vspace{-2mm}
	\subsection{\textit{\textbf{Lagrangian Droplet Dynamics}}}
	In our simulations, cloud droplets are assumed to be point particles. Therefore they are always smaller than the grid size. The liquid water component is modelled as a Lagrangian ensemble of $N$ point-like droplets. A collision is supposed to occur when the distance between droplet centers is equal or below to the sum of their radii. Collided particle coalesce. The resultant particle has a volume equal to sum of the  collided particles and keeps as identity the smaller ID number. The particle with the greater ID number is removed  form the computational domain. Collisions are assumed to be completely inelastic. We consider two different initial size distributions: a mono-disperse initial distribution of particles of size equal to $15\mu$m and an initial multi-disperse distribution of droplets of radii from 0.6 $\mu$m to 30 $\mu$m. 
	It should be noted, that similarly to what done for the condensation rate field, Eulerian flow field quantities have to be determined at the droplet position to numerically proceed with Lagrangian equations. In this concern, we must highlight that we adopt a simplified  feedback on the flow by droplets. The direct effect of the liquid droplet drag on the velocity field is neglected in the buoyancy term in the momentum equation. The feedback is therefore indirect and is confined to the coupling of the temperature field with the velocity field and the vapour mixing ratio through the condensation rate. The rationale for this position relies on the smallness of the drop Stokes numbers (drop Reynolds number much less than $1$) and liquid mass loading.
	
	The Lagrangian evolution for the i-th cloud droplet are given by
	\begin{align}
	&\frac{d {\bm X}_i}{d t} = {\bm v}_{p_i}
	\label{eq:part_momentum1}
	\\
	&\frac{d{\bm v}_{p_i}}{d t}= \frac{{{\bm u_i}({\bm X_i},t) - \bm v}_{p_i}}{\tau_i}+\left( 1-\frac{\rho_a}{\rho_w}\right){\bm g},
	\label{eq:part_momentum2}
	\end{align}
	where $\bm {v}_{p_i}$ is the droplet velocity; $\rho_w,\rho_a$, are the densities of water and air, respectively; $\bm {{u}({x}_i,t)}$ denotes flow velocity at the position of the i-th particle and $\tau_i$ is the droplet response time. This time scale is defined by the Stokes drag coefficient and is adjusted to the droplet radius dynamical evolution, $\mathrm{R}_i = \mathrm{R}({\bm X}_i,t)$. Therefore
	\begin{equation}
	\tau_i = \frac{2}{9}\frac{\rho_w}{\rho_a}\frac{\mathrm{R}_i ({\bm X}_i, t)^2}{\nu}.
	\label{eq:part_drag}
	\end{equation} 
	where $\nu$ is the air kinematic viscosity. If the droplet radius becomes smaller than the critical value where the response time is lower smaller than the numerical integration time, the droplet is removed. This situation never applies for the monodisperse population. For the polydisperse population, this droplet removal is negligible, less than 1/1000 with respect to the initial liquid water content (see Table 1, LWC = $ 0.79 g / m ^ 3 $).
In this droplet model, we neglect a few other terms that can be of importance even when the Reynolds number is below unity. In particular, we neglect Faxen's correction associated to the velocity curvature effect on the drag, the added mass, the pressure gradient term and the Basset history force. In our simulation condition, where the gas and particle density ratio is of order $10^{-3}$, these forces are really negligible, as previously shown by many studies, see among others, \cite{Armenio2001}, \cite{Bergougnoux2014}. 
	
	In this investigation, the droplet growth is governed by three processes: condensation, evaporation and full coalescence after collision. Therefore, the numerical model for the growth of the particles must be coupled with the Lagrangian tracking of each droplet.
	
	For the growth-by-condensation/evaporation, we use a model based on the Kohler theory, which includes the spontaneous growth of cloud condensation nuclei (CCN) into cloud droplets under supersaturation water vapour conditions (\cite{Kohler1936}, \cite{howell1949} \cite{pruppacher1997Book}, \cite{Seinfeld1998}). A simplified form of this model was also used by \cite{vaillancourt2001}, \cite{kumar2014}, \cite{gotoh2016}, \cite{Gotzfried2017} and \cite{Gao2018} for particles with size larger than CCN.
	
	Fundamentally, as droplets are advected by the fluid where they can grow or evaporate in response to the local vapor field, the vapor mixing ratio is coupled to the droplet growth-decay through the supersaturation $S$ which is defined in terms of the vapor mixing ratio and the saturation vapor mixing ratio as
	\begin{equation}
	S\left(\bm X_i, t\right) = \frac{q_v(\bm X_i,t)}{q_{vs}(T)} - 1.
	\label{eq:supersat}
	\end{equation} 
	
	The saturation vapor mixing ratio $q_{vs}(T)$ at the droplet position is obtained from the Tetens formula (\cite{Tetens1930}):
	
	\begin{equation}
	q_{vs}(\bm X_i, t) = \frac{e_s(T)}{R_v \rho_0 T} = \epsilon_0 \frac{610.78}{\rho_0 T} exp [17.63 \frac{T - 272.16}{T - 35.86}]  
	\label{eq:Tetens}
	\end{equation} 
	
	\noindent where $e_s$ is the saturation pressure, and  $\epsilon_0=R_a/R_v \sim 0.62$  is the ratio between the gas constants for dry air and water vapor, $R_a$ and $R_v$, respectively.  For temperature above $273.16$ K, see also \cite{Monteith2008}.
	
	The curvature of the droplet surface induces the so called Kelvin effect on the evaporation rate. The  bonding strength  between  water molecules lying on the droplet surface and its neighbors is lowered  by the surface curvature. Therefore when the curvature is high, because the radii are small,  the probability that  water molecules may leave the liquid phase  becomes higher.  This increases  the evaporation rate. 
	
	Furthermore, aside water droplets, the atmosphere contains  many other kinds of solid, or soft matter, or liquid, particles. Some of these are hydrophilic and water soluble. The effect of soluble CCN on the water evaporation rate is called the Raoult effect. The Kelvin and Raoult effects, the curvature and the solute effects, can be included in the model for the droplet growth. We follow \cite{Hudson1996} and \cite{Shi2011} and \cite{Saito2018} and write:
	
	\begin{equation}
	\mathrm{R}_i\frac{d{\mathrm{R}_i}}{d{t}}= K_s\left(S-\frac{A}{\mathrm{R}_i}+\frac{B r_d^3}{{\mathrm{R}_i^3}}\right)
	\label{eq:radius_growth}
	\end{equation} 
	
	\noindent Here, the diffusion coefficient $K_s$ has been introduced above in relation to the condensation rate field $C_d$, see equation (\ref{eq.cond-rate}). The constant terms $A$ and $B$ represent the curvature (surface tension) and solute effects, respectively and $r_d$ the dry particle radius. Term $A$ directly depends on the surface tension of water ($\sigma_w = 72.75  \cdot 10^{-7} [$ J cm$^{-2}]$), and of course also on the density, the gas constant for water vapor and the local temperature of the air phase. While, $B$, aside the water and molecular weight of water, depends on the mass of the solute particle, its molecular weight, and the total number of ions the solute molecule dissociates into. Here, we follow \citet*{Saito2019} and assume that the solute dissolved in each drop is an inorganic hygroscopic substance like ammonium sulfate, sulphuric acid or lithium chloride which have a solubility parameter $B$ close to $0.7$ and an accumulation mode with modal diameters $r_d$ in the range from 10 to 50 nanometers (fine mode, observed North Atlantic marine air masses, see \cite{Ovadnevaite2017} and \cite{Hudson1996} and \cite{Jensen1984}, \cite{Flossman1985}). 
	For an air phase temperature nearly constant and close to 281 K we have $A = 1.15 \cdot 10^{-7} $ cm and $B = 0.7 \cdot 10^{-18}$ cm $^3$. 
	
	\vspace{-2mm}
	\subsection{\textit{\textbf{Initial and boundary conditions for the flow and scalar fields}}}
	
	The interaction between  two homogeneous isotropic time decaying turbulence fields differing in only one property, the kinetic energy level, produces the simplest anisotropic turbulent flow. The simplicity of this flow lies in the absence of the average velocity gradient, 
	which means that there is no production of turbulent kinetic energy and no mean convective transport. All interaction is the result of the fluctuating pressure and velocity fields. The two interacting flows are identical apart from the kinetic energy content, which sets a ratio (and thus a gradient) of kinetic energy across the layer. Since it can be shown that the integral length scale of a turbulence field can be independent of its kinetic energy, it is possible to obtain, numerically, an inhomogeneity in the kinetic energy of two HIT fields while maintaining homogeneity in the length scale,  \cite{Batchelor1953}. 
	
	The computational domain is a parallelepiped  where periodic boundary conditions in all directions are imposed, see Figure\ref{figure1}, panel (a). In this nominally infinite domain the Navier-Stokes and passive scalar equations are solved with a fully dealiased (3/2 rule) Fourier-Galerkin pseudospectral method. Time integration is performed using a fourth order explicit Runge Kutta scheme. A parallelised version of the code for the velocity field is presented in \cite{iovieno2001}, for details see also section Software (Incompressible Turbulent Flows) in the web pages  \it{ www.polito.philofluid.it}. 
	\normalfont
	The initial conditions are generated by building a homogeneous isotropic velocity field within a volume $ 2 \pi \cdot 2 \pi \cdot 2 \pi $, see \cite{Wray1998}. To create the initial condition, the velocity field is repeated creating a $ 4 \pi \cdot 2 \pi \cdot 2 \pi$  domain. In one side of the domain, each velocity component is multiplied by a constant, thus creating a ratio of energy between the fields, but keeping similar spectra and thus introducing no ratio of integral scales. 
	
	A hyperbolic tangent function is then used to smooth the interface and to define the initial mixing layer. This transition layer represents 1/40 of the $4\pi$ domain. The matched field is 
	
	\begin{equation}
	\bm u(\bm x)=\bm u_{1}({\bm x})p(x_3) + \bm u_{2}({\bm x})(1-p(x_3))
	\label{u0}
	\end{equation}
	
	\begin{equation}
	p(x_3)=\frac{1}{2}\left[1+\tanh\left(a\frac{x_3}{L} \right)\tanh\left(a 
	\frac{x_3-L/2}{L}\right)  \tanh\left( a\frac{x_3-L}{L} \right)\right]
	\label{peso}
	\end{equation}

	\noindent where the suffixes $1,2$ indicate high and low energy sides of the cloud interface model, respectively. Direction $x_3$ is the in-homogeneous direction and $L$ is the width of the computational domain in the $x_3$ direction.
	Constant $a$ in (\ref{peso}) determines the initial mixing layer thickness $\Delta$, 
	conventionally defined as the distance between the points with normalized energy values 0.25 and 0.75 when the low energy side is mapped to zero and the high energy side to one. When $ a = 12 \pi $ the initial ratio $\Delta/L$ is about $0.026$, a value that has been chosen so that the initial thickness is large enough to be resolved but small enough to have large regions of homogeneous turbulence during the simulations.
	
	
	
	The same technique is used to generate the periodical part $T'$ of temperature field  
	$$T'(x_3,0)= \Delta T \cdot \left[tanh \left(55\left(\frac{x_3}{L}-\frac{1}{2}\right)\right)-\frac{2 x_3 }{L}+1\right]$$
	and of the water mixing ratio field, which are taken as non fluctuating fields at the initial time instant. See in Figure \ref{figure1}, in the right side of panel(a),  a generic dimensional representation  of the mean values of the temperature, water mixing ratio and root mean square of the air velocity. In this regards, see also Fig.\ref{fig:initial_conditions}, where average values along the inhomogeneous direction $x_3$ of the kinetic energy, {\em LWC},  ${\mathcal B}$, $ S$ and $\rm \epsilon $ are shown at different stages along the temporal evolution.
	
	In Figure \ref{figure1}, panel (b), we show the 3D kinetic energy spectra of the high energy homogeneous turbulent region of our system (dark blue line, ranging from $k_3 = 25 $ to $k_3 = 1570$ [m$^{-1}$]). This region  represents the small portion of cloud interacting with the clear air lying on top of it. In the figure, this spectrum is compared with a few 3D spectra obtained by infield measurement campaigns carried out in the lower atmosphere: that is in the range from a few decades of meters (over pine and hardwood forests) to a few kilometers of altitude (cirrus and aerosol lidar measurements) and extending over Earth surfaces with linear dimension of the order of the atmospheric turbulence macroscale, see \cite{BOUKABIONA2001135}, \cite{Katuletal1998}, \cite{lothon2009doppler}, \cite{radkevich2008scaling}.  
	
	In Tables \ref{tab:Table1} and \ref{tab:Table2}, the reader can find the parametrization used in the present Direct Numerical Simulations. The relevant physical and thermodynamics constants are gathered in Table 1, while domain specifications, computational grid structure, turbulence scales, field control parameters and water droplet population information are presented in Table \ref{tab:Table2}. Ensemble average realizations where obtained under exactly identical physical conditions, by simply rotating the box of the initial HIT velocity field in the high energy cloud region (homogenous by definition in this study). In such a way, the statistical properties of the initial condition remain unchanged. For instance: run 1 -- initial condition as obtained from the output of the computation synthetic divergence-free field (see last paragraph of page 4, new version), run 2 – initial condition equal to the run 1 field but now rotated by 90 degrees around the $x_3$ axis, run 3 - initial condition equal to the run 1 field but now rotated by 90 degrees around the $x_2$ axis, etc.

	\begin{table}[bht!]
		\caption{\textbf{List of thermodynamics constants and flow field parameters and their corresponding values in the present DNS}}
		\centering
		\resizebox{0.99\textwidth}{!}{
			\begin{tabular}{lccc}
				\toprule
				\textbf{Quantity} & \textbf{Symbol}	& \textbf{Value}  & \textbf{Unit}\\
				\midrule
				Latent heat of evaporation  & $\mathcal{L}$ & $2.48\cdot10^{6}$  &$\rm{J\,kg^{-1}}$\\
				Heat capacity of the air at constant pressure  &$c_p$  & $1005$ &$\rm{J\,kg^{-1}\,K^{-1}}$\\
				Gravitational acceleration & ${g}$ & $9.81$ & m/s$^2$\\
				Gas constant for water vapour & $R_v$  & $461.5$  &\rm{J\,kg$^{-1}$\,K}\\
				Gas constant for air & $R_a$ & $286.7$& \rm{J\,kg$^{-1}$\,K}\\
				Diffusivity of water vapour &$\kappa_v$ & $2.52\cdot10^{-5}$ &$\rm{m^2\,s^{-1}}$\\
				Thermal conductivity of dry air & $\kappa$ &$2.5\cdot10^{-2}$  &$\rm{J\,K^{-1}\,m^{-1}\,s^{-1}}$\\
				Density of liquid water &$\rho_l$ &$1000$ &$\rm{kg\,m^3}$\\
				Dry air density, altitude 1000 m & $\rho_0$& $1.11$  & $\rm{kg\,m^{-3}}$\\
				Reference kinematic viscosity  & $\nu$ & $1.399\cdot10^{-5}$  & $\rm{m^2\,s^{-1}}$\\
				Entire domain average temperature & $T_0$ & 281.16 & $\rm{K}$\\
				Temperature in cloud region & $T_1$ & $282.16$  & $\rm{K}$\\
				Temperature in clear air region & $T_2$ & $280.16$  & $\rm{K}$\\
				Background temperature gradient & $G$ & -2 & $\rm{K/m}$\\
				Diffusion coefficient in eq.s 8 and 14 & $K_s$ & $8.6 \cdot 10^{-11}$ & m$^2$ s$^{-1}$\\
				Accumulation diameter & $r_d$ & $0.01\cdot10^{-6}$ & {m}\\
				Kelvin droplet curvature constant & $A$ & $1.15\cdot 10^{-7}$ & {cm}\\
				Raoult solubility parameter for inorganic hygroscopic\\ substances, like ammonium sulfate, lithium chloride, ... & $B$ & $0.7\cdot10^{-18}$ &$ \rm{cm^3}$\\
				Initial relative humidity inside cloud & $S$ (cloud) & 1.02 & - \\
				Initial relative humidity inside clear air & $S$ (clear air) & $0.7$ & - \\
				Saturation vapor mixing ratio at $T_1$ &$q_{vs}$(cloud) & $0.79\cdot10^{-2}$ &$\rm{kg\,m^{-3}}$\\
				Saturation vapor mixing ratio at $T_2$ &$q_{vs}$(clear air) & $0.69\cdot10^{-2}$ &$\rm{kg\,m^{-3}}$\\
				Water saturation pressure at $T_0=281$ &$e_s$& $1.061$ & kPa\\
				Molar mass of air  & $m_{air}$  &$28.96$ &$\rm{kg\,mol^{-1}}$\\
				Molar mass of water  & $m_{v}$  &$18$ &$\rm{kg\,mol^{-1}}$\\
				Initial liquid water content  & $LWC$ & $7.9\cdot10^{-4}$ & $\rm{kg/m^3}$\\
					\bottomrule
			\end{tabular}}
		\label{tab:Table1}
	\end{table}
\FloatBarrier

	\begin{table}[bht!]
		\caption{\textbf{List of parameters for the unstable cloud-clear interface direct numerical simulation hosting the monodisperse and polidisperse distribution of water droplets}}
		\centering
		\resizebox{0.99\textwidth}{!}{
			\begin{tabular}{lccc}
				\toprule
				\textbf{Quantity}& \textbf{Symbol}	& \textbf{Value}  & \textbf{Unit}\\
				\midrule
				Simulation domain size & $L_{x_1}\cdot L_{x_2}\cdot L_{x_3}$ & $0.512\cdot0.512\cdot1.024$ & $\rm{m^3}$\\
				Simulation domain discretization &$N_1\cdot N_2\cdot N_3$ &$512\cdot512\cdot1024$ &~\\
				Simulation grid step & $\Delta x$  & $0.001$  & m \\
				Initial and final Kolmogorov time & $\tau _{\eta}$ & $3.75\cdot 10^{-2}, 0.27$ & s\\
				Initial and final Kolmogorov scale in the cloud & $\eta$ & $0.6, 2.2$ & \rm{mm}\\
				Root mean square of velocity fluctuation in cloud region& $u_{rms}$ & $0.1125$ & $\rm{m/s}$\\
				Initial particle response time at $R_0=15 \mu$m & $\tau _p$ & $3.6 \cdot 10^{-3}$& s\\
				Initial large eddy turn over time & $T_l={\bm\tau}$ & $0.42$ & s\\
				Initial droplet radius for\\ monodisperse distribution  & $\rm{R}_{in}$ & 15 & $\rm{\mu m}$\\
				Minimum droplet radius for\\ polydisperse distribution  & $\rm{R}_{in-p,min}$ & 0.6 & $\rm{\mu m}$\\
				Maximumm droplet radius for\\ polydisperse distribution  & $\rm{R}_{in-p,max}$ & 30 & $\mu m$\\
				Total number of initial droplets (monodisperse population) & $N_{tot-m}$ & $8\cdot10^6$ & -\\
				Total number of initial droplets (polydisperse population) & $N_{tot-p}$ & $10^7$ & -\\
				Simulation time step & $\Delta t$  &$ 3.8\cdot 10^{-4}$ & $\rm{s}$\\ 
				Initial energy ratio & $E_{cloud}/E_{clear\ air}$  &$6.7$ & -\\
				Initial integral scale & $l$ & $0.048$ & m\\
				Initial Taylor micro-scale Reynolds no.  & $Re_{\lambda}$ & 42 & -\\
				Reynolds number based on domain dimension & $Re_{L}$& 5000 & -\\
				Brunt-V\"ais\"al\"a amplification factor, where $N=(-g \Delta  T T_0^{-1} L_{x_3}^{-1})^{0.5}$,\\ unstable stratification & $N^2$ & $-0.068$ & s$^{-2}$\\
				\bottomrule
		\end{tabular}}
		\label{tab:Table2}
	\end{table}
	\FloatBarrier

	\normalsize
	\vspace{-2mm}
	\subsection{\textit{\textbf{Monodisperse and polydisperse droplets initial distributions}}}
	
\begin{figure}[bht!]
		\centering
		\includegraphics[width=0.47\textwidth]{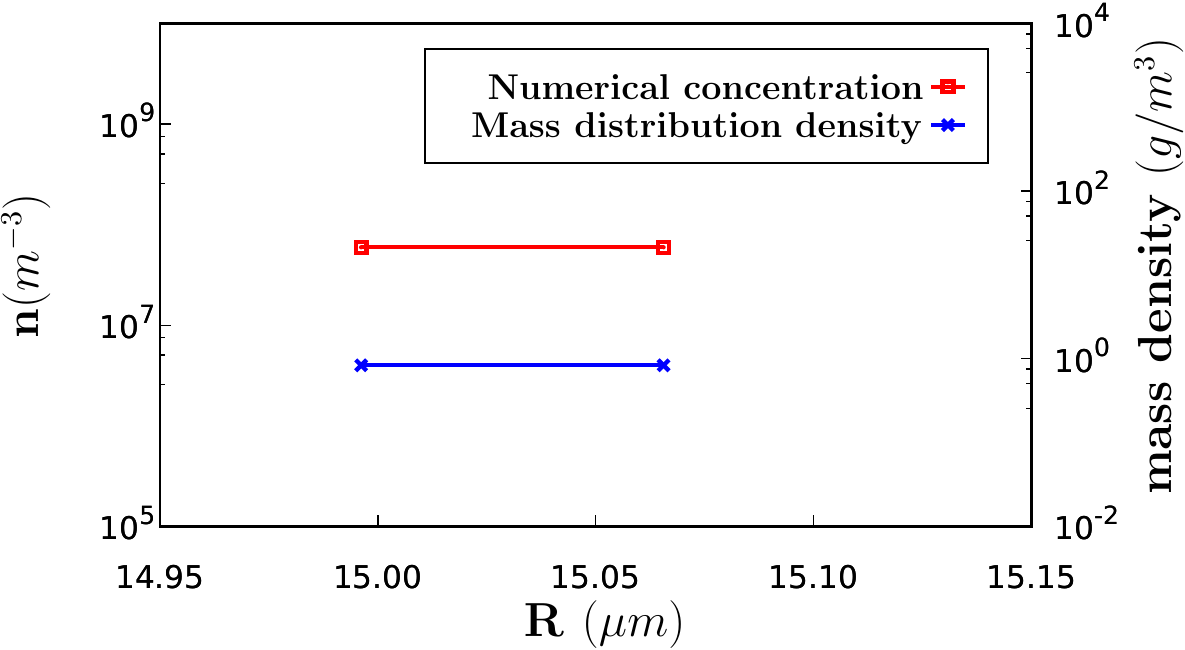}
		\hspace{3mm}	
		\includegraphics[width=0.47\textwidth]{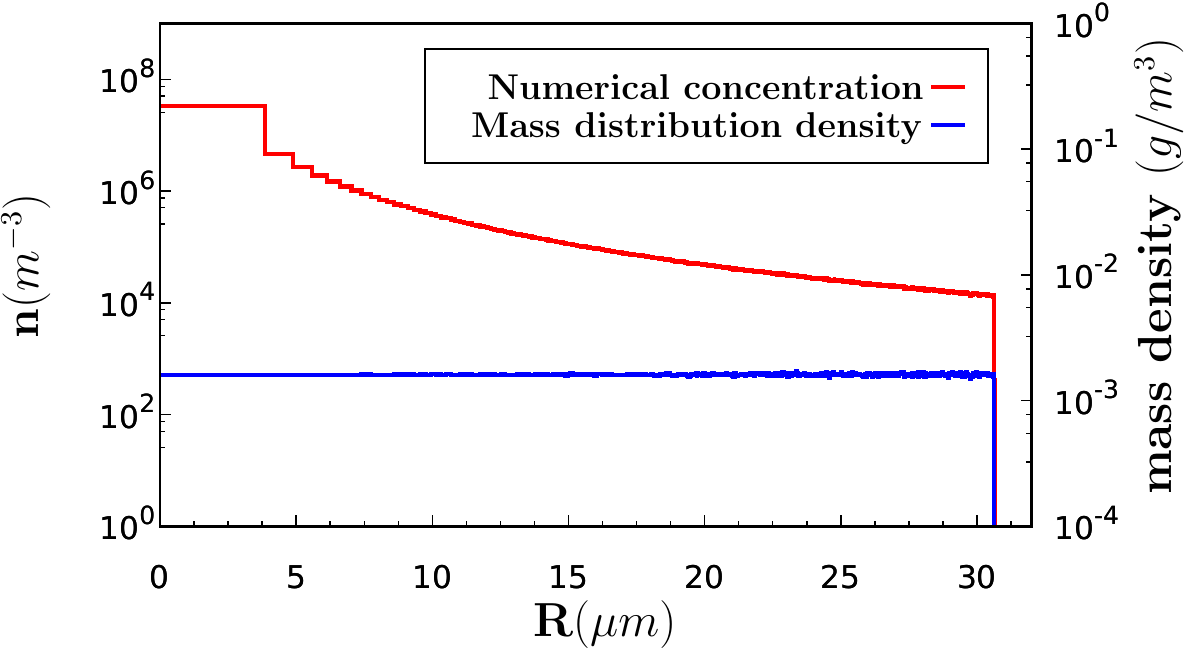}
		\caption{\bf Monodisperse (left panel, $8 \cdot 10^6$ particles) and Polydisperse (right6rt panel, $10^7$ particles) drop size distributions; for both distributions the initial value of total liquid content is $LWC_0 = 0.8 \rm{g/cm}^3$}
		\label{fig:init_size_distrib}
	\end{figure}

	 We compare the extremes between possible population size distributions of water drops: a monodisperse versus a polydisperse population with uniform mass per class of radii. Droplets are initially placed in the cloud only, i.e. in the region where turbulent energy is higher, see fig.s \ref{figure1} and \ref{fig:initial_conditions}. The initial spatial distribution is random and uniform. 	The two  distributions  are showed in figure \ref{fig:init_size_distrib}. 
		The choice was made because, in the literature, a typical form of the size distribution in warm  natural clouds to refer with is not yet available  and perhaps will not be in the near future.
	The monodisperse distribution, a drop size selected distribution, presents a small number of collisions given the fact that equal drops do not collide unless the local spatial variation of the turbulent air velocity are sufficient to give neighbouring drops different velocities leading to collision. See, for simplicity, the classical  theories where turbulence is treated as steady, homogeneous and isotropic, with a small eddies length scale at least one order of magnitude larger than the drop size, \cite{marshall1954}, \cite{saffman1955}. The other way around, inside a polydisperse drop size distribution, the collision rate is high because different inertial drops show a different motion relative to the air and this is even more so because of gravity. 
	On the other hand, it is recognized that the existence of a unique functional shape for the distribution size is still questioned on many grounds: different and competing mechanism for droplets nucleation, growth and removal are present  in different context of cloud regions and cloud lives, see for instance the \cite{Chandrakar2020}. 
	Furthermore, since we wish to model a realistic cloud-clear-air boundary temporal evolution we are out of the ideal conditions, based on statistical steadiness in time and spatial homogeneity, that at the moment are the hypotheses that can only lead to a theoretical treatment. See, for instance, the recent approach based on the principle of maximum entropy (\cite{Liu1998} and \cite{wu2018}) or the approach based on a Langevin equations representing the stochastic condensation-evaporation (\cite{McGraw2006}; \cite{chandrakar2016}; \cite{siewert2017} and \cite{Saito2019}).  
	
	As mentioned above, the two populations are evolving inside the inherently turbulent interface layer between the small portion of the warm cloud and the clear air on top of it. The turbulent layer feels the unstable stratification (with a Brunt-V\"ais\"al\"a fluctuation growth factor $N^2$ equal to $- 0.0687$) which induces in both cases a velocity transient amplification which is followed by a free temporal decay, see panel (a) in Fig. \ref{fig:initial_conditions}. 
	
	No forcing is set on the system which aims at modeling a realistic small cloud perturbation localized near the cloud boundary. The presence of a turbulence energy gradient is sufficient for Gaussian departure due to the anisotropy effects, and intermittency of velocity fluctuation and velocity derivative statistics, see Fig.s \ref{fig:velocity_derivative_3} and \ref{fig:supersaturation_vapor}.
	The turbulence energy gradient quickly leads the small scales of the turbulence out of isotropy and induces a pressure transport not negligible with respect to the turbulent velocity transport \cite{Tordella2008}; \cite{Tordella2011}; \cite{Tordella2012}. All these aspects are active along with the transient evolution of the cloud/clear-air system and affect the drop collision rate in a way that has not yet been explored in literature so far. In particular, since in this situation the background airflow penetration inside the region of low turbulence is maximum, it is interesting to observe what happens to the droplet collision rate and penetration throughout the interfacial layer and into the clear-air portion of the  system.

	\vspace{2mm}
	
	\FloatBarrier
	
	\normalsize	
	
	\vspace{-3mm}
	\section{\textit{\textbf{Results}}}
	
\begin{figure}[bht!]
		\centering
		\begin{subfigure}{0.3\textwidth}
			\includegraphics[width=\linewidth]{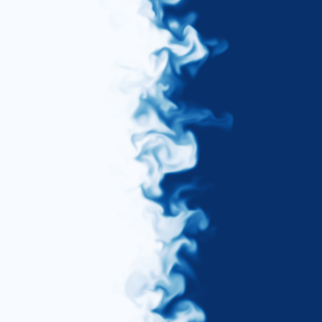}
		\end{subfigure}\hspace*{\fill}
		\hspace{0.03\textwidth}
		\begin{subfigure}{0.3\textwidth}
			\includegraphics[width=\linewidth]{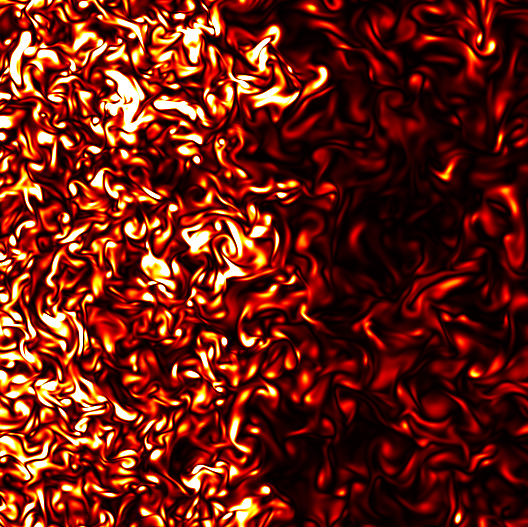}
		\end{subfigure}
		\hspace{0.03\textwidth}
		\begin{subfigure}{0.3\textwidth}
			\includegraphics[width=\linewidth]{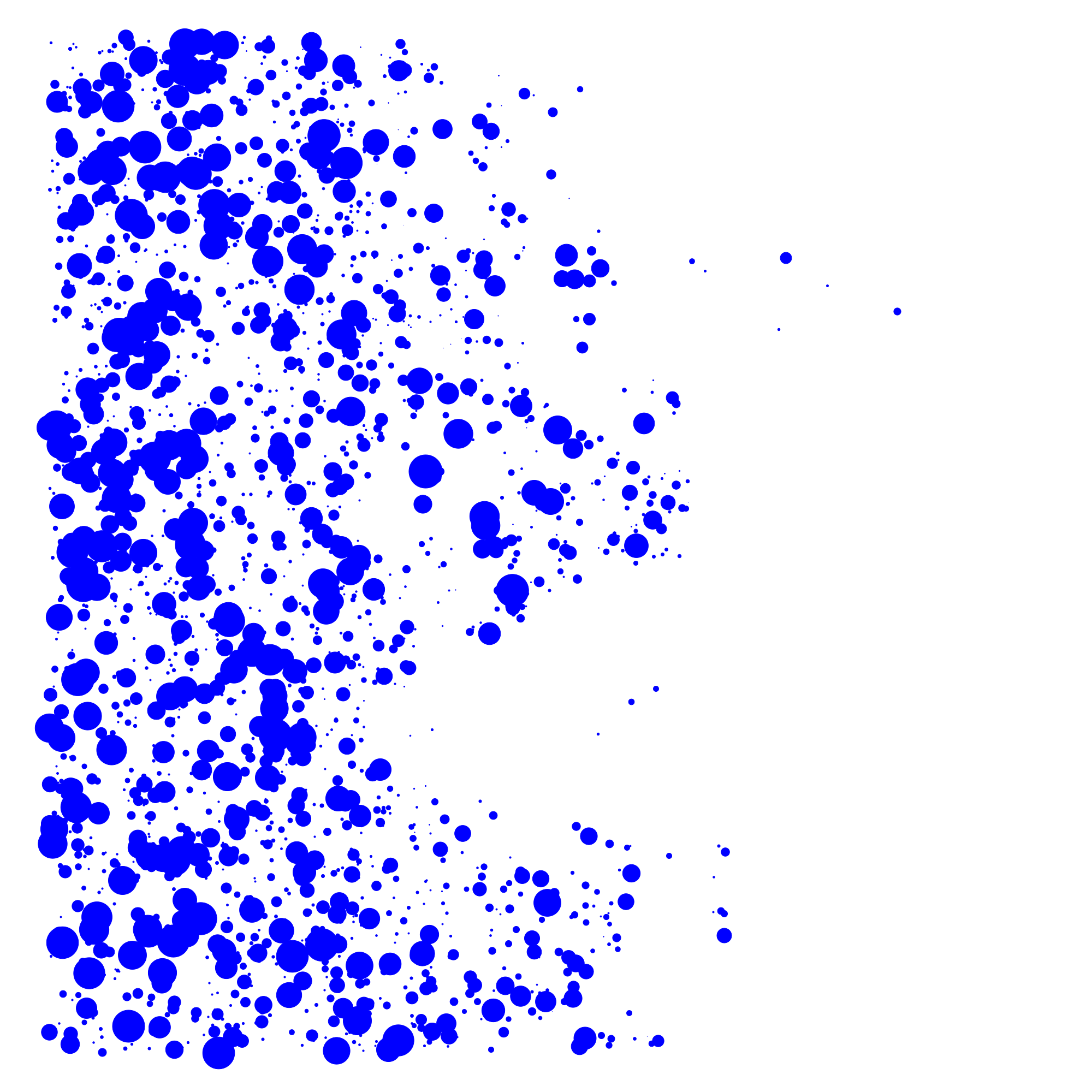}
		\end{subfigure}
		
		\smallskip
		\begin{subfigure}{0.3\textwidth}
			\includegraphics[width=\linewidth]{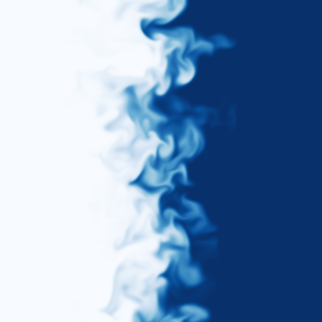}
		\end{subfigure}\hspace*{\fill}
		\hspace{0.1cm}
		\begin{subfigure}{0.3\textwidth}
			\includegraphics[width=\linewidth]{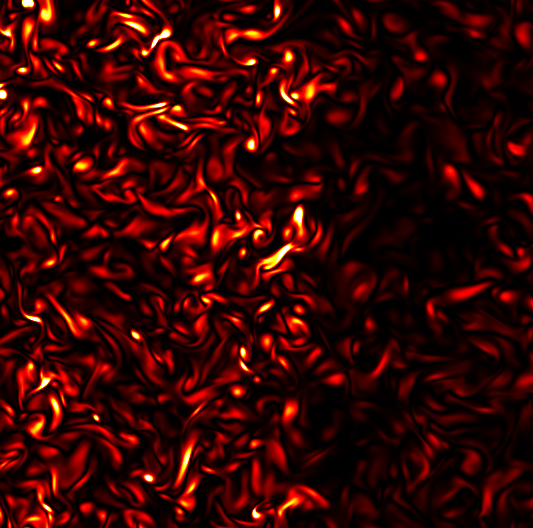}
		\end{subfigure}
		\hspace{0.03\textwidth}
		\begin{subfigure}{0.3\textwidth}
			\includegraphics[width=\linewidth]{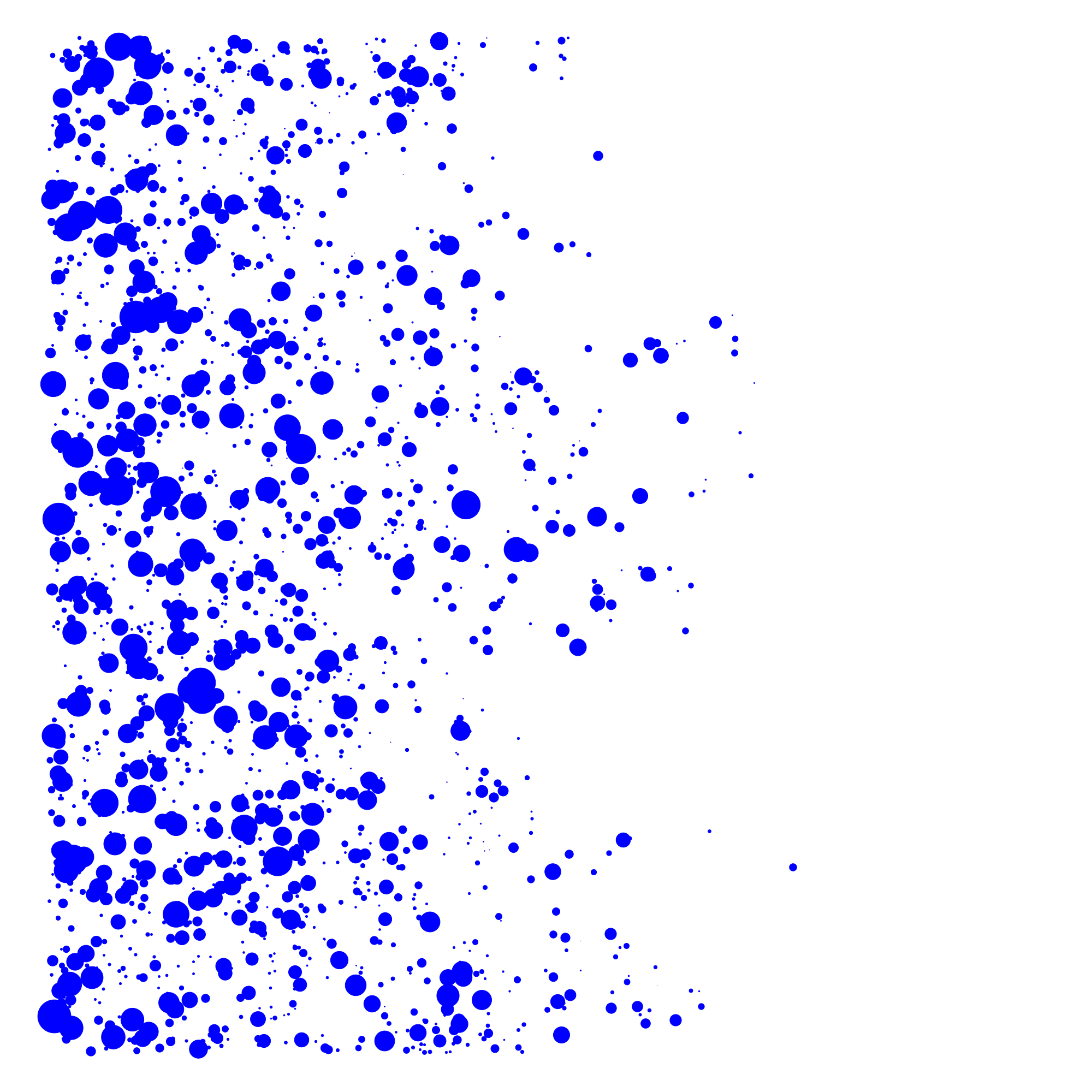}
		\end{subfigure}

		\smallskip
		\begin{subfigure}{0.3\textwidth}
			\includegraphics[width=\linewidth]{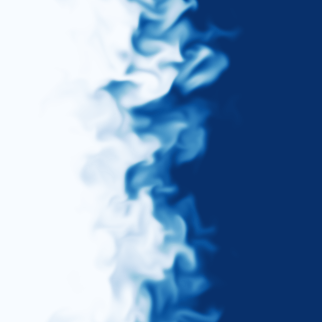}
		\end{subfigure}\hspace*{\fill}
		\hspace{0.1cm}
		\begin{subfigure}{0.3\textwidth}
			\includegraphics[width=\linewidth]{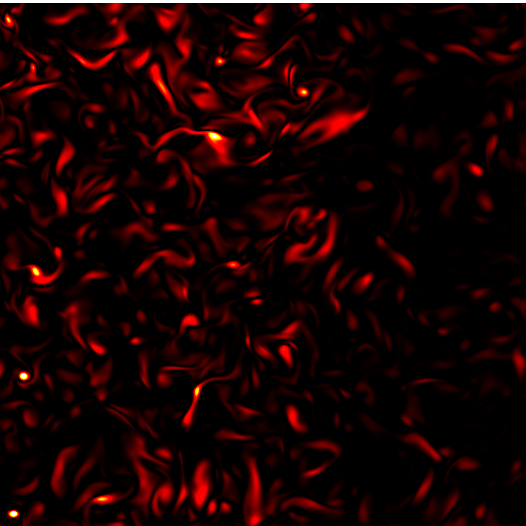}
		\end{subfigure}
		\hspace{0.03\textwidth}
		\begin{subfigure}{0.3\textwidth}
			\includegraphics[width=\linewidth]{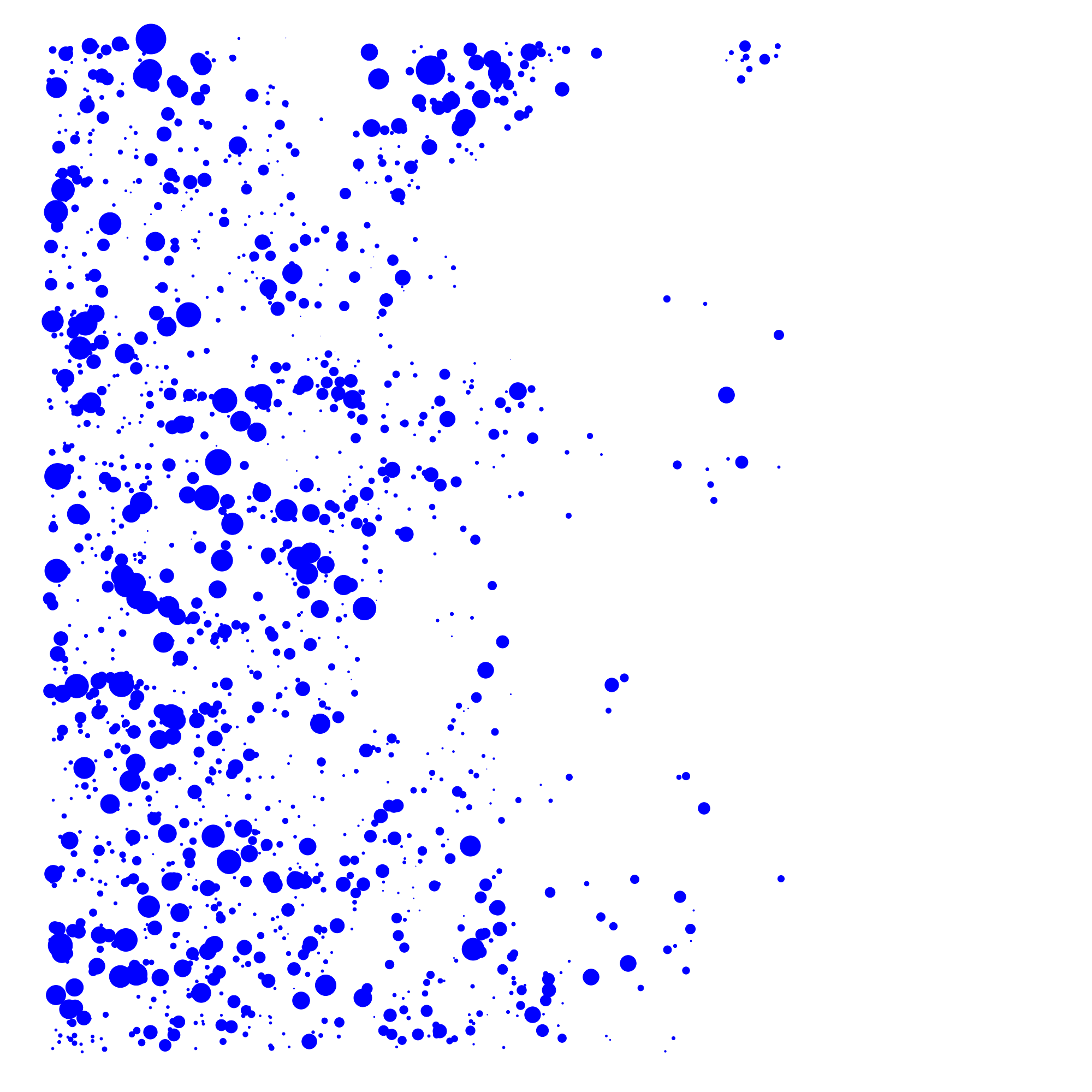}
		\end{subfigure}
		
		\begin{subfigure}{0.3\textwidth}
			\includegraphics[width=\linewidth]{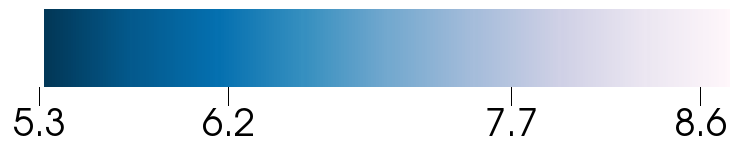}
		\end{subfigure}\hspace*{\fill}
		\hspace{0.03\textwidth}
		\begin{subfigure}{0.3\textwidth}
			\includegraphics[width=\linewidth]{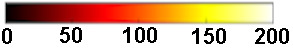}
		\end{subfigure}\hspace*{\fill}
		\hspace{0.03\textwidth}
		\begin{subfigure}{0.3\textwidth}
			\includegraphics[width=\linewidth]{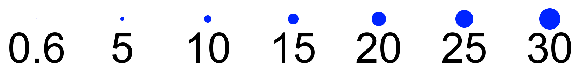}
		\end{subfigure}\hspace*{\fill}
		
		\caption{\bf Visualization of fields inside the turbulent shearless layer in between the cloud and clear-air portions of the simulation: water vapor (left, legend values in kg/m$^3$), enstrophy (middle, values in sec$^{-2}$ ) and droplets (right, diameters in arbitrary units, polydisperse population). From top to bottom, snapshots at 3, 6 and 9 eddy turnover times.} \label{fig_vap_enst}
	\end{figure}

\begin{figure}[bht!]
		\centering
		\begin{subfigure}[t]{0.47\textwidth}
			\includegraphics[width=\linewidth]{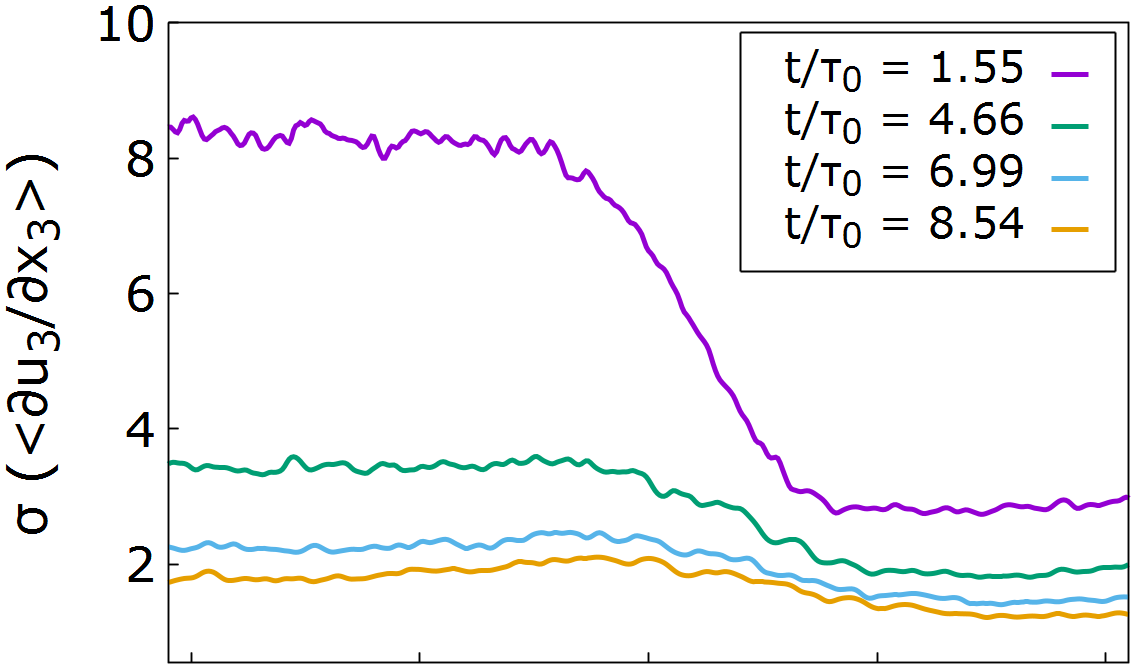}
		\end{subfigure}
		\hspace{0.03\textwidth}
		\begin{subfigure}[t]{0.47\textwidth}
			\includegraphics[width=\linewidth]{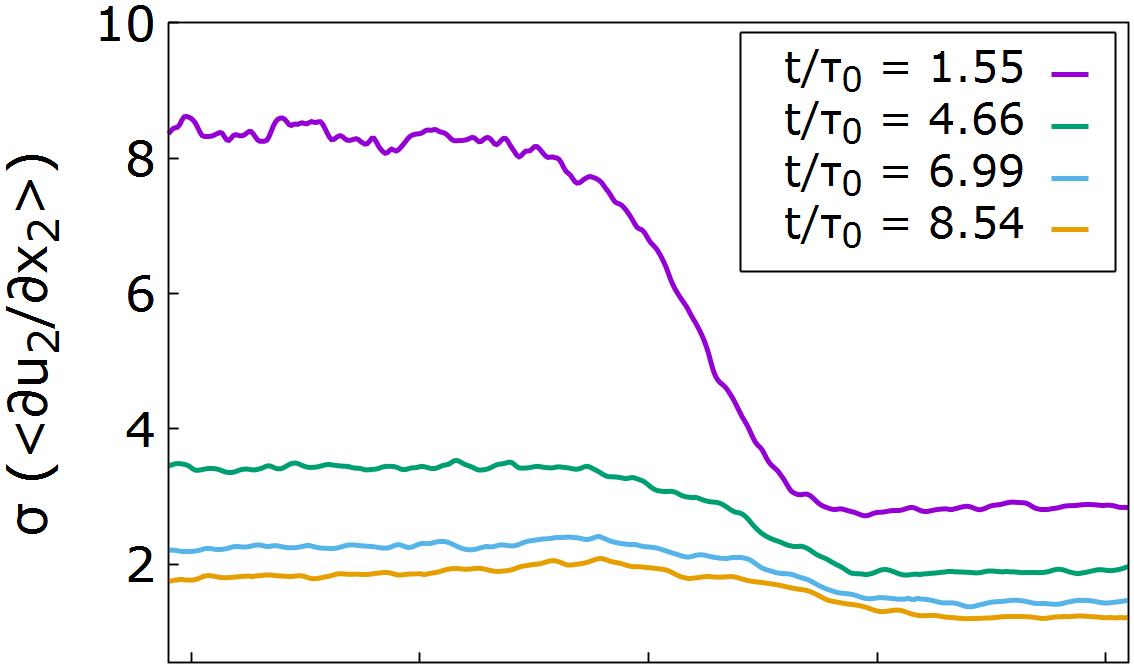}
		\end{subfigure}
		
		\vspace{0.3cm}
		
		\begin{subfigure}[t]{0.47\textwidth}
			\includegraphics[width=\linewidth]{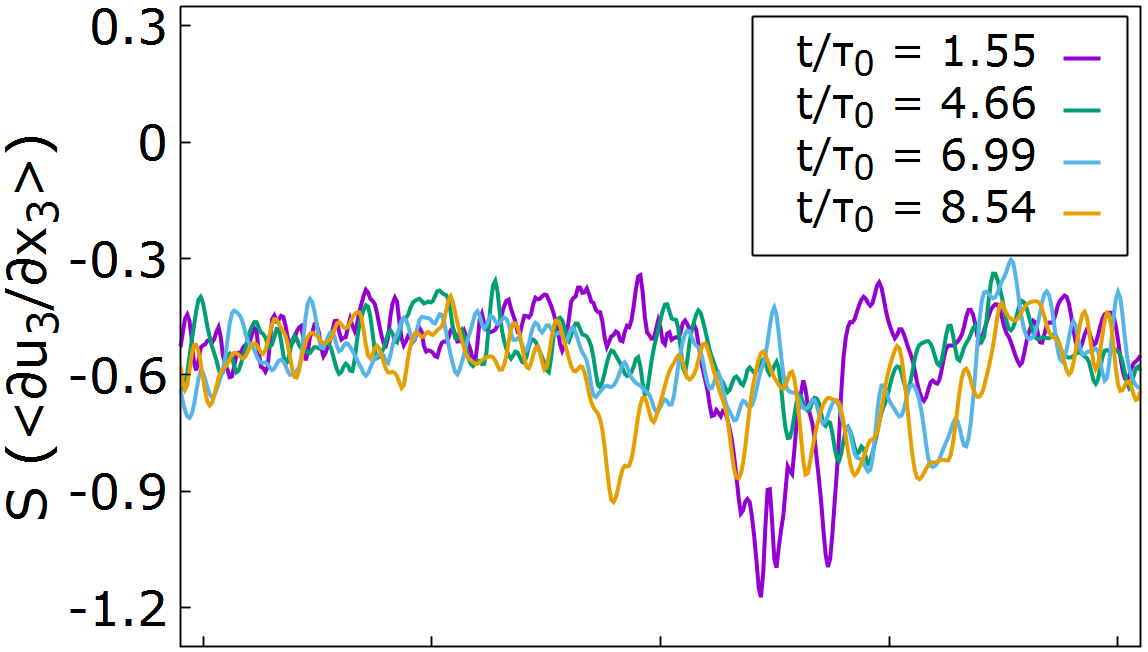}
		\end{subfigure}
		\hspace{0.03\textwidth}
		\begin{subfigure}[t]{0.47\textwidth}
			\includegraphics[width=\linewidth]{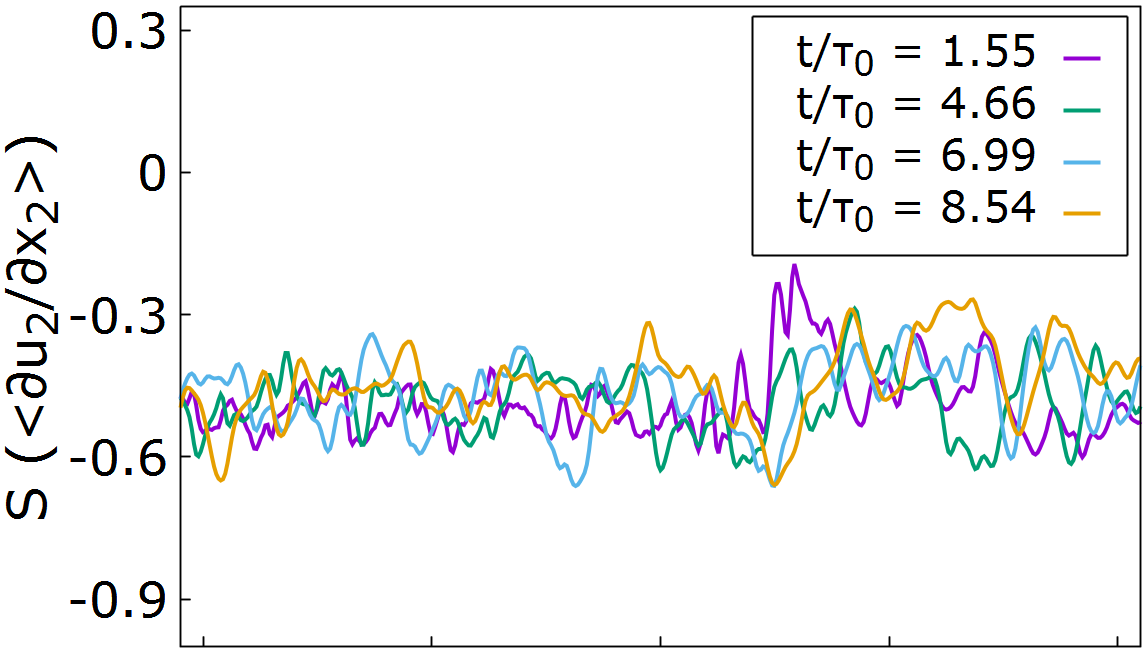}
		\end{subfigure}
		
		\vspace{0.3cm}
		
		\begin{subfigure}[t]{0.47\textwidth}
			\includegraphics[width=\linewidth]{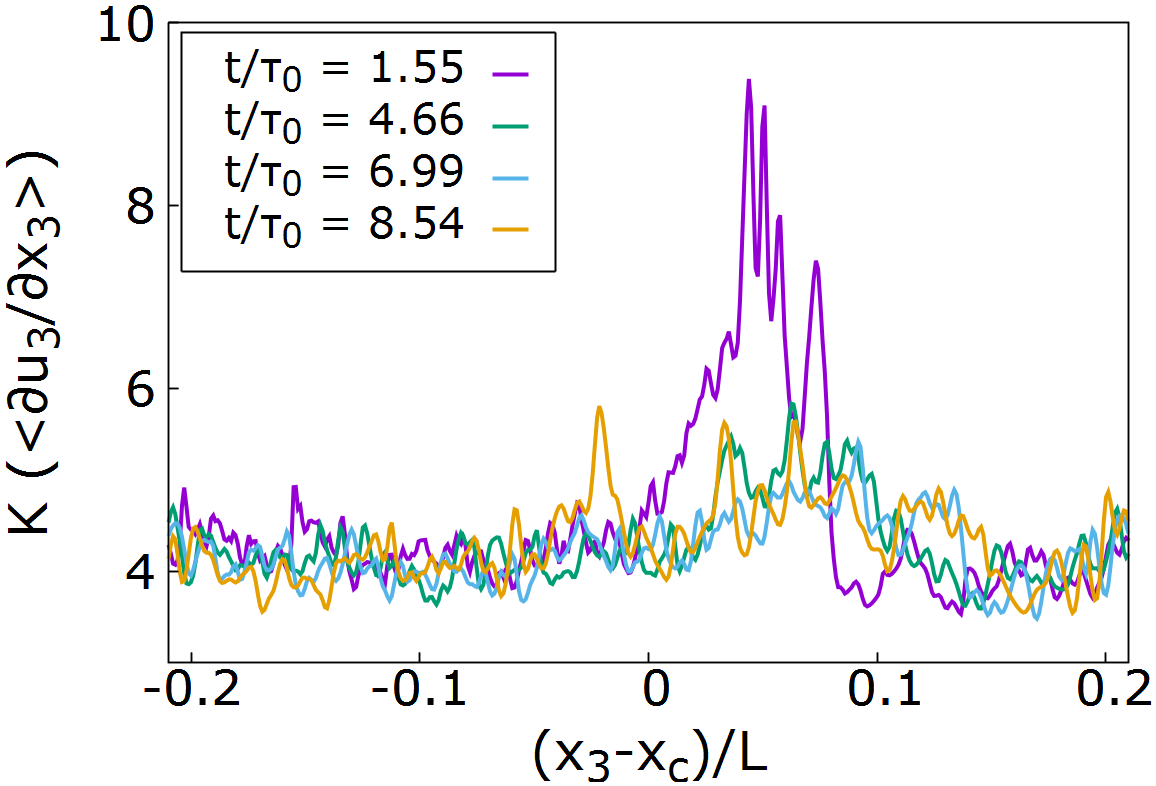}
		\end{subfigure}
		\hspace{0.03\textwidth}
		\begin{subfigure}[t]{0.47\textwidth}
			\includegraphics[width=\linewidth]{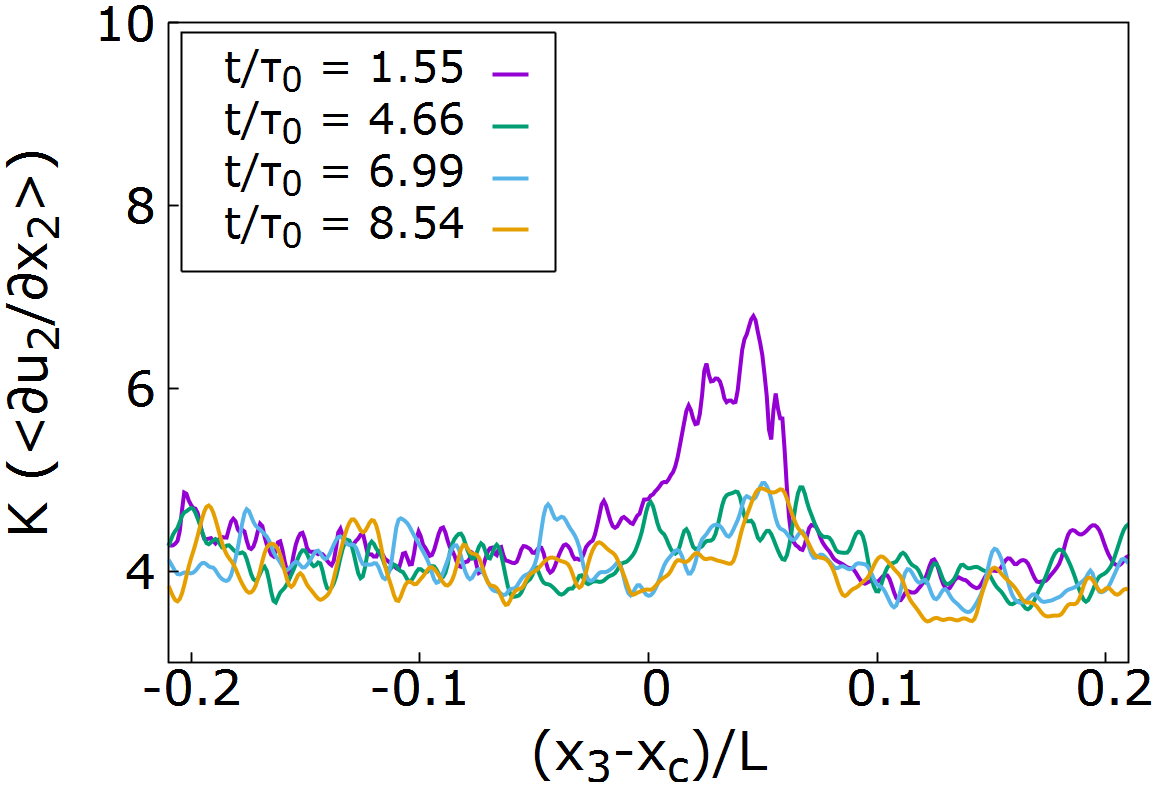}
		\end{subfigure}
		\caption{\bf Statistical moments of the fluctuation of the velocity  longitudinal $x_3$ derivative. Left, derivative in the direction across the interface. Right, derivatives along the direction parallel to the interface. The different behaviour highlights the intrinsic anisotropy of the fine scales of the turbulence inside this shear free layer.}
		\label{fig:velocity_derivative_3}
	\end{figure}

\begin{figure}[bht!]
		\centering
		\begin{subfigure}[t]{0.47\textwidth}
			\includegraphics[width=\linewidth]{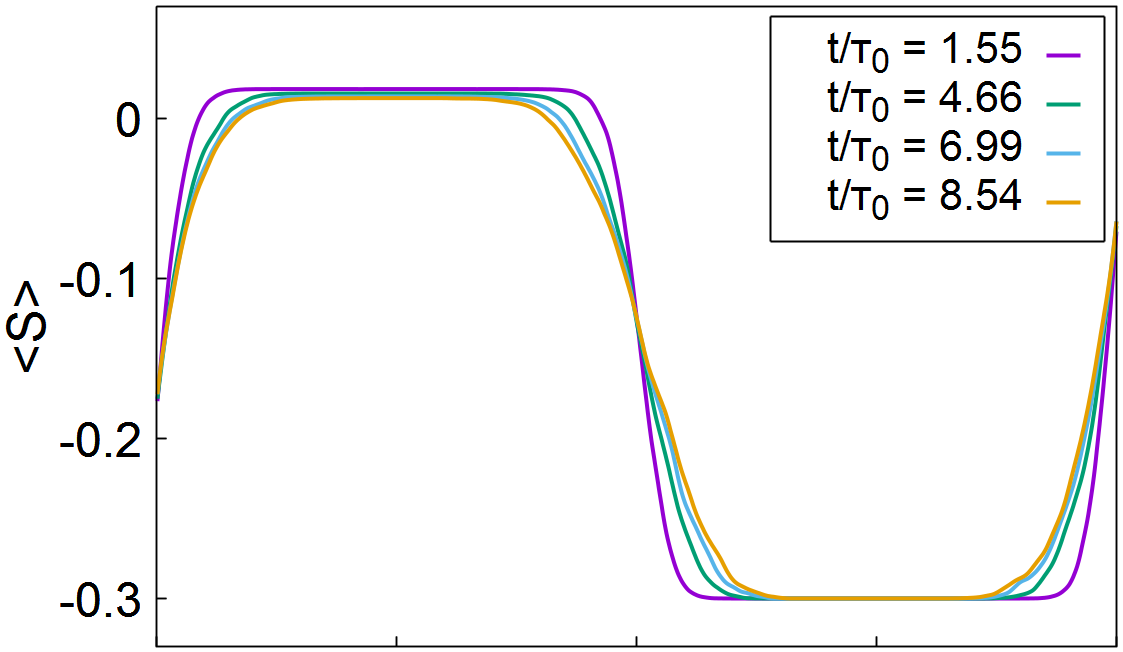}
		\end{subfigure}
		\hspace{0.03\textwidth}
		\begin{subfigure}[t]{0.47\textwidth}
			\includegraphics[width=\linewidth]{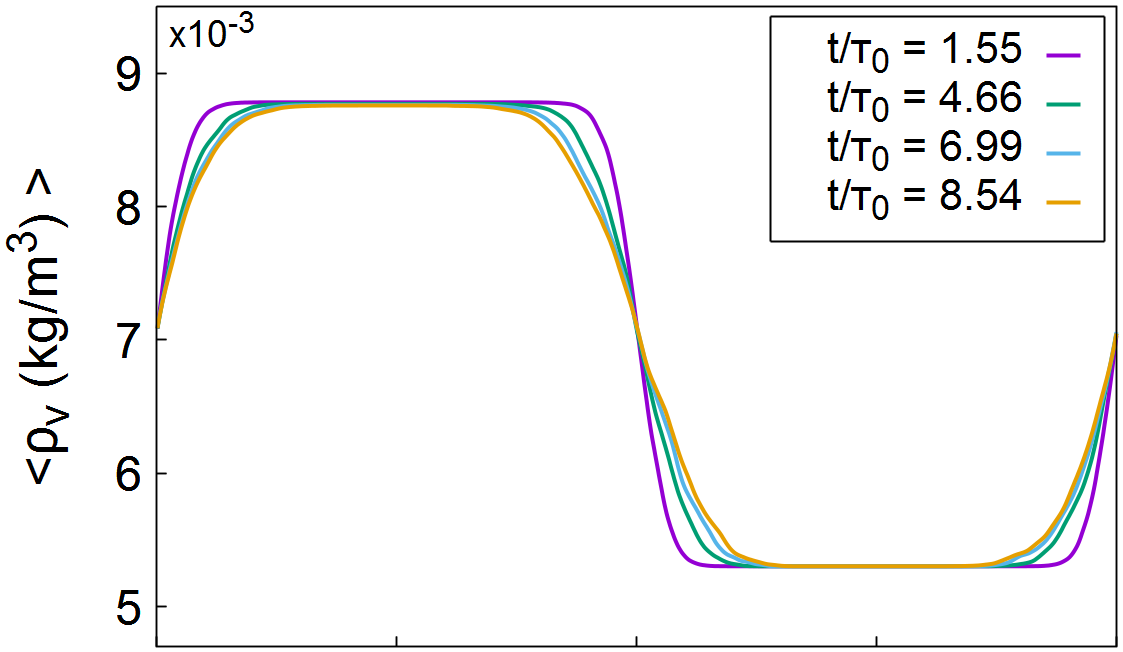}
		\end{subfigure}
		
		\vspace{0.5cm}
		\begin{subfigure}[t]{0.47\textwidth}
			\includegraphics[width=\linewidth]{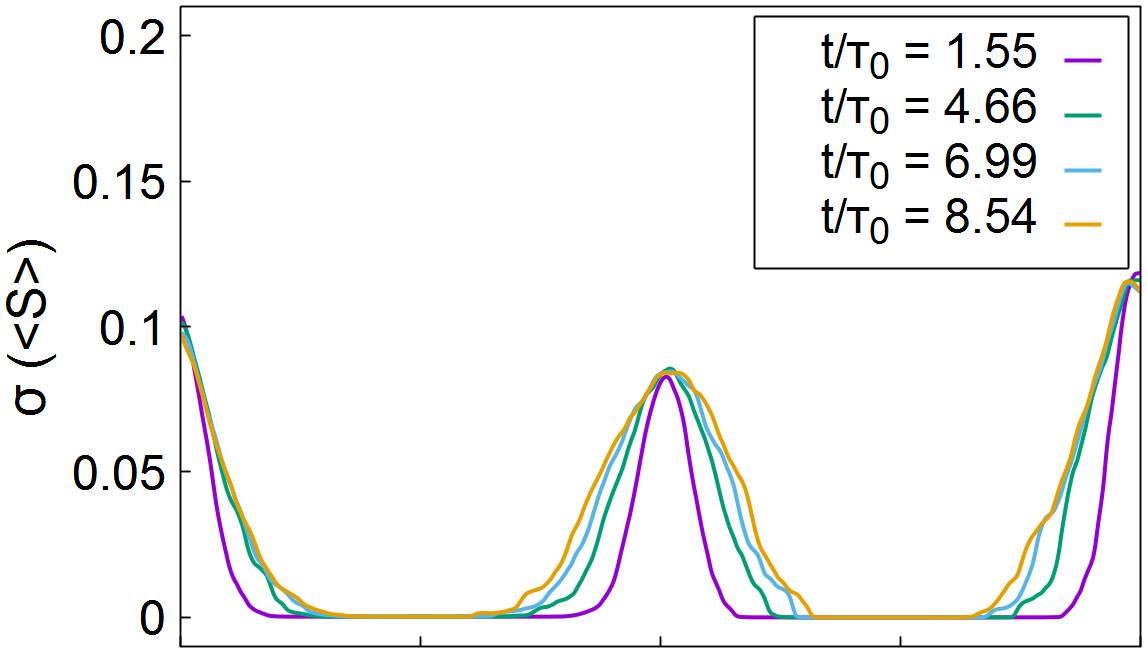}
		\end{subfigure}
		\hspace{0.03\textwidth}
		\begin{subfigure}[t]{0.47\textwidth}
			\includegraphics[width=\linewidth]{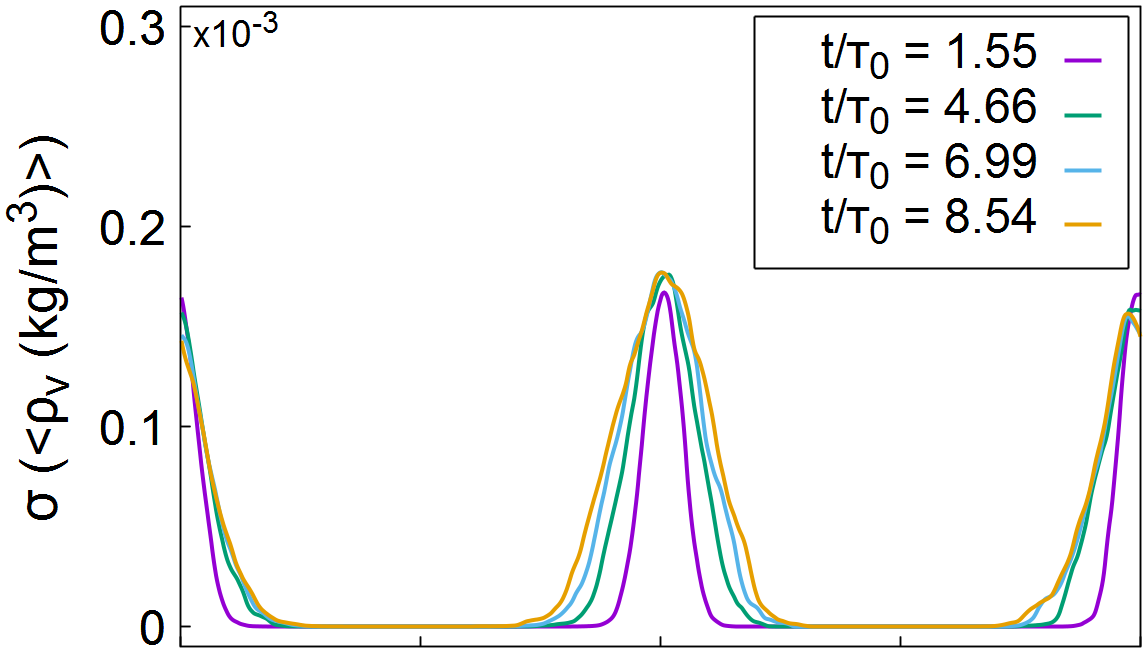}
		\end{subfigure}
		
		\vspace{0.3cm}
		
		\begin{subfigure}[t]{0.47\textwidth}
			\includegraphics[width=\linewidth]{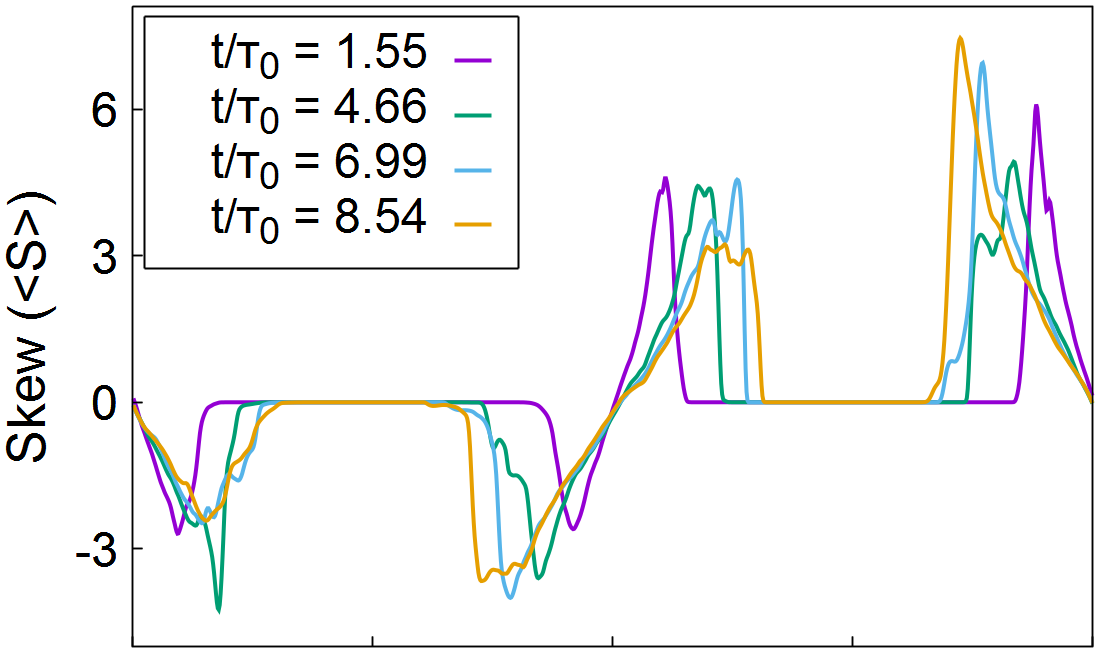}
		\end{subfigure}
		\hspace{0.03\textwidth}
		\begin{subfigure}[t]{0.47\textwidth}
			\includegraphics[width=\linewidth]{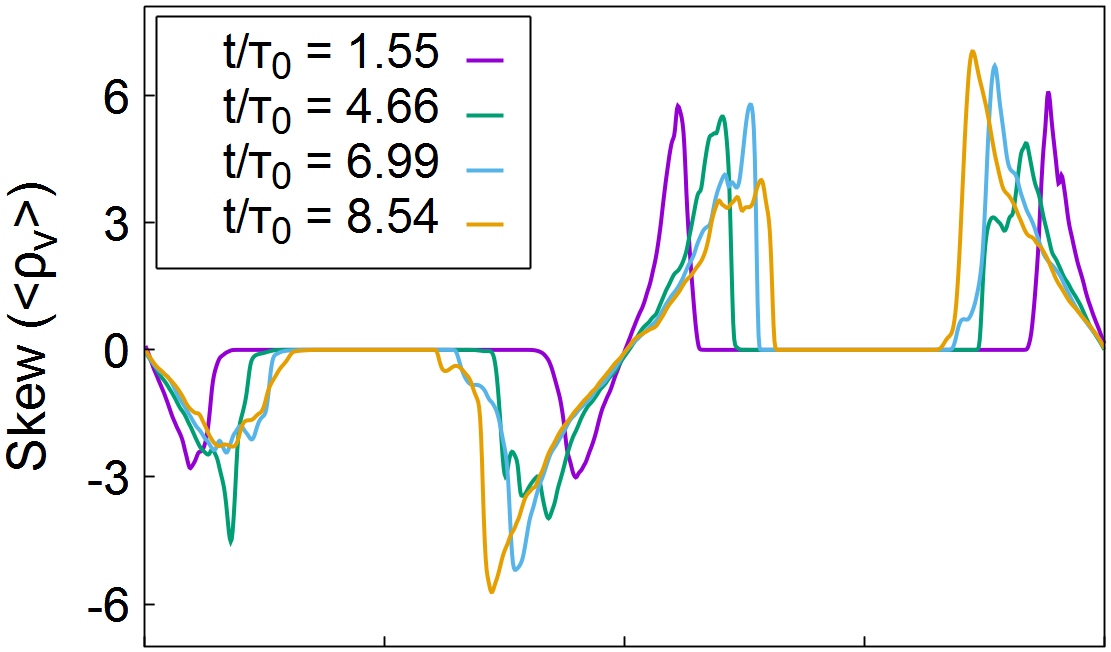}
		\end{subfigure}
		
		\vspace{0.5cm}
		\begin{subfigure}[t]{0.47\textwidth}
			\includegraphics[width=\linewidth]{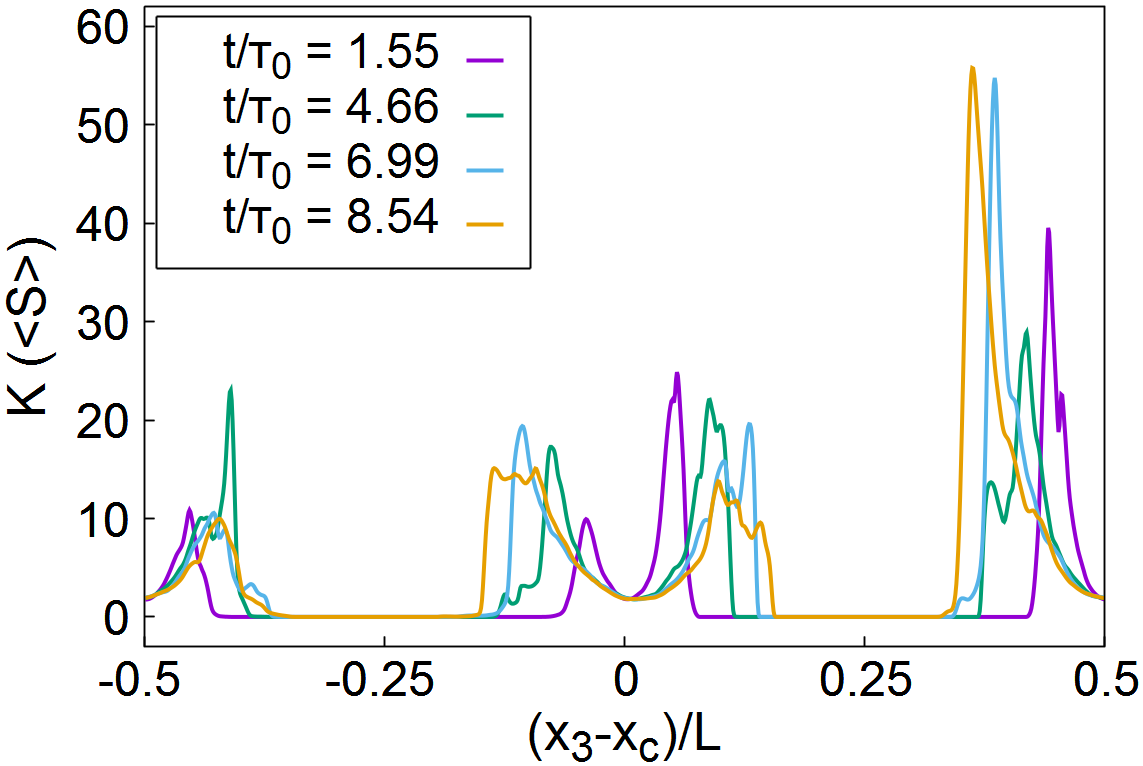}
		\end{subfigure}
		\hspace{0.03\textwidth}
		\begin{subfigure}[t]{0.47\textwidth}
			\includegraphics[width=\linewidth]{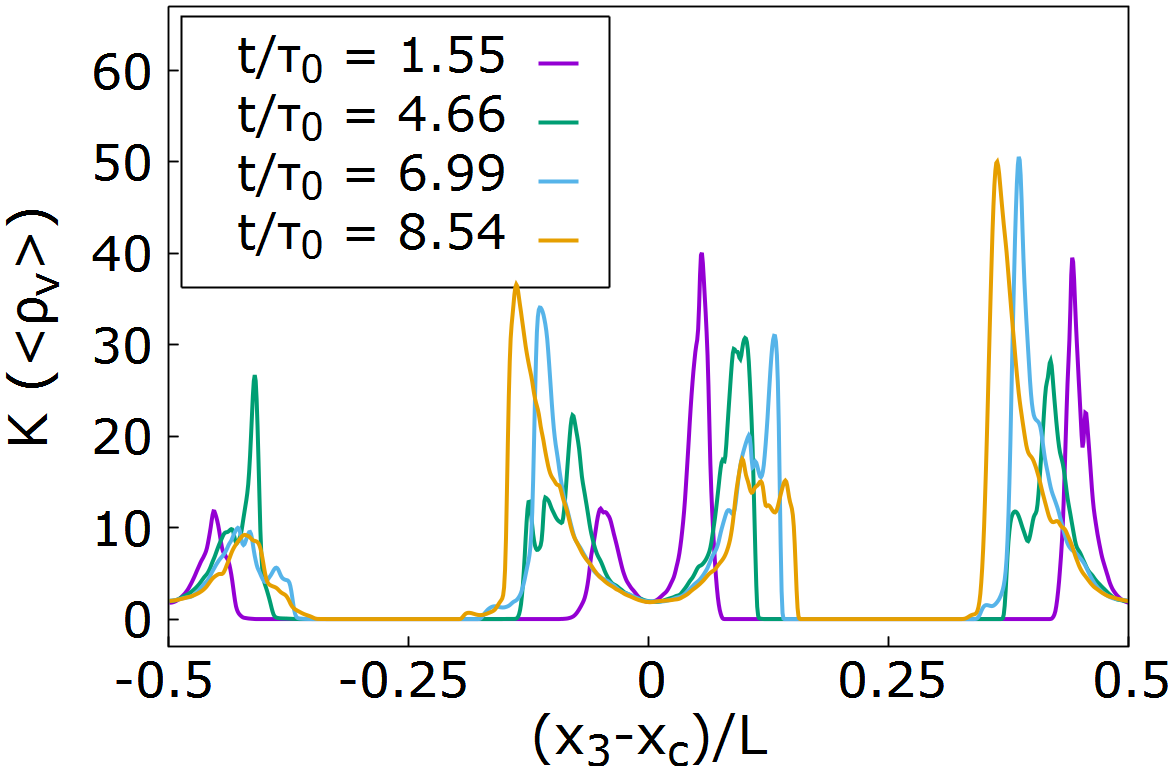}
		\end{subfigure}
		\caption{\bf Statistical moments of the supersaturation and water vapor density fluctuations. Mono-disperse drop size distribution, unstable and time decaying cloud clear-air interaction. A practically identical situation holds for the poly-disperse distribution. When keeping constant the total liquid water content (LWC), this is due to the fact that the kind of distribution barely influence the background velocity and scalar fields.}
		\label{fig:supersaturation_vapor}
	\end{figure}

	In a comparative way with respect to the two kinds of drop populations, in the following two subsections, we describe results concerning drop size  growth (positive in case of condensation, negative in case of evaporation) and the modification of their distributions along transient observed up to 10 eddy turn over times. 	For three different transient stages, figure \ref{fig_vap_enst} visualize an inner slice of the computational domain normal to the mixing layer. Visualization highlights the mixing layer in-homogeneity, its time growth, the water vapor concentration,  the velocity enstrophy  decay and concurrent small scale dissipation, and the droplet spatial distribution. 
		
	
	\vspace{-2mm}
	\subsection{\textit{\textbf{ Droplet size distribution temporal evolution. Condensation, evaporation, collision-coalescence.}}}
	
\begin{figure}[bht!]
		\centering
		\begin{subfigure}[t]{0.47\textwidth}
			\caption{cloud region} \label{fig:B004_cloud_num_PDF}
			\includegraphics[width=\linewidth]{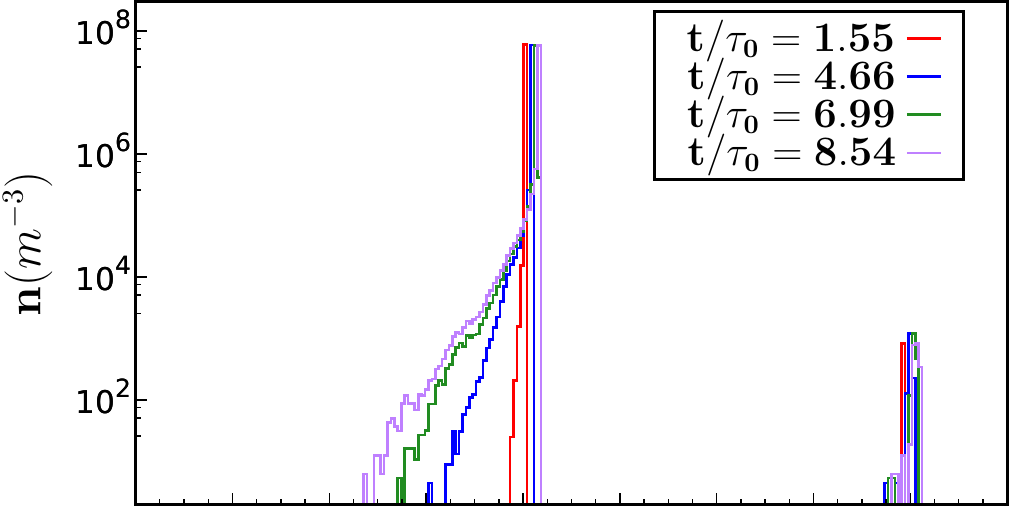}
		\end{subfigure}
		\hspace{0.03\textwidth}
		\begin{subfigure}[t]{0.47\textwidth}
			\caption{interface with the clear-air} \label{fig:B004_mixing_num_PDF}
			\includegraphics[width=\linewidth]{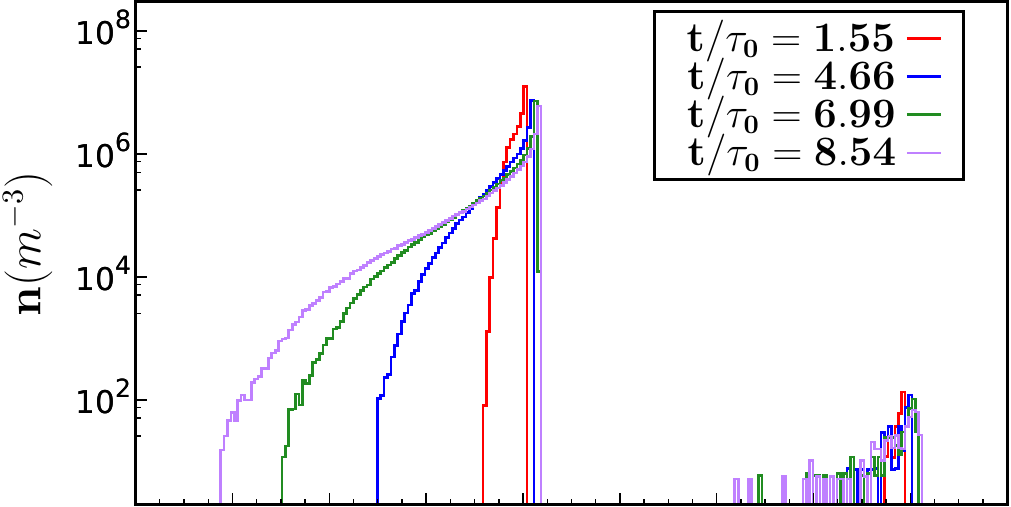} 
		\end{subfigure}
		
		\vspace{0.3cm}
		
		\begin{subfigure}[t]{0.48\textwidth}
			\includegraphics[width=\linewidth]{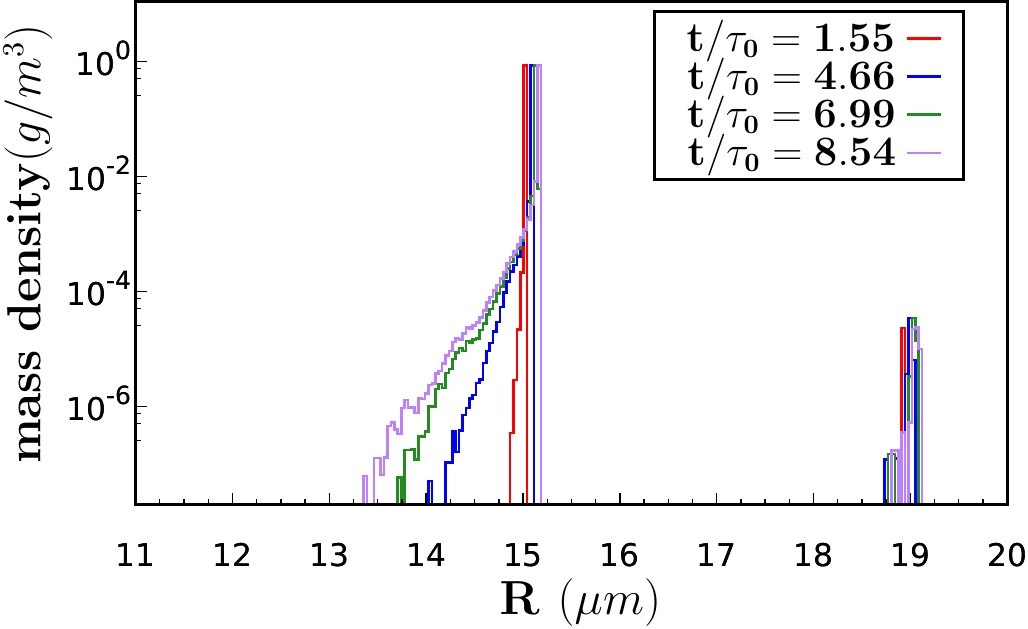} 
		\end{subfigure}
		\hspace{0.02\textwidth}
		\begin{subfigure}[t]{0.48\textwidth}
			\includegraphics[width=\linewidth]{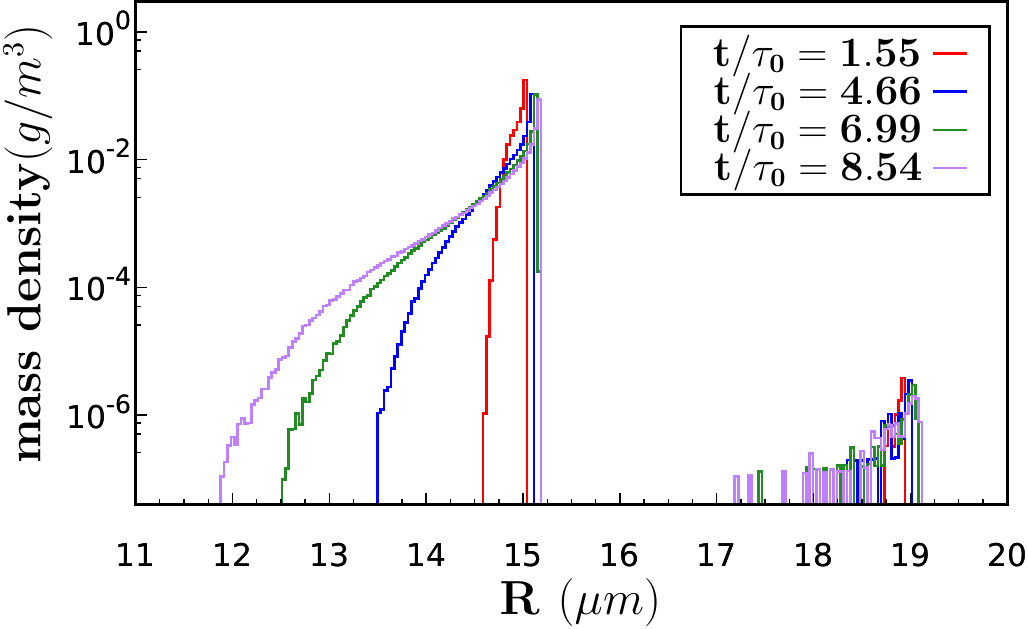} 
		\end{subfigure}
		
		\caption{\bf Water droplet size and mass distribution. Simulation of the monodisperse drop population centered around the initial value of 15 \textmu m, $8 \cdot 10^6$ droplets.  Panel (a) droplet size distribution and mass distribution as a function of radius classes for the cloud region (HIT turbulence). Panel (b): droplet size distribution and mass distribution as a function of radius classes for the cloud and clear air/interface (shearless turbulent layer). See table 1 for physical and thermodynamical parameters, see table 2 for details on the numerical simulation parameters.}
		\label{fig:mono_pdfs}
	\end{figure}

\begin{figure}[bht!]
		\centering
		\begin{subfigure}[t]{0.49\textwidth}
			\caption{cloud region} \label{fig:B003_cloud_num_PDF}
			\includegraphics[width=\linewidth]{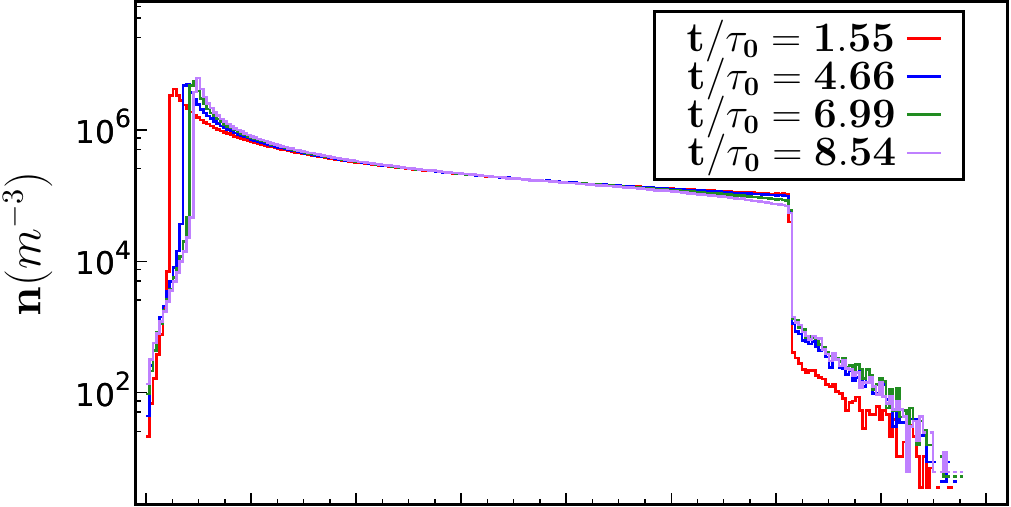} 	
		\end{subfigure}
		\hfill
		\begin{subfigure}[t]{0.49\textwidth}
			\caption{interface with the clear-air} \label{fig:B003_mixing_num_PDF}
			\includegraphics[width=\linewidth]{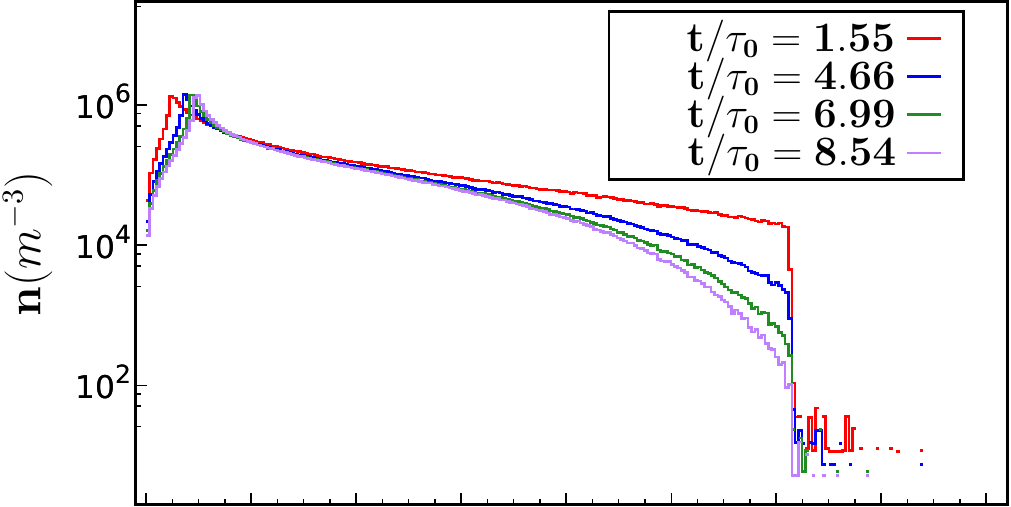} 
		\end{subfigure}
		
		\vspace{0.3cm}
		
		\begin{subfigure}[t]{0.49\textwidth}
			\includegraphics[width=\linewidth]{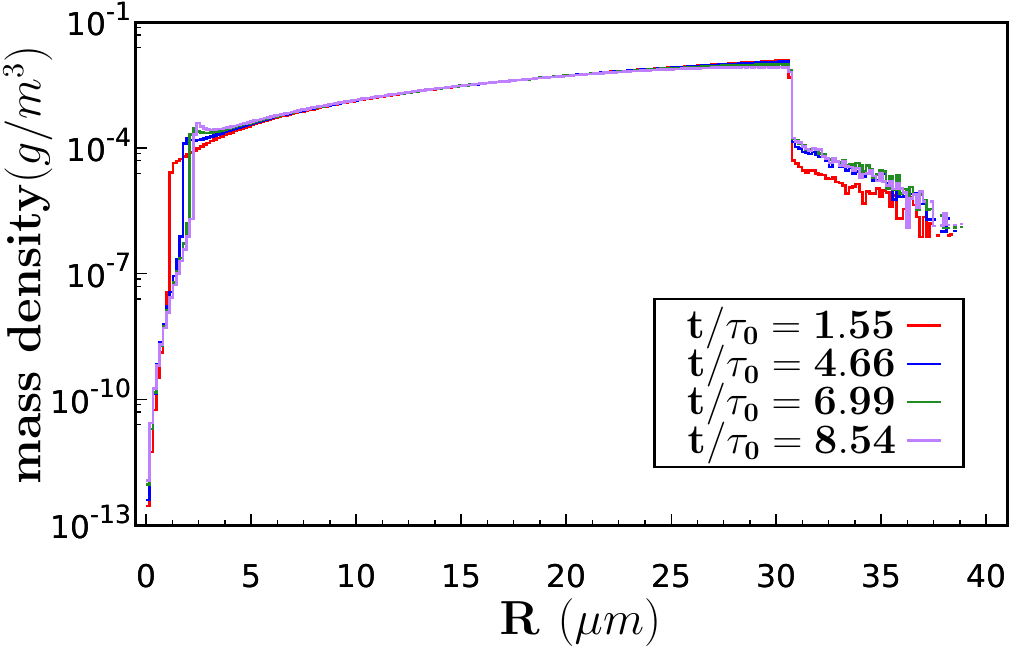} 
		\end{subfigure}
		\hfill
		\begin{subfigure}[t]{0.49\textwidth}
			\includegraphics[width=\linewidth]{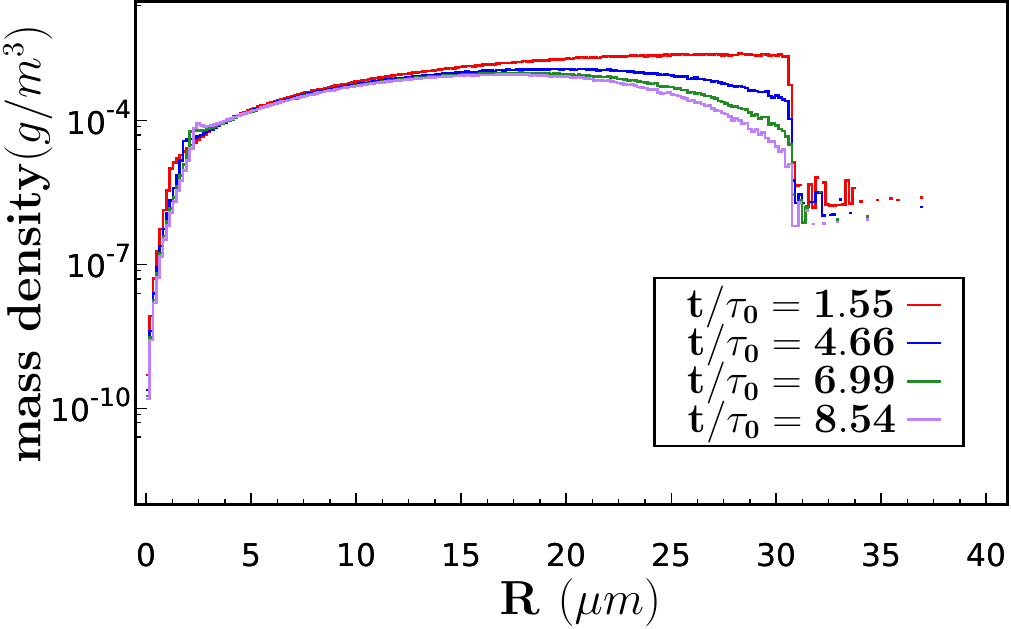} 
		\end{subfigure}
		
		\caption{\bf Water droplet size and mass distribution. Simulation of the polydisperse population with radii initially inside the range: $0.6 - 30$ \textmu m, $10^7$ droplets. Panel (a) droplet size distribution and mass distribution as a function of radius classes for the cloud region (HIT turbulence). Panel (b) Droplet size distribution and mass distribution as a function of radius classes for the cloud and clear air/interface (shearless turbulent layer). See table 1 for physical and thermodynamical parameters, see table 2 for details on the numerical simulation parameters.}
		\label{fig:poly_pdfs}
	\end{figure}

	\noindent For a few time instants inside the transient,  figures \ref{fig:mono_pdfs} and  \ref{fig:poly_pdfs} show the numerical and mass concentrations for both drop populations. In both cases, it is evident a variation of the shape of the  distribution inside the interaction layer. 
	
\begin{figure}[bht!]
		\centering
		{\bf Monodisperse droplets condensation and evaporation}
		\begin{subfigure}[t]{0.49\textwidth}
			\caption{\bf condensation, cloud region} \label{fig:B004_cloud_grow_rate_positive}
			\includegraphics[width=\linewidth]{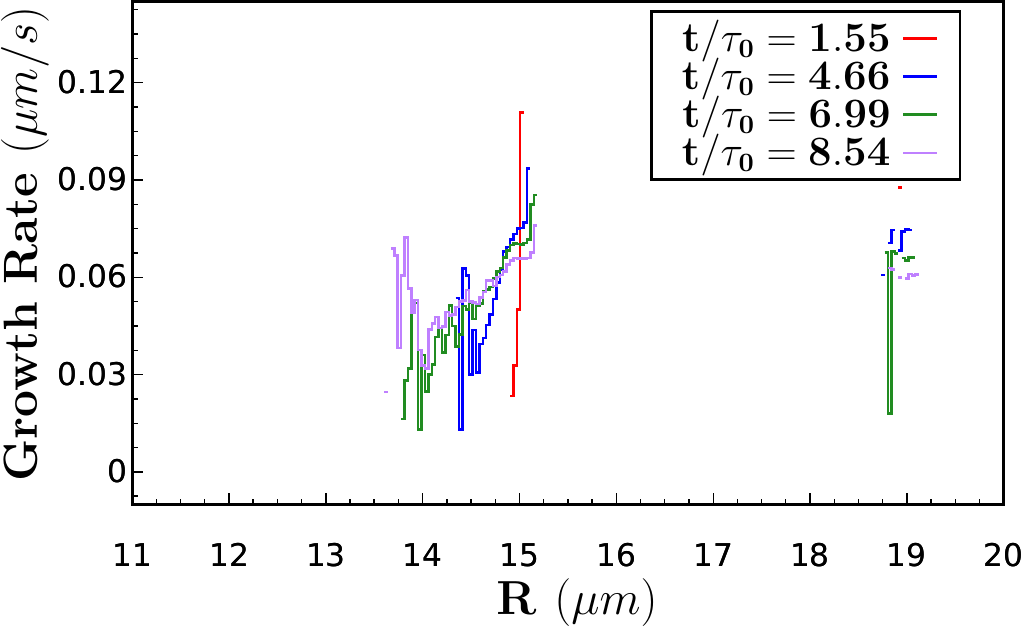} 	
		\end{subfigure}
		\hfill
		\begin{subfigure}[t]{0.49\textwidth}
			\caption{\bf condensation, interface region} \label{fig:B004_mixing_grow_rate_positive}
			\includegraphics[width=\linewidth]{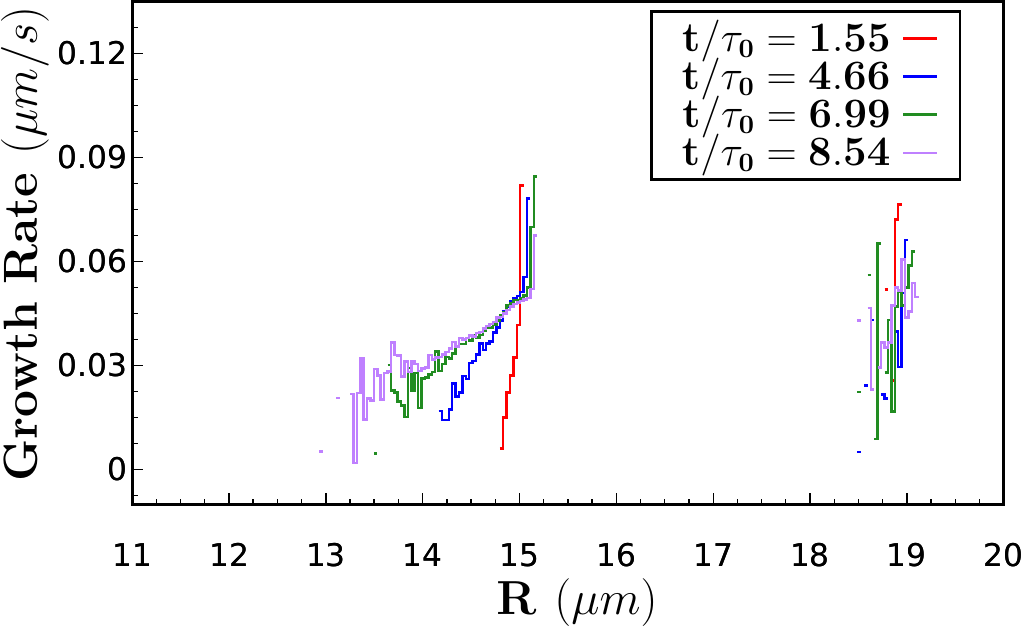} 	
		\end{subfigure}
		\smallskip
		\begin{subfigure}[t]{0.49\textwidth}
			\caption{\bf evaporation, cloud region} \label{fig:B004_cloud_grow_rate_negative}
			\includegraphics[width=\linewidth]{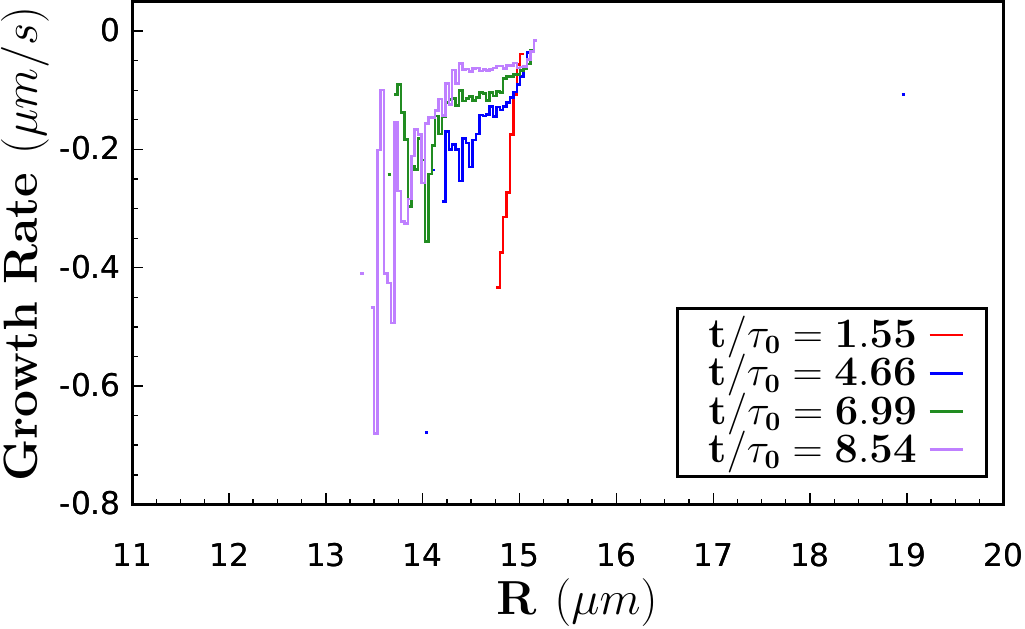} 	
		\end{subfigure}
		\hfill
		\begin{subfigure}[t]{0.49\textwidth}
			\caption{\bf evaporation, interface region} \label{fig:B004_mixing_grow_rate_negative}
			\includegraphics[width=\linewidth]{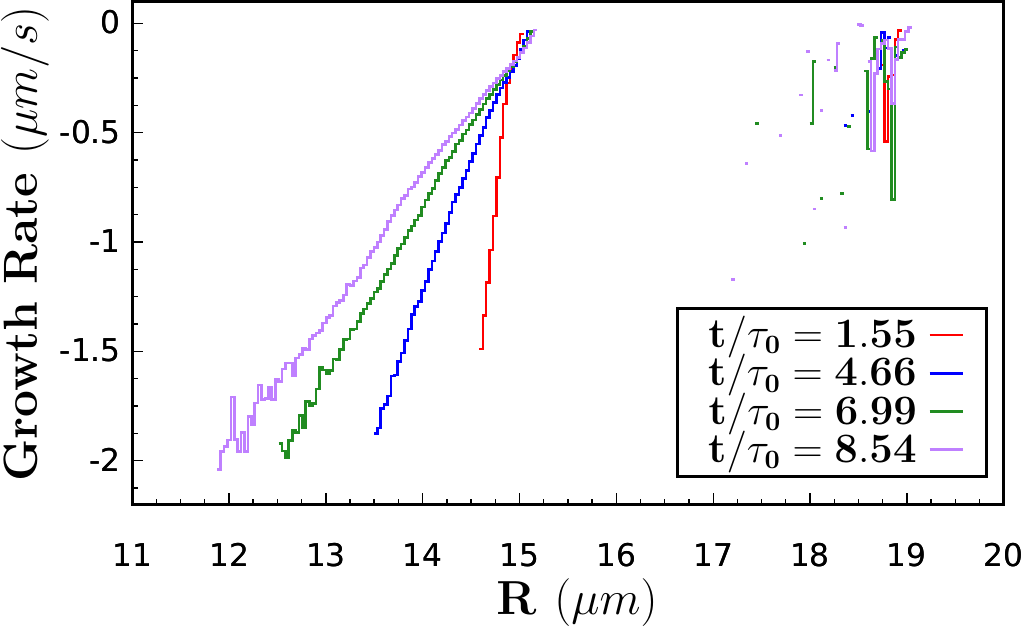} 
		\end{subfigure}
		\smallskip
		\begin{subfigure}[t]{0.49\textwidth}
			\caption{\bf cloud region: enhanced view near transient end, R=  $13.5 \div 15.5 \mu m$} \label{fig:B004_cloud_grow_rate_selected}
			\includegraphics[width=\linewidth]{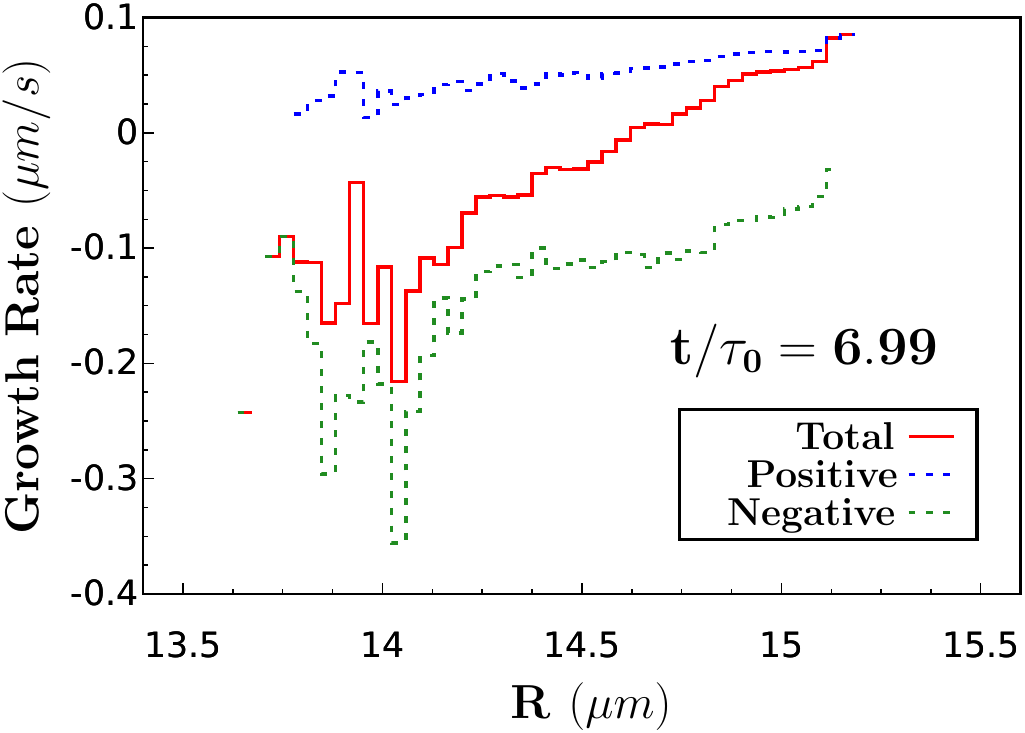} 	
		\end{subfigure}
		\hfill
		\begin{subfigure}[t]{0.49\textwidth}
			\caption{\bf interface region: enhanced view near transient end, R= $12.5:\div 15.5 \mu m$} \label{fig:B004_mixing_grow_rate_selected}
			\includegraphics[width=\linewidth]{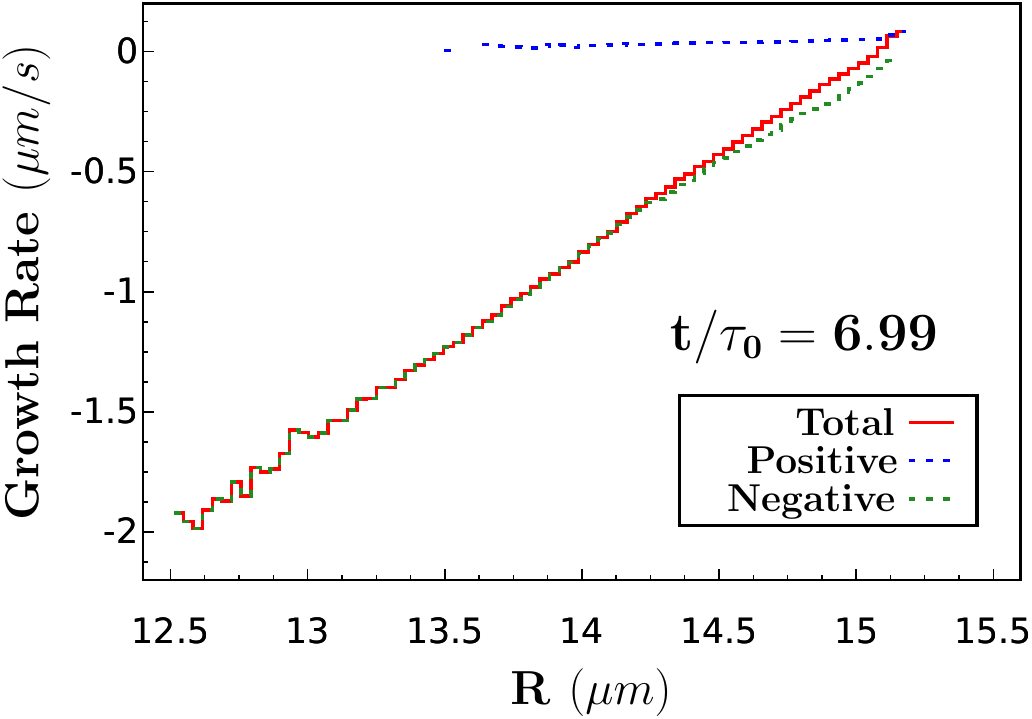}	
		\end{subfigure}
		\caption{\bf 
			Monodisperse drop size distribution, unstable and time decaying cloud clear-air interaction. Mean droplet radius growth rate over different radius classes. Top panels: positive growth by condensation; central panels: negative growth by evaporation; bottom panels: resulting mean growth rate at selected time instant, computed on the entire population of droplets.}
		\label{fig:mono_cond_evap}
	\end{figure}

	In the monodisperse case, figure \ref{fig:mono_pdfs}, the distribution progressively enlarges on the side of sizes smaller than the initial radius, which was $15$ \textmu m. At about 8.54 $\tau$, inside the interaction zone, the numerical concentration of drops of 13 \textmu m is 100 times higher than in the cloud and the minimal radius is a bit lower than 11 \textmu m, while inside the cloud the minimal radius is slightly below 13 \textmu m. In the mixing layer,  the width of the distribution part associated to coalesced droplets is much wider. It is noticeable to observe  that collisions can happen between drops of radius different from the initial 15 \textmu m, e.g. between two drops slightly below radius 13 \textmu m or one drop of 11 \textmu m and another one of nearly 13 \textmu m, while inside the cloud portion collision	happen 	almost only between droplets that both are close to 15 \textmu m, meaning that the evaporation is much more intense inside the anisotropic portion of the system. As can be quickly appreciated also by looking at panels c) and d) of figure \ref{fig:mono_cond_evap}, figure that describes the processes of condensation and evaporation concomitantly taking place in both parts of the system. We will come back to these aspects below.
	
	By looking at the polydisperse distribution, see fig. \ref{fig:poly_pdfs}, which initially includes drops randomly positioned inside the cloud region with a uniform mass in the volume classes from 0.6 to 30 \textmu m, once again, we observe a more intense dynamics inside the mixing region as compared to the cloud. Concentration highly differentiates in time inside the interface: for instance, at large radii, the ones close to 30 \textmu m, the decrease is of three order of magnitude, see panel b) of figure \ref{fig:poly_pdfs}. Either in panels a) and b), one can appreciate the enlargement of the distribution up to radii around 38 \textmu m, the maximum radius reachable from the coalescence of two droplets of 30 \textmu m. However, in the cloud region, the growth by coalescence is accompanied by a robust condensation which is marginally present in the interface region beyond radii of 30 \textmu m. 
	
\begin{figure}[bht!]
		\centering
		{\bf Polydisperse droplets condensation and evaporation}
		\begin{subfigure}[t]{0.49\textwidth}
			\caption{\bf condensation, cloud region} \label{fig:B003_cloud_grow_rate_positive}
			\includegraphics[width=\linewidth]{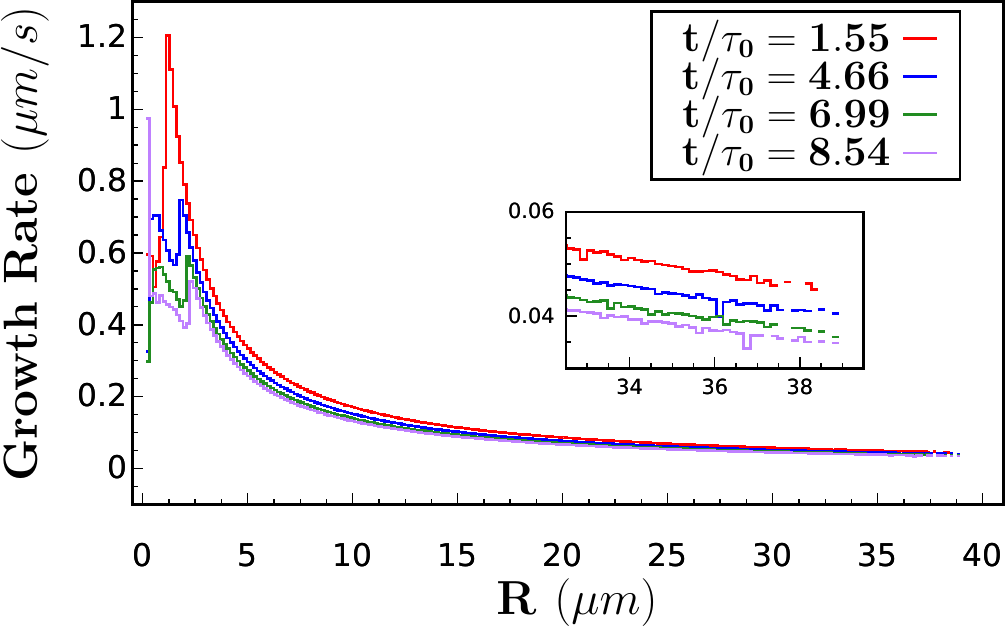} 	
		\end{subfigure}
		\hfill
		\begin{subfigure}[t]{0.49\textwidth}
			\caption{\bf condensation, interface region} \label{fig:B003_mixing_grow_rate_positive}
			\includegraphics[width=\linewidth]{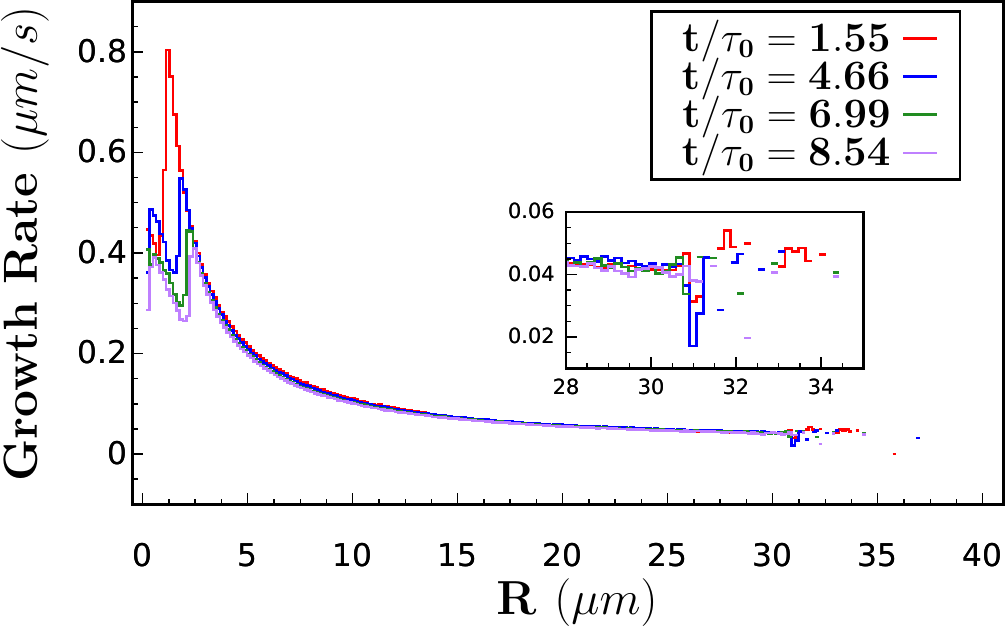} 	
		\end{subfigure}
		\vspace{0.5cm}
		\begin{subfigure}[t]{0.49\textwidth}
			\caption{\bf evaporation, cloud region} \label{fig:B003_cloud_grow_rate_negative}
			\includegraphics[width=\linewidth]{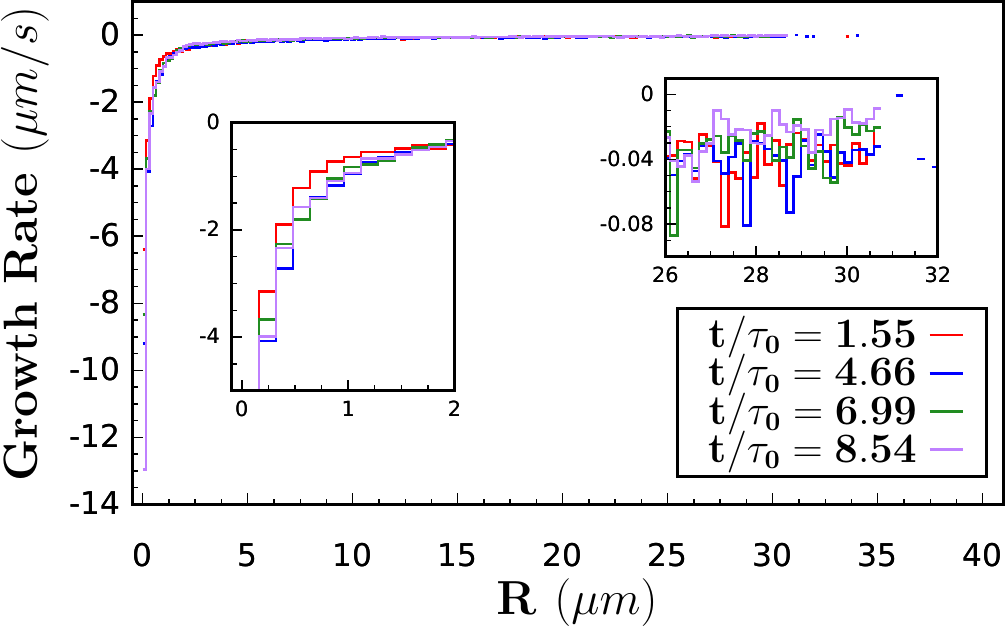} 	
		\end{subfigure}
		\hfill
		\begin{subfigure}[t]{0.49\textwidth}
			\caption{\bf evaporation, interface region} \label{fig:B003_mixing_grow_rate_negative}
			\includegraphics[width=\linewidth]{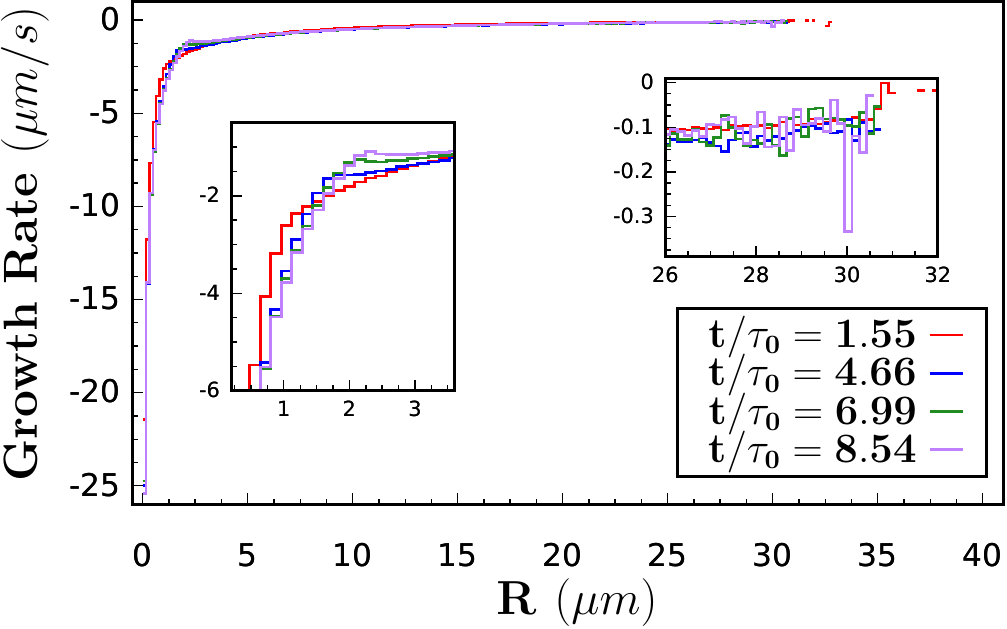} 
		\end{subfigure}
		\vspace{0.5cm}
		\begin{subfigure}[t]{0.49\textwidth}
			\caption{\bf cloud region: view near transient end} \label{fig:B003_cloud_grow_rate_selected}
			\includegraphics[width=\linewidth]{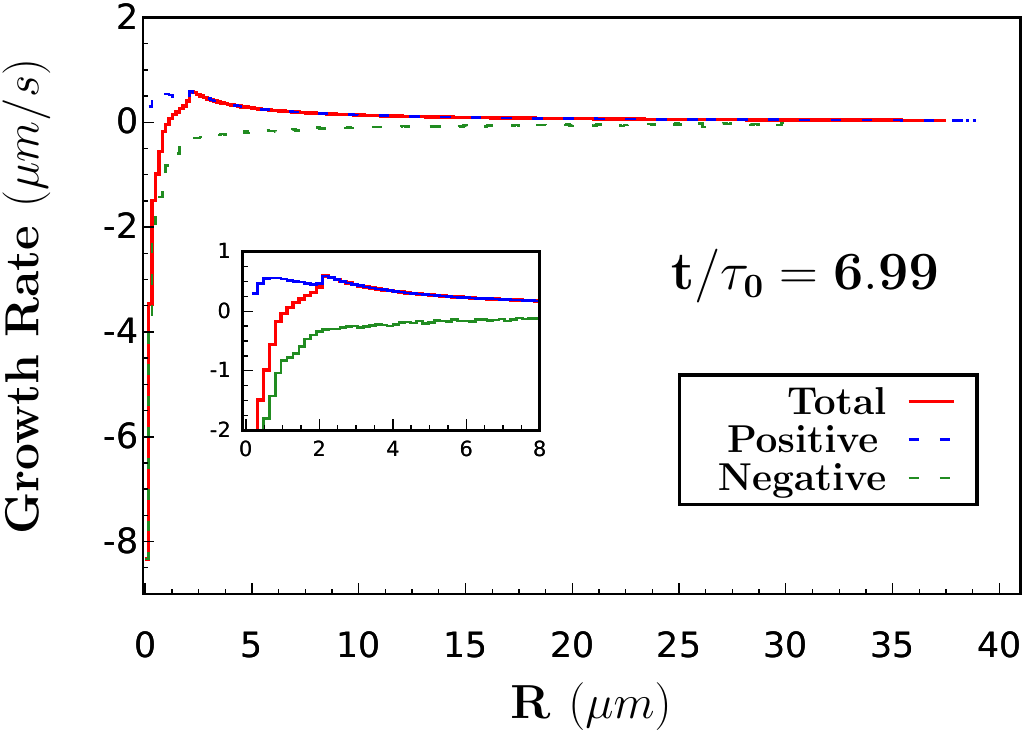} 	
		\end{subfigure}
		\hfill
		\begin{subfigure}[t]{0.49\textwidth}
			\caption{\bf interface region: view near transient end} \label{fig:B003_mixing_grow_rate_selected}
			\includegraphics[width=\linewidth]{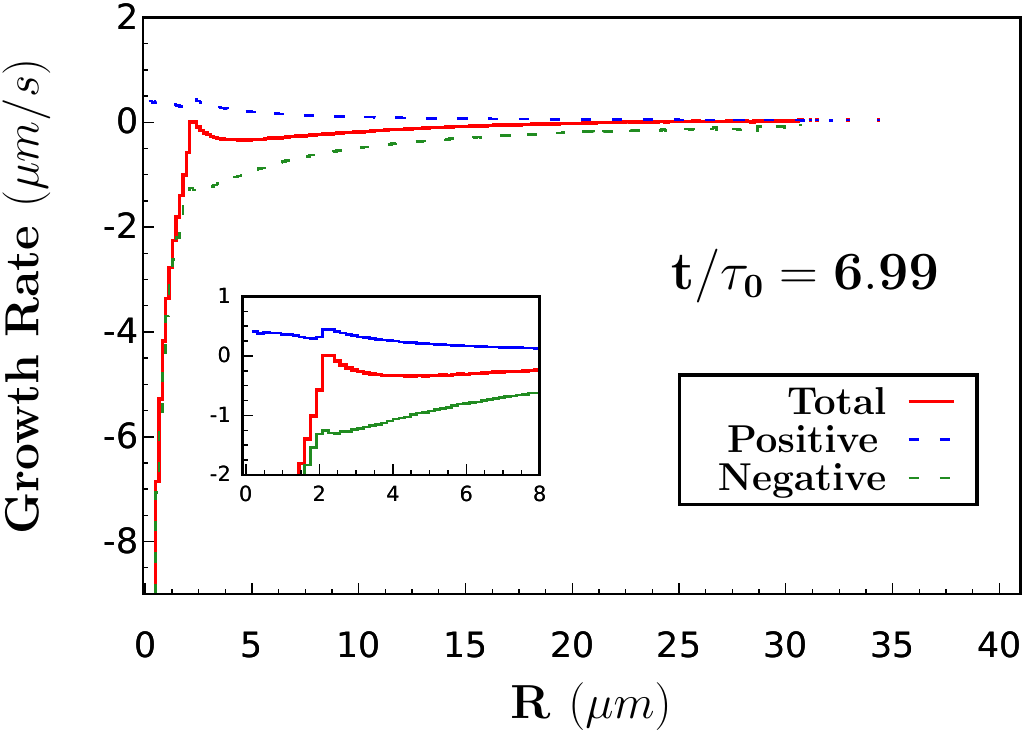}	
		\end{subfigure}
		\caption{\bf 
			Polydisperse drop size distribution, unstable and time decaying cloud clear-air interaction. Mean growth rate over different radius classes. Top panels: positive growth by condensation; central panels: negative growth by evaporation; bottom panels: resulting mean growth rate at selected time instant, computed on the entire population of droplets.}
		\label{fig:poly_cond_evap}
	\end{figure}

	Coming to figures \ref{fig:mono_cond_evap}, \ref{fig:poly_cond_evap}, we can discuss the different weight  that  condensation and evaporation  have in the temporal evolution of the system. From top to bottom, these figures present data on the positive growth of the ray (condensation), on the negative growth (evaporation) and on their combined effects in a given instant near the end of the transient at about 7 eddy rotation times. 
	
	Let us consider first, the mono-disperse population dynamics. In the left side of figure \ref{fig:mono_cond_evap}, one can see that inside the cloud portion the condensation action is present but milder than the evaporation, about 10 times less. Counter-intuitively, condensation is proportionally more intense on collided-coalesced drops, see right side in  panel a). Also, we observe a small range of radii ($15.125 - 15.25$ \textmu m) where condensation-evaporation balance perfectly, see panels c) and e). Now, by considering the interaction region, panel b), d), f), we can observe  the highest level of condensation for droplets close to 15 \textmu m and for the collided-coalesced droplet with  radii close 18.9 \textmu m. However, at these radii values, evaporation  balances condensation. Furthermore, evaporation becomes in time very important and generates drops as small as 11.8 \textmu m after 8.5 $\tau_0$, even if then kinetic energy inside the system is falling down by 18 times in the cloud region and by 6 times in the clear-air region. Notice that inside the shearless layer, evaporation is immediately active on collided particle, a thing which does not happen inside the cloud region. Overall, in the interface region, evaporation and collision prevail over condensation.
	
	In the polydisperse case, see fig. \ref{fig:poly_cond_evap}, the situation is different. In the cloud region, all along the transient,  condensation is prevailing on evaporation only in the radii range $2-6 \mu m$. Inside the interface layer, instead, evaporation always prevails on condensation. Even if, a near balance is reached at a radius of $3 \mu m$. 
	A more intense evaporation is active on smaller drops where the curvature effect (the negative Kelvin term in the radius growth rate equation (\ref{eq:radius_growth})) plays an important role. Here, we observe an evaporation rate about three times more intense than the condensation rate. One thing worth noticing is that in the interface both condensation and evaporation rates evolve non linearly in time, reaching the maximum around five eddy turn over times.
	
		\begin{table}[bht!]
		\caption{\textbf{Droplet size distribution trends during the transient decay inside the cloud and the interfacial layer}}
		\centering
		\resizebox{0.99\textwidth}{!}{
			\begin{tabular}{lccc}
				\toprule
				\textbf{CLOUD}	\\		
				\textbf{Quantity} & \textbf{Fit Law}	 & \textbf{Unit}\\
				\midrule
				\hspace{20mm}{\bf Initial Monodispersion}\\
				\midrule
				Standard deviation & $\sigma_{CM}(t) = 0.015 \, t/\tau_0 + 0.05$   & $\rm{\mu m}$\\
				Width $w$ at the $0.03\%$ of the probability density peak 
				&$w_{CM}(t) = 0.047 \,  t/\tau_0 - 0.006$  &  $\rm{\mu m}$\\
				\midrule
				\hspace{20mm}{\bf Polydispersion (initial uniform mass)}\\
				\midrule
				Standard deviation & $\sigma_{CP}(t) = -0.19 \, t/\tau_0 + 19.69$   & $\rm{\mu m}$\\
				Width at the $3\%$ of the probability density peak 
				&$w_{CP}(t) = 26.47 - 2 \, \exp (0.11  t/\tau_0) $  &  $\rm{\mu m}$\\
				\bottomrule
		\end{tabular}}
		\resizebox{0.99\textwidth}{!}{
			\begin{tabular}{lccc}
				\textbf{INTERFACIAL MIXING}	\\		
				\textbf{Quantity} & \textbf{Fit Law}	 & \textbf{Unit}\\
				\midrule
				\hspace{22mm}{\bf Initial Monodispersion}\\
				\midrule
				Standard deviation & $\sigma_{IM}(t) = 0.23 \, t/\tau_0 + 0.003$   & $\rm{\mu m}$\\
				Width $w$ at the $0.03\%$ of the probability density peak 
				& $w_{IM}(t) = 0.28  \, t/\tau_0 - 0.02$  &  $\rm{\mu m}$\\
				\midrule
				\hspace{22mm}{\bf Polydispersion (initial uniform mass)}\\
				\midrule
				Standard deviation & $\sigma_{IP}(t) = -0.74 \,  t/\tau_0 + 17.94$   & $\rm{\mu m}$\\
				Width at the $3\%$ of the probability density peak 
				& $w_{IP}(t) = 16.62 - 17.23 \, \exp (0.67  t/\tau_0) $  &  $\rm{\mu m}$\\
				\bottomrule		
	\end{tabular}}
\vspace{2mm}
\small
Suffices: $CM$ cloud mono, $CP$ cloud poly, $IM$ interface mono, $IP$ interface poly
			\label{tab:Table3}		
	\end{table}
	
	\normalsize
	\vspace{-2mm}
	\subsection{\textit{\textbf{A comment on the droplet size distribution structure.}}}
	
	
	It is important to remember that the system longterm state is that of a residual turbulence intensity spatially nonuniform  where a general time reduction of the collision rate should be expected. It must be also noted that the information conveyed by the drop size distributions (figures \ref{fig:mono_pdfs}, \ref{fig:poly_pdfs}) are not sufficient to highlight the quantitative details of the condensation-evaporation processes which are instead visible from the analysis shown in previous figures \ref{fig:mono_cond_evap}, \ref{fig:poly_cond_evap}. 
	
	However, size distribution shape  variation are useful to get an overall view of the population evolution. In the following figures \ref{fig:mono_pdf_structure} and \ref{fig:poly_pdf_structure}, we focus on the distribution shape, width, position and value of the maximum for radii range where condensation and evaporation dominate. We neglect the coalescence between large particles that leads to radii larger than 18 $\mu$m in the monodisperse case and larger than 31 $\mu$m in the polydisperse case.  
	
	In the monodisperse case,  see figure \ref {fig:mono_pdf_structure}, both inside the cloud region and the interface layer, size distributions are highly skewed. See panels a) and b), where they are shown in the last part of the observed transient. The distribution width is greater in the mixing layer than inside the cloud region, at 7.8 $\tau_0$ the standard deviation is 11 times larger. We measured the time scale of the drop size standard deviation growth and, to enrich the information on the shape, we measured it also at a given percentage of the probability density peak ($0.03 \%$), see panels c) and d). Inside the interface, with respect to the cloud region,  the growth of standard deviation is 15 times faster, while  at the $0.03 \%$ of the probability density peak, the growth is 6 times faster. The drop radius at the distribution peak slightly grows in time, more in the interface than in the cloud; while the value of the  peak decreases more rapidly (4.5 times) in the interface, see panels e) and f) in fig. \ref{fig:mono_pdf_structure}.
	
	In the polydisperse case, see figure \ref{fig:poly_pdf_structure}, trends are reversed. The concentration distributions are now skewed in the opposite way, see panels a) and b), where the distribution shape is shown near the end of transient, again at about 7.8 eddy turn over times. The width of distributions shrinks in time, more quickly (about 4 times) inside the interface region, see panels c) and d) where we include the information on the exponential variations of the distribution width at a concentration corresponding to the $3\%$ of the probability density peak value.  In the mixing, the exponential decay is 6 time faster. At distribution peak, the drop radius grows in time more or less in the same way both inside the cloud and the mixing region, while the value of the related concentration grows in the cloud and remains nearly  constant in the interface.
	
	By equating the time variations of the standard deviations of the monodisperse and polydisperse size distributions, see table \ref{tab:Table3} and the captions of  figures \ref{fig:mono_pdf_structure} and  \ref{fig:poly_pdf_structure}, we can estimate the time required by the two populations to reach a same width under the evaporation and condensation processes. {\bf In the cloud region},  which is a homogeneous isotropic time decaying  turbulence, {\bf the estimate is about} ${\bm {100 \tau_0}} $. Note, at this time, the turbulence intensity will be reduced to about one hundredth of its initial value. {\bf In the interface region}, which is an anisotropic and very intermittent,    {\bf the estimate is of about $\bm{18.5 \tau_0}$, i.e. more than 5 times faster.  A remarkable acceleration of these processes is therefore observed in the shear-free mixing layer separating the cloud from the sub-saturated  environmental air.}
	
	This result is somehow counter intuitive since it is observed despite the fact that beyond the temporal decay of the turbulence, present in the whole system, the interface also hosts the spatial decay of the kinetic energy. We explain this behavior in terms of the turbulence small scale anisotropy and intermittecy peculiar of the interface layer. This is characterized by a large departure from the typical values of the isotropic condition of the longitudinal velocity derivative moments in the directions across and parallel to the layer. It has been demonstrated that longitudinal derivatives in the energy gradient direction are more intermittent, while the intermittency is milder in the orthogonal directions, \cite{Tordella2011}. This structure of anisotropy is such that the skewness departure from isotropy reduces the contraction of fluid filaments parallel to the mixing layer and enhances that of the filaments orthogonal to it.  A possible  interpretation is that filament contraction across the interface enhances the collision rate and the local  supersaturation, thus enhancing  condensation of coalesced droplets, while the concomitant relative expansion of fluid filaments parallel to it enhances the evaporation. On the other hand, the large scales of turbulence cannot greatly influence the evolution within the mixing layer. They vary little in this type of simulation. First, because the computation domain is fixed. Second, because outside the mixing region both the ratio of the large scales and the ratio of the kinetic energies slowly vary in time, \cite{Tordella2006}.

	\begin{figure}[bht!]
		\centering
		{\bf Monodisperse population size distribution}
		\includegraphics[width=0.99\textwidth]{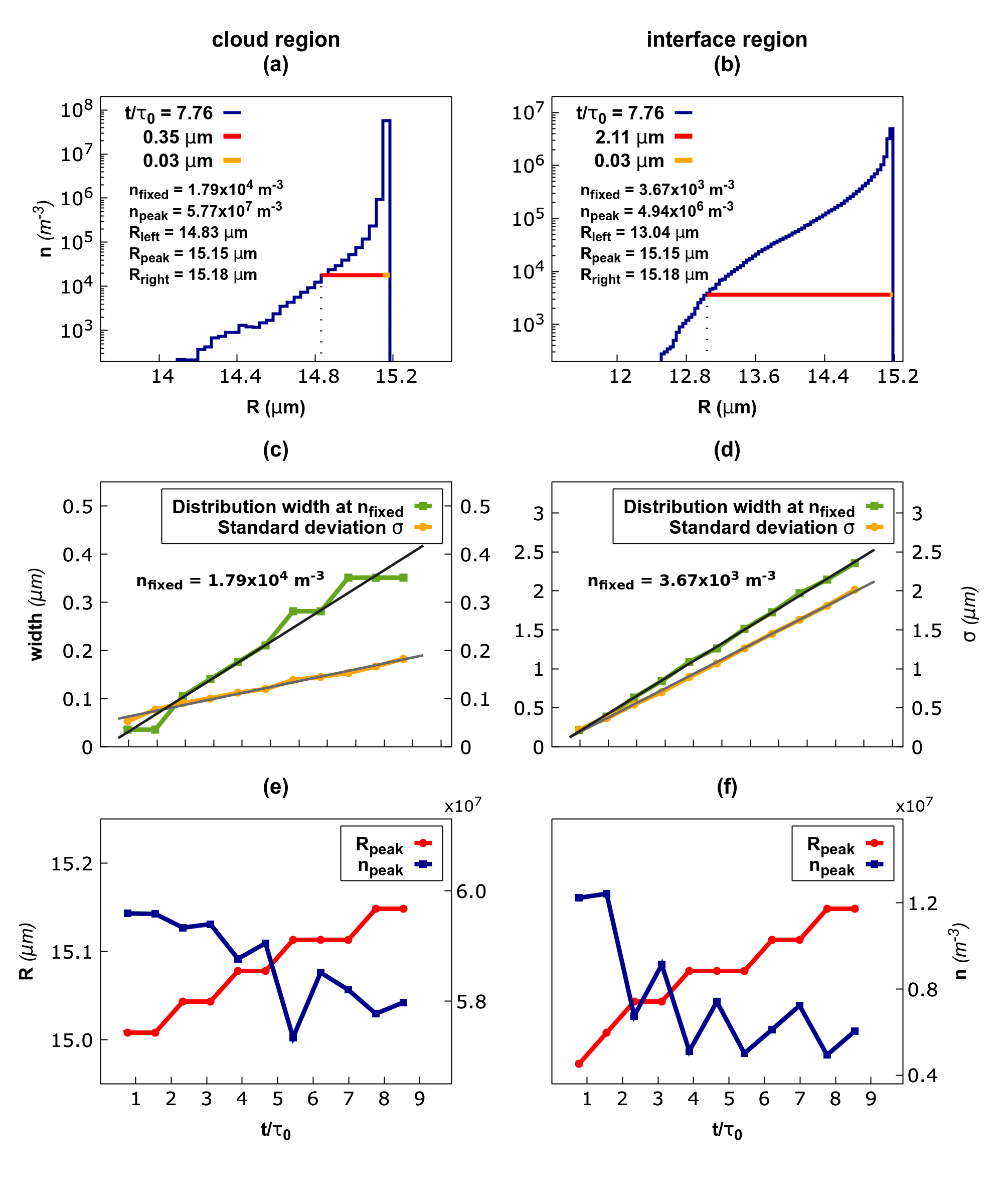}
		\caption{\bf Mono-disperse drop size distribution, unstable and time decaying cloud clear-air interaction. Distribution characteristics. From top to bottom: left and right part of the distribution with respect to peak value for selected time instance(a,b); change of the distribution width over time (green) and its fit(black, $0.047\;(t/\tau_0)-0.006$ in cloud and $0.28\;(t/\tau_0)-0.02$ in mixing), standard deviation of the distribution over time (orange) and its fit (gray, $0.015\;(t/\tau_0)+0.05$ in cloud and $0.23\;(t/\tau_0)+0.003$ in mixing) (c,d); change of peak distribution value (blue) and corresponding radius class (red) over time (e,f).}
		\label{fig:mono_pdf_structure}
	\end{figure}
	
	\begin{figure}[bht!]
		\centering
		{\bf Polydisperse population size distribution}
		\includegraphics[width=0.99\textwidth]{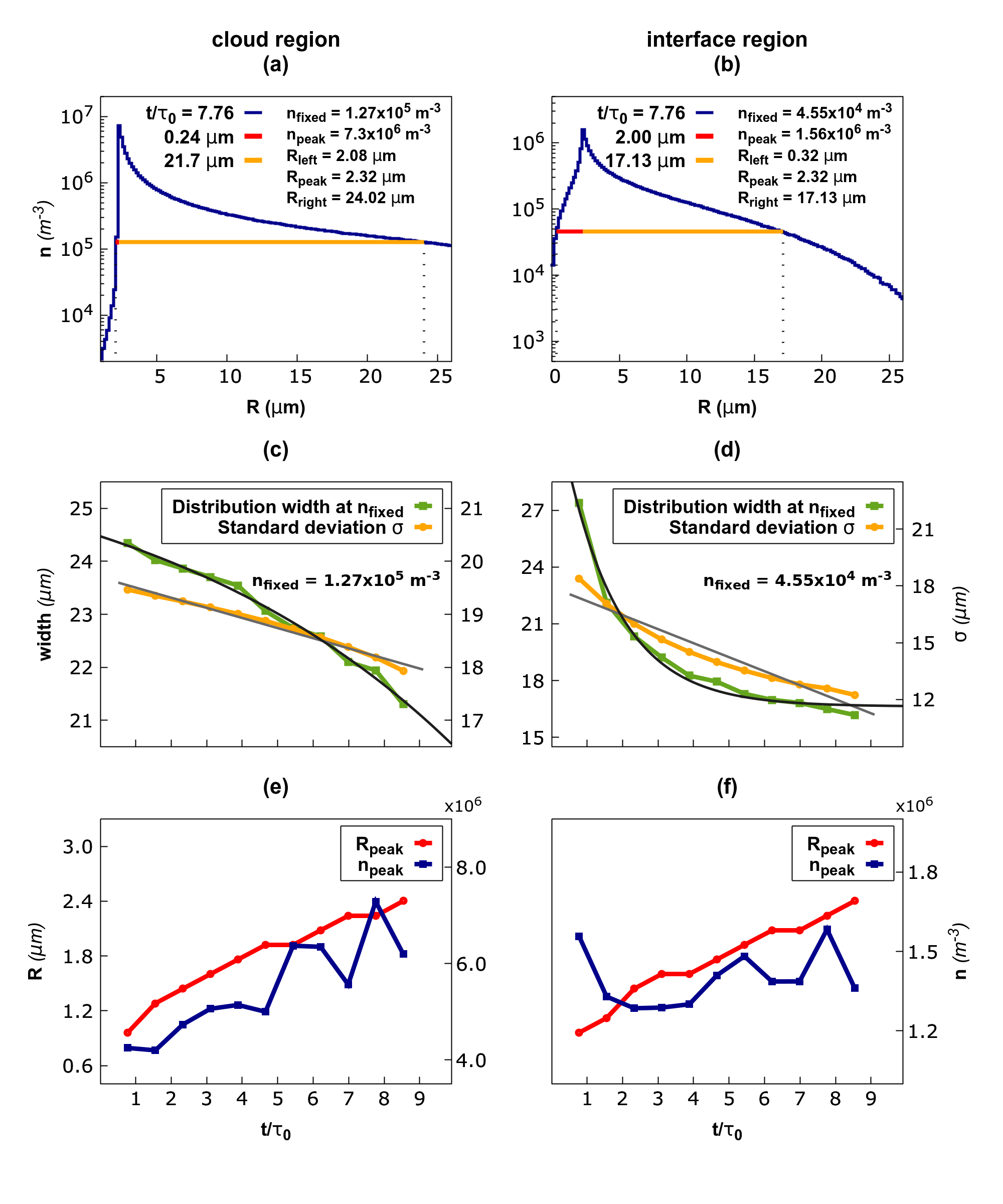}
		\caption{\bf Poly-disperse drop size distribution, unstable and time decaying cloud clear-air interaction. Distribution characteristics. From top to bottom: left and right part of the distribution with respect to peak value for selected time instance(a,b); change of the distribution width over time (green) and its fit(black, $26.47-2\exp\:(0.11\:(t/\tau_0))$ in cloud and $16.62-17.23\:\exp\:(-0.67\:(t/\tau_0))$ in mixing), standard deviation of the distribution over time (orange) and its fit (gray, $-0.19\;(t/\tau_0)+19.69$ in cloud and $-0.74\;(t/\tau_0)+17.94$ in mixing) (c,d); change of peak distribution value (blue) and corresponding radius class (red) over time (e,f).}
		\label{fig:poly_pdf_structure}
	\end{figure}

	\vspace{-3mm}
	\section{\textit{\textbf{\textcolor{blue}{On the feasibility of an approximate determination of the collision coalescence kernel within both homogeneous and inhomogeneous time decay turbulences}}}}

		A collision kernel function is the factor  within the aggregation integral term present in a typical Population Balance Equation (PBE) of drops of water, i.e. a model equation that aims to describe the dynamics of droplet size distributions, \cite{Kostoglou1994}, \cite{vanni1999}, \cite{Aiyer2019}. In the economy of this work, the main focus is not on the kernel question. We simply want to open the discussion on problems that arise when we consider realistic inhomogeneous turbulence conditions in a phase of rapid decay not matching the time scales typical of the micro-physics of droplet population.
		In case of substantial flow temporal transient where the intensity of turbulence decays of more than $90\%$ in less than 10 eddy turn over times, this implies that kernel depends also on  the kind of initial droplet size distribution.	In fact, on the one hand, the action of turbulence in favoring the collision of droplets is fading, on the other hand, there is not enough time for the population to reach the asymptotic state relevant to the set of physical parameters operating in the system (super / sub-saturation, temperature stratification, total liquid water content, Reynolds number). This longterm is in fact reached in $18.5 \tau_0$ in the interface and in $100 \tau_0$ in the cloudy region, see the estimates presented in Section 3.2. More, it must always be remembered that turbulence in lukewarm clouds has a global time scale of nearly 100 seconds, only. Therefore speaking of asymptotic conditions can be meaningless. Their dynamics must be conceived as a continuous succession of transients, one different from the other.

		Given this overall picture, it would be appropriate to extend the concept of collision kernel by recognizing its explicit temporal and spatial dependence. Regarding the spatial dependence, it must also be recognized that within the cloud-clear air interface the turbulence while spatially decaying also manages to significantly accelerate the evaporation and collision processes compared to what happens inside the cloud, here represented as homogeneous.
	
	In literature, the common scenario for studies of turbulent flows laden with solid  particles or liquid droplets is the steady-state homogeneous isotropic turbulence. In this situation the kernel is not time dependent for two reasons: - because the turbulence, as well the related control parameters, is steady, - because any kind 
	initial drop size distribution has the time to get the asymptotic configuration pertaining to the, super or sub, water vapor saturation,  and to the liquid water content condition present in the 
	system. 	Related contributions, including the context of atmospheric cloud physics, are numerous and by now a few of them became historically important. For reviews on the subject, readers can  refer to \cite{Wang2009}, \cite{Grabowski2013}, \cite{Devenish2012}.

	The variations in particle concentration are far greater than would be expected from statistical considerations. This raises serious concerns about the utility of statistical models to represent particle-laden turbulent flows,  \cite{Eaton1994}). 
	
	The turbulent process for which we measure the collision kernel tries to mimic a real small initial  perturbation of the interface cloud clear-air which includes  a mild unstable stratification.Collisions are viewed as geometric since the  Stokes' drag  was included  in the momentum equation of the particle, however, droplet - droplet local aerodynamic interactions are not included. We assume a collision efficiency equal to unity. Another minor simplification is that the coalescence efficiency, which is defined as the ratio of the number of actual merged drops and the total number of collisions, is taken equal to unity (\cite{Woods1965} and \cite{Beard2002}). IThe initial liquid water content $(LWC)$ is 0.8 g~m$^{-3}$, a  close value to the typical adiabatic value found in cumulus clouds. 
	We computed the collision kernel from our simulation as: 
	\begin{equation}
	\Gamma_{FS, SPP}(\rm{R}_1,\rm{R}_2;  t, \mathcal{V}_r)=\frac{N_\mathrm{coll}}{n_1 n_2}\frac{\mathcal{V}_r}{(t_2-t_1)},
	\end{equation}
	
\noindent where $FS$ means flow structure, $SPP$ means relevant set of physical parameters $N_\mathrm{coll}(\rm{R}_1,\rm{R}_2, t\in[t_1,t_2])$ is the number of collisions between droplets of radius $\rm R_1$ and $\rm R_2$, occurred during a selected time window $[t_1, t_2]$ and within a selected spatial region of volume $\mathcal{V}_r=L_1\times L_2\times\Delta x_3$. In the denominator, $n_1$ and $n_2$ are the numbers (counters) of {\it \bf all droplets} within the class size where $\rm{R}_1$  and $\rm{R}_2$ belong, for the same temporal range and spatial volume. See, for example, equation (\ref{eq.T}) in \cite{vanni1999}. All counters are obtained from a uniform radii discretization.
	
\begin{figure}[bht!]
		\centering
		\begin{subfigure}[t]{1.0\textwidth}
			\includegraphics[width=\linewidth]{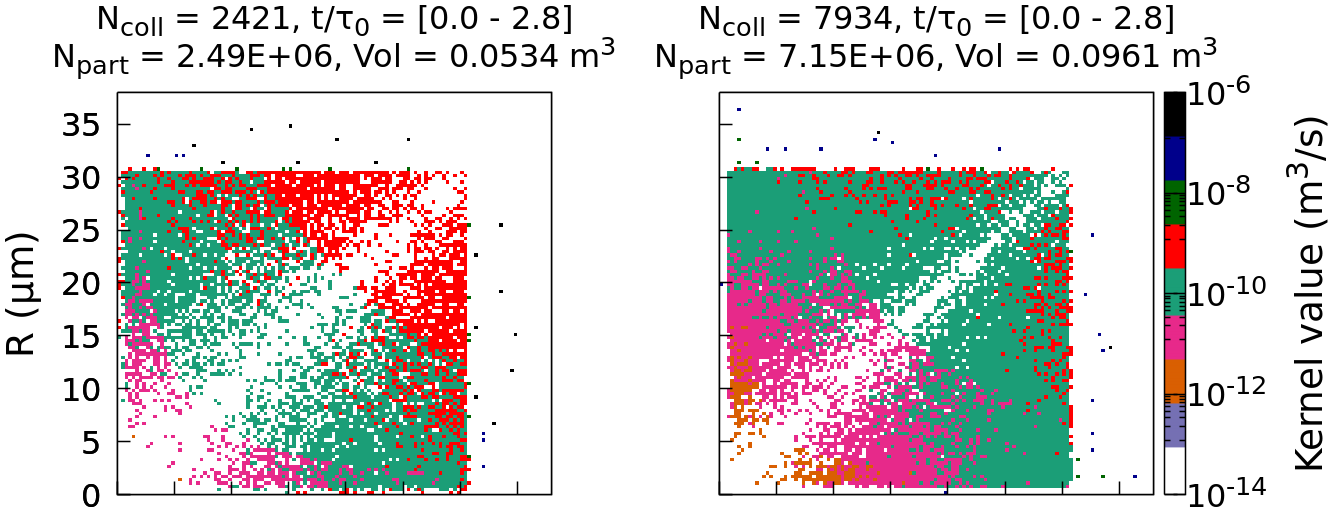} 	
		\end{subfigure}
		
		\vspace{0.3cm}
		
		\begin{subfigure}[t]{1.0\textwidth}
			\includegraphics[width=\linewidth]{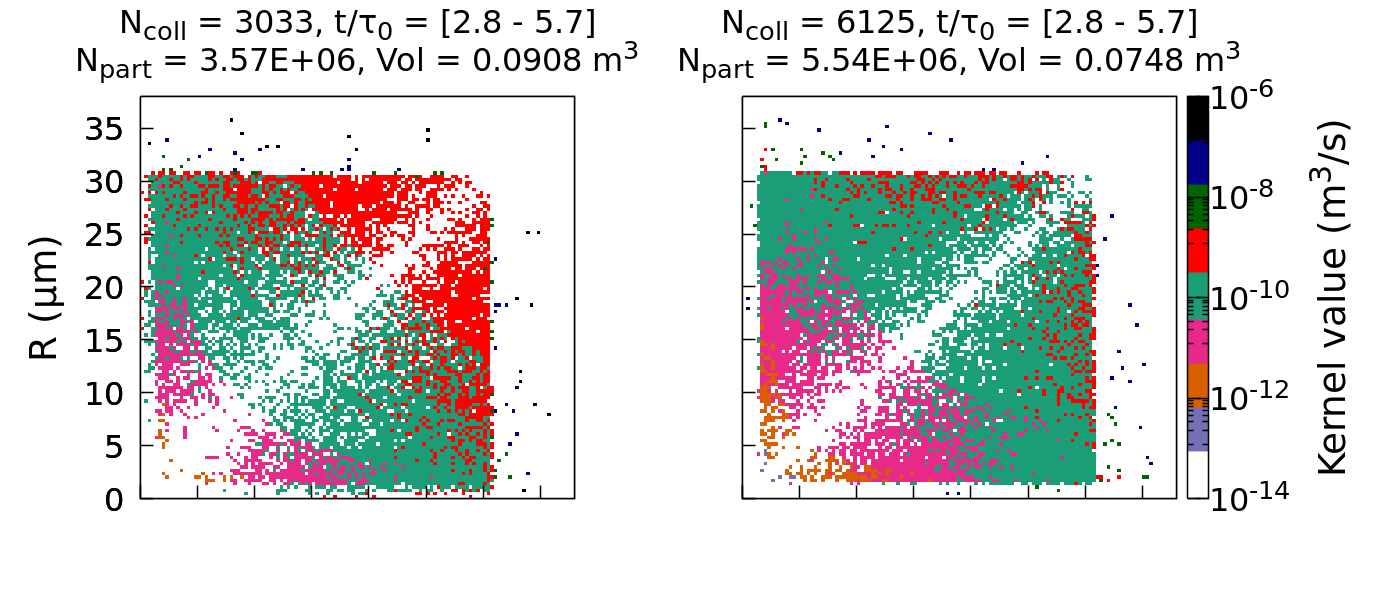} 	
		\end{subfigure}
		
		\vspace{0.3cm}
		
		\begin{subfigure}[t]{1.0\textwidth}
			\includegraphics[width=\linewidth]{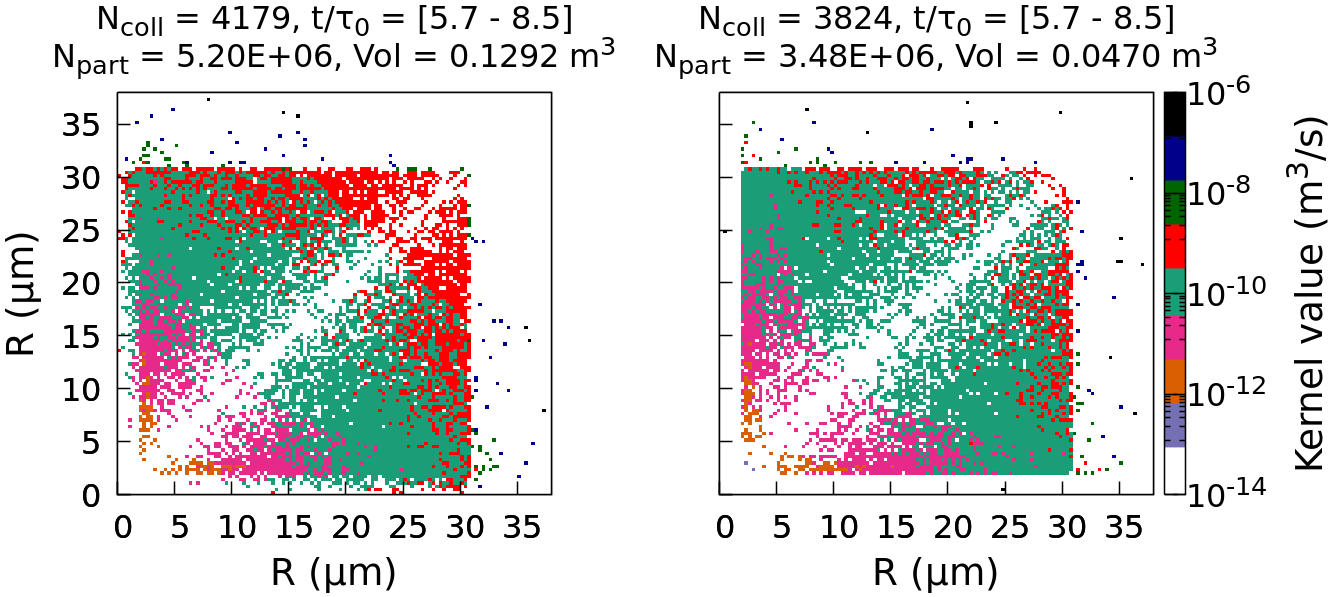} 	
		\end{subfigure}
		\caption{\bf Polydisperse drop size distribution, unstable and time decaying cloud clear-air interaction. Comparison of kernel value evolution inside the cloud-clear air interface (left) and the homogeneous cloud region (right). Ensemble average obtained over three realizations of simulation data, mean evolution over time intervals as long as one third of the entire observed decay. Collision radii subdivided into 256 classes.}
		\label{fig:kernel_poly_time_x3}
	\end{figure}

	Let us start the results description by considering in figure \ref{fig:kernel_poly_time_x3}  the collision kernel values for the polydisperse population computed inside time intervals as wide as one third of the transient decay.  In this case, given the concomitant presence of very different droplets, the volume ratio between the largest to smallest is of the order of 1.25$\cdot 10^5$, thus the number of collision will be large. Out of $10^7$ total droplets, we in fact observe about 5$\cdot 10^4$ collisions over about 10 physical time scales.In the left column we have the interface values, in the right column the cloud values, drop radii are classified in 256 ranges. The top panels, the first third of the transient, own about 10400 collisions on 10 million of drops. About one fourth of collisions take place inside the interface. One can appreciate that practically anywhere inside the pixelated matrix the values of the kernel values inside the interface are higher than in the cloud portion. Kernel value levels are not sharply contoured. In fact, we can see wide portions of the matrix where an intense and discrete (pixelated) merging of values that differ by one or more orders of magnitude is observed. This remain true for the other two thirds of the transient where the main difference is the increase of the number of collisions inside the interface at the expenses of the number of collisions inside cloud. At the end of the transient, bottom panels,  the collisions inside the interface are more numerous that inside the cloud (4179 versus 3824). In figure \ref{fig:kernel_poly_time_x3}, outside the initial drop radii area $[0 - 30] \mu m $ x $[0 - 30 ] \mu m $, we can see points (pixels) that represent drops resulting from a possible  double or triple sequence of collisions, see also figures \ref{fig:poly_pdfs} and \ref{fig:poly_cond_evap}. Values here are maximal ($1 \cdot 10^6$). 
	More interesting, however, is the situation inside the interface region which is expanding both in the simulation and in the real system. Here, notwithstanding the intense energy decay, see figure \ref{fig:initial_conditions}, the absolute number of collisions inside the interface layer grows, while the volume density of collision slightly decay of nearly a $30 \%$. We will come back later on this aspect by commenting on the collision correlation with the velocity and passive scalar fluctuation intermittency. 
	
\begin{figure}[bht!]
		\centering
		\begin{subfigure}[t]{1.0\textwidth}
			\includegraphics[width=\linewidth]{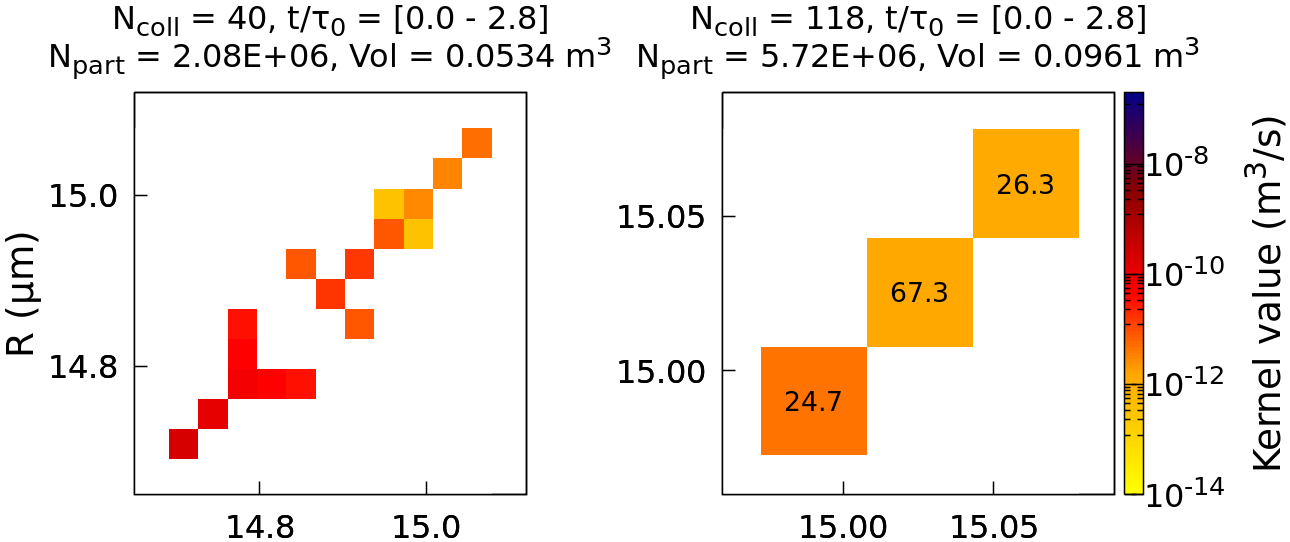} 	
		\end{subfigure}
		
		\vspace{0.3cm}
		
		\begin{subfigure}[t]{1.0\textwidth}
			\includegraphics[width=\linewidth]{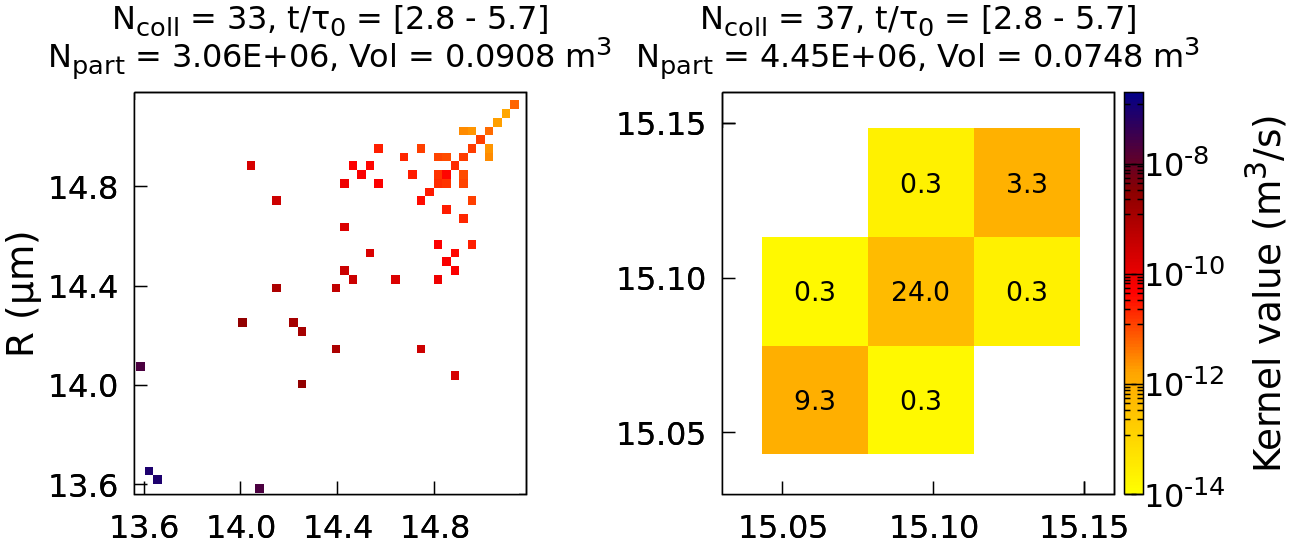} 	
		\end{subfigure}
		
		\vspace{0.3cm}
		
		\begin{subfigure}[t]{1.0\textwidth}
			\includegraphics[width=\linewidth]{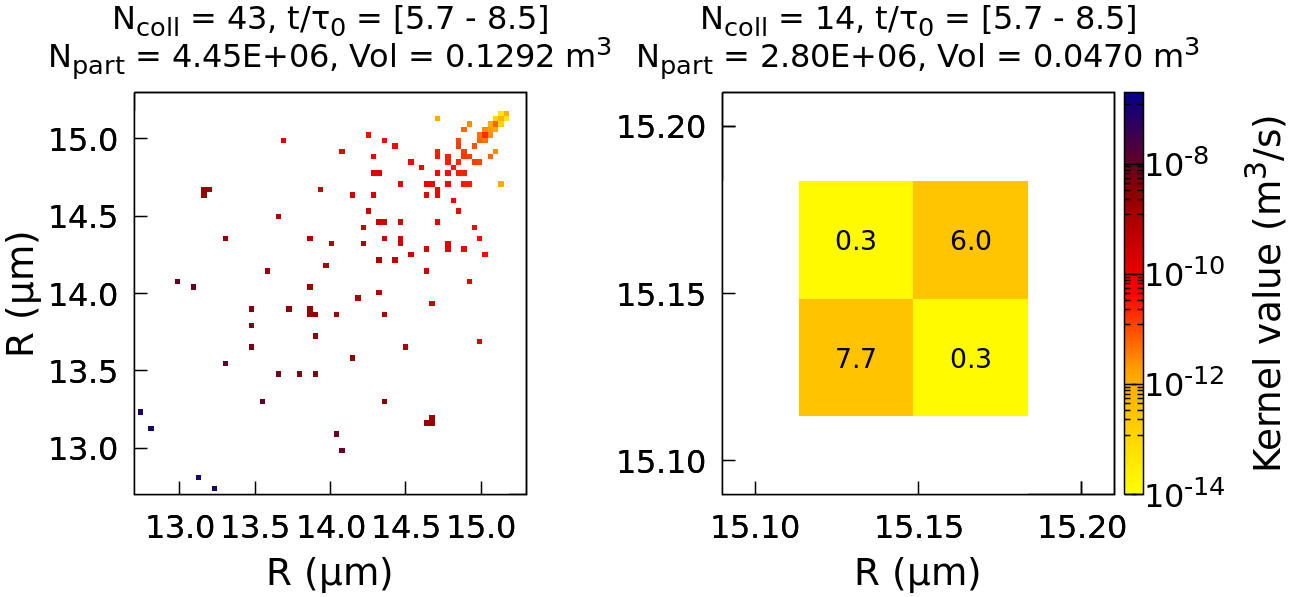} 	
		\end{subfigure}
		\caption{\bf Mono-disperse drop size distribution, unstable and time decaying cloud clear-air interaction. Comparison of kernel value evolution inside the interface region (left) and the cloud region (right). Ensemble average obtained over three realizations from  simulation data.}
		\label{fig:kernel_mono_time_x3}
	\end{figure}

	Considering now the situation for the initially monodisperse drop population, we observe a dramatically lower number of collisions - a thing attended given that initially drops are identical. See figure \ref{fig:kernel_mono_time_x3}, where the total number of collision along the entire transient is about 400 out of the 7 million of drops introduced in the system to reach the total liquid water content for  warm cloud ($LWC = 0.8 $  g/$m^3$). Inside the cloud region, the number density of collisions decays of the $76\%$ along the transient.  This happens concurrently with the decay of the kinetic energy of $ 92 \% $. The absolute number remains instead constant inside the expanding interface region where drops undergo a rapid evaporation. This corresponds to a decay of the $50 \%$ in term of the number density concurrently with a $86\%$ decay of kinetic energy. 
	Actually, the information that can be derived from this analysis is the diagonal and lateral spreading on the radii range where information is available. From this set of simulations and the actual ensemble averaging over three samples, as preliminary information, we can deduce a diagonal spreading of about $18\%$ per eddy turn over time, and a lateral spreading of $25\%$. To put forward a simulation campaign leading to an ensemble averaging based on a number of collision events of the order of a few thousands, a number of realizations of the order of 100-200 is needed.
	
	By observing the temporal evolution of the polydisperse population within shorter intervals, the kernel morphology disclosed by the 256 radii classes computation appear to be layered. 
	The peak values are concentrated in the lateral corners where the collisions take place between the smallest and the largest droplets. Intermediate values pertain to collision between large drops. Minimal values to collisions between small drops. Zero probability for collisions among same radius drops, for any radius value. This trend apply both to the interface and the cloud regions. But in proportion values inside the cloud homogeneous region are lower, in general, by less than one order of magnitude. A reasonably sufficient number of realizations to get a statistical base of a few $10^4$ events would be 10-20.

	Eventually, we would like to briefly discuss the previous results compared to the very popular theory of Saffman and Turner (1955) \cite{saffman1955}, hereinafter referred to as the ST model. This model is still a reference of general interest in the field of the engineering of multi-phase turbulent flow systems. The Saffman and Turner model holds for a background turbulence which is steady state, homogeneous and isotropic. A situation thus far from the system conditions we are studying here. That is a situation characterized by an  unstable density stratification and  transient decay of an inhomogeneous and anisotropic shearless turbulence  which is mimicking the interaction between a warm cloud portion and the clear air bounding it. Anyway, at present, the literature does not present kernel statistics for collisions hosted by an anisotropic turbulence in temporal decay and thus this kind of comparison can be useful to highlight differences between a near ergodic and a fully non ergodic system.
	
	\begin{figure}[bht!]
		\centering
		\begin{subfigure}[t]{0.8\textwidth}
			\includegraphics[width=\linewidth]{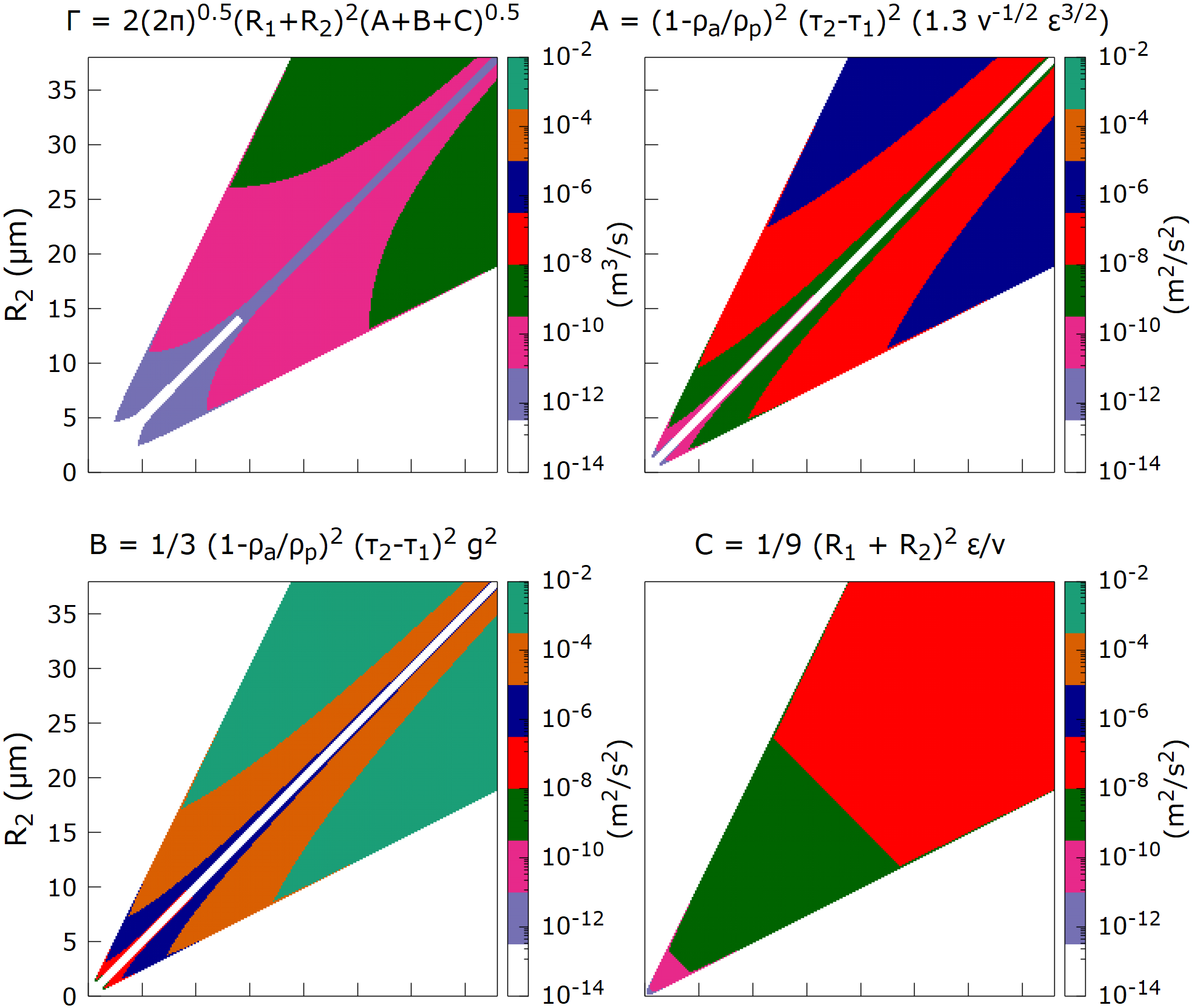} 			\end{subfigure}
				\vspace{0.3cm}
		\begin{subfigure}[t]{0.8\textwidth}
			\includegraphics[width=\linewidth]{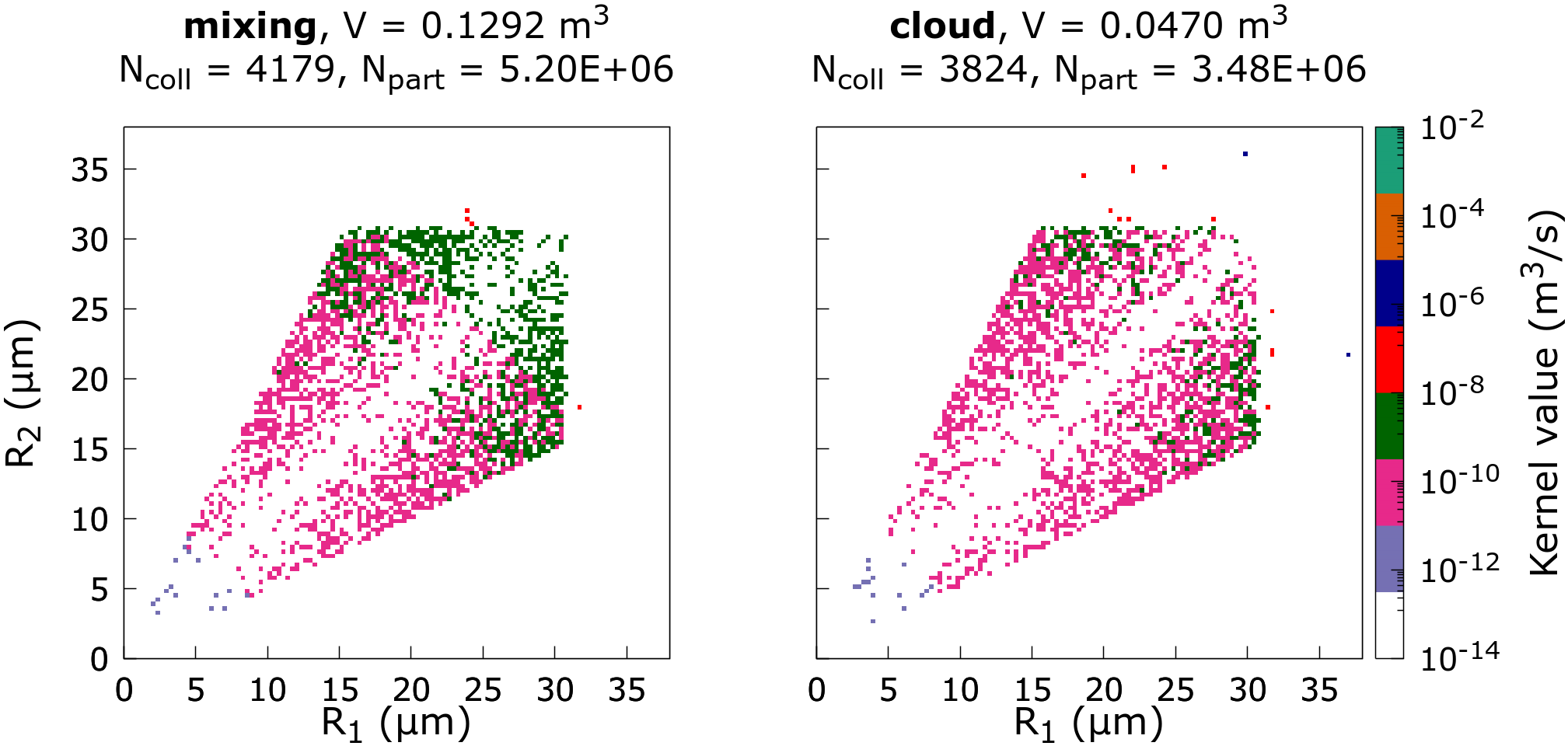} 	
		\end{subfigure}		
		\caption{\textbf{Comparison between the Saffman and Turner model (valid for steady state HIT: $\epsilon = 10\;cm^2/s^3$, $R_1, R_2\in[0, 38]\;\mu m$, $R_1/R_2 <= 2$ and $R_2/R_1 <= 2$) and our simulation (unsteady, inhomogeneous, with anisotropic small scale) on collision kernel values in a transient lapse where the dissipation has a comparable value, $t/\tau_0 \in[5.7, 8.5]$. Top left panel, kernel as deduced from  eq.10 and related not numbered eq.s in \cite{saffman1955}. Top right and middle panels, the three ST terms:  droplet motion relative to the air, droplet relative motion due to gravity, droplet motion with the air, respectively.  The portion of the $R_1, R_2$ graph where the model is valid is only considered. Bottom panels: kernel values for the  polydisperse population. Left, mixing interface, right, cloud region.}}
		\label{fig:kernel_saffman_vs_poly}
	\end{figure}
	
	The comparison is presented in figure \ref{fig:kernel_saffman_vs_poly},  where the  three contributions inside the ST model are contrasted, namely,  i) collision rate due to different particle inertia because of the action of the turbulent acceleration, term A, ii) action of gravity, term B, and  iii) collision rate due to the spatial variation of turbulence air velocity, term C.
	It should be recalled that ST model is not parameterized with the Reynolds number, which is anyway hypothesized very large. It can be noticed that in this model, for $\epsilon = 10$ cm$^2/$m$^3$ and an air temperature of $280$ K, the collision between drops moving with the air, term C, is playing a minor role with respect to terms A and B.  The two bottom panels of figure \ref{fig:kernel_saffman_vs_poly} show the comparison of the ST model  with the kernel computed for the polydisperse droplet population case. 
	The comparison is done inside a portion of the transient where the dissipation value is not far from the value inserted in the ST model. One can appreciate a difference in the kernel values, that in the simulation are generally lower (from a few persents and up to about 90-100\%) than the model (see the top left panel in fig. \ref{fig:kernel_saffman_vs_poly}). The shape of the kernel is different, similar more to a band than a butterfly morphology. However, in view of a future study, fully dedicated to obtaining quantitatively accurate values and morphology of the collision kernel under spatially non-homogeneous and time decay conditions, we believe that it would be necessary to conduct a large campaign of simulations. Aiming to obtain ensemble averages on a large set of samples.	 In the case of the widely polydispersed population in which the probability of droplet collision was conditioned to be high, we estimate a need of  ensemble averages based on 10-20 samples,  a number that should produce statistics based on a number of collision events larger than $ 10 ^4$. In the opposite case of monodisperse population, the number of samples for ensemble averages to attain statistics on  about $ 10 ^ 3 $ collision events should be of the order 100. 
	
	\vspace{-2mm}
	\subsection{\textit{\textbf{Small scale turbulent velocity fluctuation and collision count correlation.}}}
	
\begin{figure}[bht!]
		\centering
		\begin{subfigure}[t]{0.8\textwidth}
			\caption{Pearson's correlation index between air flow intermittency and collision count inside cloud} \includegraphics[width=\linewidth]{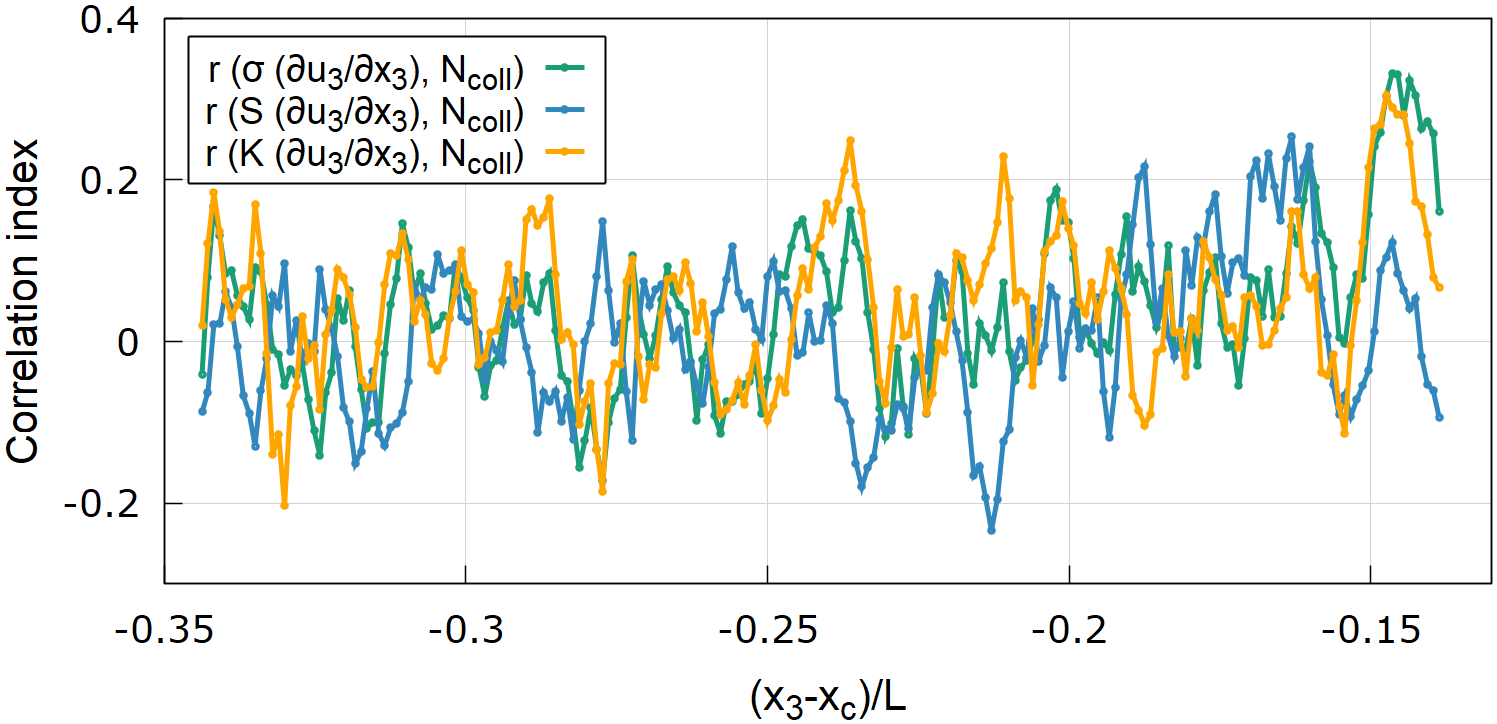}
		\end{subfigure}
		
		\vspace{0.1cm}
		
		\begin{subfigure}[t]{0.8\textwidth}
			\caption{Pearson's correlation index between air flow intermittency and collision count inside cloud interface} \includegraphics[width=\linewidth]{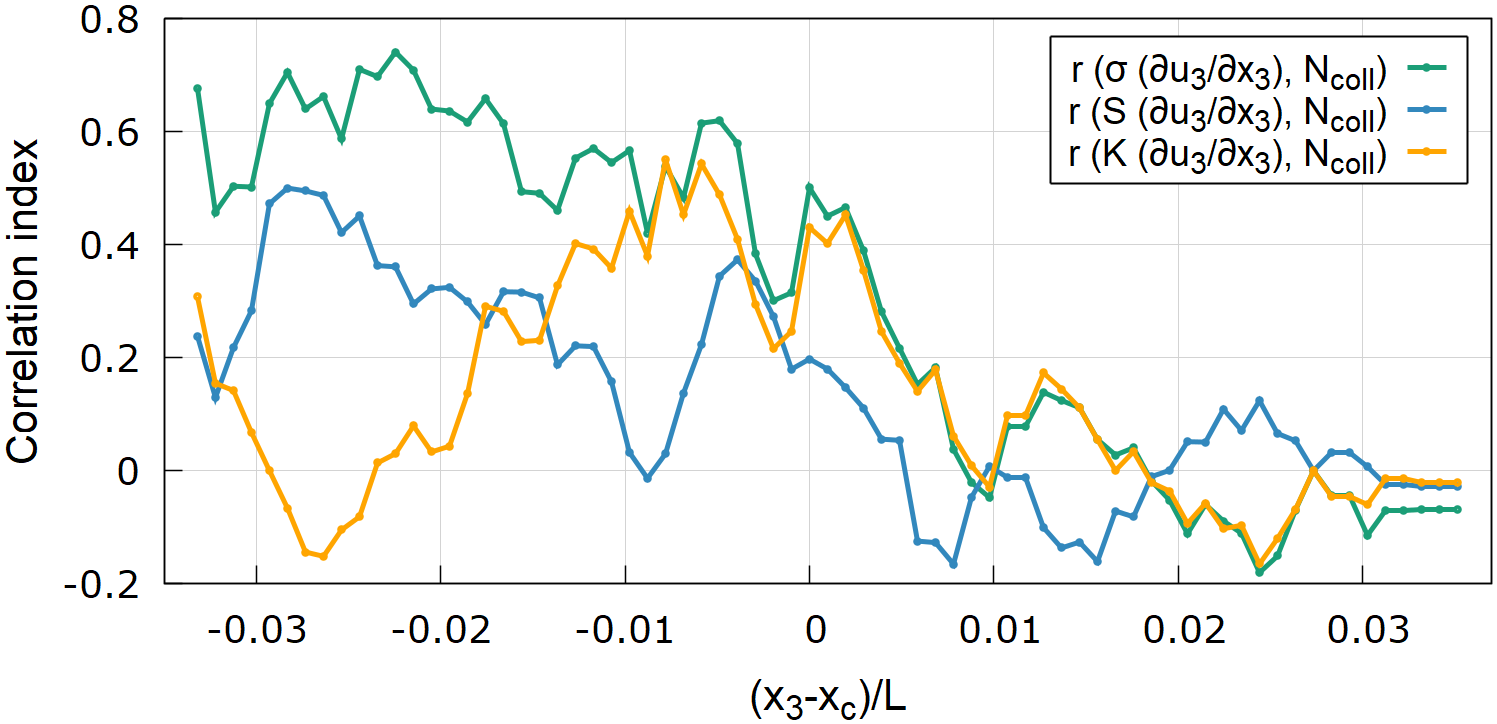}
		\end{subfigure}
		
		\vspace{0.1cm}
		
		
		\begin{subfigure}[t]{0.48\textwidth}
			\includegraphics[width=\linewidth]{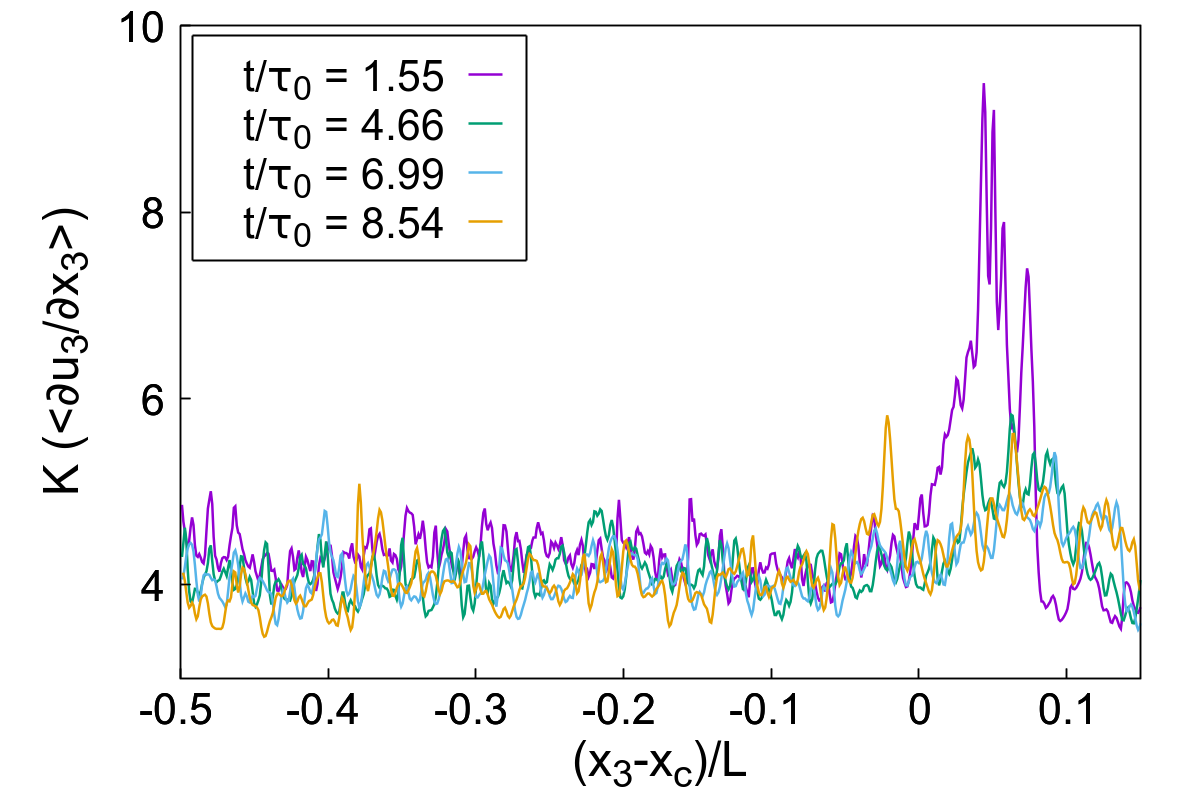}
		\end{subfigure}
		\hfill
		\begin{subfigure}[t]{0.48\textwidth}
			\includegraphics[width=\linewidth]{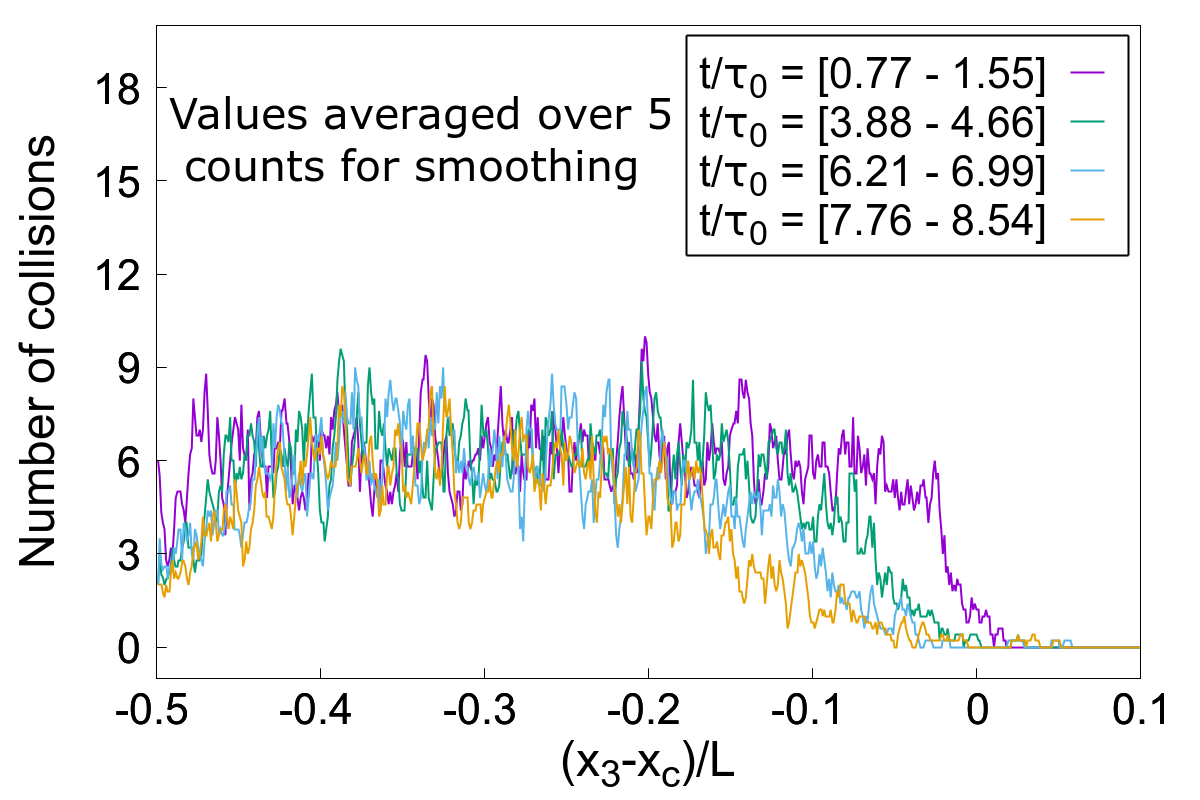}
		\end{subfigure}

		\caption{\textbf{Pearson's correlation index,  $r_{\Delta t}(x_3)= \sum_{i=1}^{N_t} (x_i - \overline{x(x_3)})(y_i - \overline{y(x_3)})/ 
				\sqrt{\sum_{i=1}^{N_t} (x_i - \overline{x(x_3)})^2} \sqrt{\sum_{i=1}^{N_t} (y_i - \overline{y(x_3)})^2}$, between  small scale intermittency of the turbulent velocity field and the droplet collision count ($N_{coll}$).  The correlation is showed via turbulence small scale anisotropy related quantities: standard deviation, skewness and kurtosis of the longitudinal derivative fluctuation $\partial u_3/ \partial x_3$.  $\Delta t$ is the observed transient length, $N_t$ is the number of turbulent velocity and droplet collision fields recorded along the transient, in this case $N_t=33$. Bottom panels: distribution along $x_3$ of the velocity longitudinal derivative Kurtosis and of the collision count. 
				}}
		\label{fig:corr_vel_coll}
	\end{figure}

	 To explain physically the observed acceleration of evaporation-condensation and collision inside the cloud interface, it is important to verify the correlation between the fine scale of the turbulence and the collision count for the the poly-disperse population transient evolution where we have a high rate of collision events. 
	 It should be noted, that in the shear-free transient decay, the turbulence large scale remains almost unchanged while inertial and dissipative small scale are widening because both kinetic energy and dissipation are decaying. The mixing layer width is concomitantly  growing  and a measure of the turbulence penetration in the sub-saturated ambient is given by the displacement of the maxima of the velocity field skewness and kurtosis. A  measure of the intermittency and anisotropy of the smallest scale in the mixing is obtained in terms of velocity derivative statistics, in particular in terms of the longitudinal derivative statistics. 
	We computed the variation along $x_3$, the direction across the interface, of the  correlation index (Pearson's product-moment correlation index) in the temporal window observed during the simulation. The correlation is shown in figure \ref{fig:corr_vel_coll}.  Inside the cloud region, the correlation oscillates about  zero but in the mixing it reaches the  value of  0.5 for both the velocity derivative skewness and kurtosis of the longitudinal component across the interface. The correlation value rises to  about 0.8 for the velocity derivative standard deviation. This result highly  support the interpretation that the relative fluid filaments compression across the interface foster collision among droplets. 
	and  highlight how water droplet growth by coalescence due to collision can still take place at the cloud border.
		
	\vspace{-3mm}
	\section{\textit{\textbf{Conclusions}}}

	The contribution of this study consists in having considered one of the possible shear-free transient interactions between cloud and clean air that are commonly present in the natural sequence of stages lasting  about 100 seconds that mark the rhythm of a cloud life. The natural system anisotropy including that of the small scale of the turbulence is taken into account. Two different types of water droplet populations have been considered. The case of a population containing drops initially having the same diameter, a situation in which collisions are in fact unlikely, and the case of a population of drops with very different radii (polydisperse, with uniform mass per class of radii),  which, on the contrary, is biased to host many collision events. In both cases we have included in the computational domain a number of drops of the order of 10 million, which matches the real liquid water content of warm clouds.
	
	The important clue we got from both the monodisperse and polydisperse population simulations is that the unsteady turbulence mixing confining the cloud region hosts a remarkable acceleration of the droplet population dynamics. In particular, the droplet evaporation and collisional activity is enhanced. In a time span where the kinetic energy of the air flow hosting the cloud is dropping of the 90\%, the collision activity reduces by the 40\% inside the cloud but rises by the 25\% in the interaction mixing with the clear air. For the initially monodisperse population, in the mixing layer, the size-distribution of the drop numerical density shows a growth of standard deviation 15 times faster of that in the cloud region. The drop radius of the distribution peak slightly grows in time, more in the interface than in the cloud; while the value of the  concentration peak decreases 4.5 more rapidly in the interface than in the cloud. In the polydisperse case,  trends are reversed. The concentration distributions are now skewed in the opposite way and the width of distributions shrinks in time, more quickly (about 4 times) inside the interface region than in the cloud. The drop radius of the distribution peak slightly grows in time, in the same way in both  regions; while the value of the peak grows in the cloud and stays almost constant in the interface.
	
	The observed acceleration of the population dynamics in the interface, the rapid differentiation of the size of the droplets due to the different weight that evaporation,  condensation and collision have in the highly intermittent mixing region can, at least in part, explain the rapid increase in the size of droplets that is observed in some formations of cumulus clouds, in particular the maritime ones, and is considered capable of locally inducing rainfall, \cite{Mason1961}, \cite{Li2020}. 
	
	These findings are observed despite the fact that beyond the temporal decay of the turbulence, present in the whole system, the interface also hosts the spatial decay of the kinetic energy. In this flow system, the large scales of turbulence vary little, because the computational domain is fixed and because the ratio of the large scales and the ratio of the kinetic energies between the cloudy and ambient air regions slowly vary in time, \cite{Tordella2006}. An inference can be made where the accelerated dynamics is associated to the small scale anisotropy and intermittecy peculiar of the interfacial layer. In fact, here, the small scale structure is characterized by a large departure of the longitudinal velocity derivative moments from those typical of isotropy. The longitudinal derivative in the energy gradient direction is more intermittent, while the intermittency is milder in the orthogonal directions.  The structure of the anisotropy is such that the skewness departure from isotropy reduces the contraction on fluid filaments parallel to the mixing layer and enhances that of the filaments orthogonal to it, \cite{Tordella2011}. We thus hypothetize that flow filament contraction across the interface enhances the droplet collision rate while the  relative stretching of fluid filaments parallel to it enhances  evaporation. Of course, this picture must also be associated with the high degree of non-Gaussianity of the supersaturation and density of water vapor within the interfacial layer.
	
	In a condition where the total water liquid content matches that of a warm cloud, our collision kernel analysis has shown a clear dependence on time and spatial regions where collisions take place. Thus an extension of the  concept of collision kernel is required for a transient  and inhomogeneous system in which turbulence is decaying faster than the proper  time scales of the aqueous phases involved. It is interesting to observe that, due to the dynamical acceleration inside the interface, an asymptotic state for the population droplet-size distribution  could be reached more easily inside the interface than inside the decaying cloudy region. For asymptotic state we mean the longterm state of the droplet population associated to a given structure of the background turbulent air flow, supersaturation, stratification and total water liquid content.	Our observation suggests that it may be more feasible to determine the kernel function within the interfacial region. Although it must be taken into account that in physical reality the boundaries of the clouds do not reach true asymptotic states as they are subject to a continuous sequence of transitory phases that are different from each other. And therefore the search for long-term statistics would not be very meaningful.

	The comparison with the Saffman-Turner model, valid for a population in conditions of stationary and isotropic turbolence, is partly positive. By placing ourselves in a condition where the number of droplets corresponds to the physical water liquid content of warm clouds and the dissipation of turbulent energy -  the only dynamical parameter present in that model - has a same value both in the model and in the numerical simulation, we observe  kernel values  below those of  Saffman-Turner (from a few percents and up to 90\%). The morphology is also different mainly inside the mixing region where a band structure more than a butterfly shape is visible. 
			
	\vspace{-3mm}

	\section{Acknowledgements}
	\noindent This project has received funding from the Marie-Sklodowska Curie Actions (MSCA ITN ETN COMPLETE) under the European Union's Horizon 2020 research and innovation programme. Grant agreement n°675675,  \url{http://www.complete-h2020network.eu}.\\
	We acknowledge the CINECA award HP10CA7H4X, under the ISCRA initiative, for the availability of high performance computing resources and support. Computational resources were also provided by HPC@POLITO, a project of Academic Computing within the Department of Control and Computer Engineering at the Politecnico di Torino (https://hpc.polito.it).
	
	\vspace{3mm}
	
	{\bf References}

	\bibliography{references-IJMF}
	
\end{document}